\documentclass{iopjournal}

\usepackage{amsmath}
\usepackage{multirow}
\usepackage{float}
\usepackage{amssymb}

\def\bea{\begin{eqnarray}}
\def\eea{\end{eqnarray}} 
\def\be{\begin{equation}}
\def\ee{\end{equation}} 
\def\nn{\nonumber}
\def\Re{\text{Re}}
\def\Im{\text{Im}}
\def\tev{\ensuremath{\mathrm{Te\kern -0.1em V}}}
\def\gev{\ensuremath{\mathrm{Ge\kern -0.1em V}}}
\def\mev{\ensuremath{\mathrm{Me\kern -0.1em V}}}

\newcommand{\hs}{\hspace{.4mm}}

\newcommand{\GeV}{\,\mathrm{GeV}}

\def\c9b2sll{\ensuremath{C_9^{32,22}}}
\def\c10b2sll{\ensuremath{C_{10}^{32,22}}}
\def\c9pb2sll{\ensuremath{C_9^{\prime\,32,22}}}
\def\c10pb2sll{\ensuremath{C_10^{\prime\,32,22}}}

\def \cqloneT {{$[\widetilde{c}_{ql}^{(1)}]^{ ij,22}\,$}}
\def \cqlthreeT {{$[\widetilde{c}_{ql}^{(3)}]^{ij,22}\,$}}

\def \czT {{$[\widetilde{c}_{z}]^{ij}\,$}}

\def \cqlonebsT {{$[\widetilde{c}_{ql}^{(1)}]^{ 32,22}\,$}}
\def \cqlthreebsT {{$[\widetilde{c}_{ql}^{(3)}]^{32,22}\,$}}
\def \cdlbsT {{$[\widetilde{c}_{dl}]^{32,22}\,$}}
\def \cqebsT {{$[\widetilde{c}_{qe}]^{32,22}\,$}}
\def \cdebsT {{$[\widetilde{c}_{de}]^{32,22}\,$}}
\def \czbsT {{$[\widetilde{c}_{z}]^{32}\,$}}
\def \czpbsT {{$[\widetilde{c}_{z^{\prime}}]^{32}\,$}}

\def \cqlonesdT {{$[\widetilde{c}_{ql}^{(1)}]^{21,22}\,$}}
\def \cqlthreesdT {{$[\widetilde{c}_{ql}^{(3)}]^{21,22}\,$}}

\def \czsdT {{$[\widetilde{c}_{z}]^{21}\,$}}

\begin{document}

\articletype{Paper} 

\title{Correlated $b \to s$ and $s \to d$ Rare Semileptonic Transitions \\[2mm] in the Standard Model Effective Field Theory}

\author{Nilakshi Das$^1$\orcid{0000-0002-5824-3094}, Rusa Mandal$^{1,*}$\orcid{0000-0001-8053-6881
} and Praveen S Patil$^{1}$\orcid{0000-0003-0793-4108}}

\affil{$^1$Department of Physics, Indian Institute of Technology Gandhinagar, Gandhinagar, India}


\affil{$^*$Author to whom any correspondence should be addressed.}

\email{rusa.mandal@iitgn.ac.in, nilakshi.das@iitgn.ac.in and praveen.patil@iitgn.ac.in}

\keywords{Standard Model effective field theory, flavor physics, new physics, rare $B$ decays}

\begin{abstract}
The persistent anomalies observed in $b \to s\,(\ell^+\ell^-,\,\nu\bar{\nu})$ transitions continue to provide strong motivation for exploring possible extensions of the Standard Model (SM). Motivated by these discrepancies, we present a comprehensive analysis of semileptonic flavor changing neutral current processes within the Standard Model Effective Field Theory (SMEFT), encompassing both $b \to s\,(\mu^+\mu^-,\,\nu\bar{\nu})$ and $s \to d\,(\mu^+\mu^-,\,\nu\bar{\nu})$ transitions. We perform a combined fit to  $b \to s\,(\mu^+\mu^-,\,\nu\bar{\nu})$ observables, allowing the relevant dimension-six Wilson coefficients to be complex. We find that the four-fermion operators involving left-handed quark and lepton doublets provide the preferred description of the current $b \to s$ data, while the electroweak operator modifying the $Z$-boson couplings also plays an important role in improving the fit.
We show that flavor-universal SMEFT couplings lead to strongly enhanced rare semileptonic kaon decay branching ratios, in conflict with current experimental bounds and thus motivating the implementation of Minimal Flavor Violation. In particular, we demonstrate that flavor-symmetric frameworks based on $U(3)^5$ and $U(2)^5$ naturally restore the required CKM hierarchies and bring the predicted kaon observables into agreement with present data. 
We further analyze the differential distributions with respect to the dineutrino invariant mass squared $q^2$, as well as the reconstructed variable $q^2_{\mathrm{rec}}$, in $B \to K^{(*)}\nu\bar{\nu}$ decays, demonstrating their sensitivity to different new physics operators. In addition, we investigate the impact of complex Wilson coefficients on $\mathcal{CP}$ asymmetries in $B \to K^{(*)}\mu^+\mu^-$ decays and find that percent-level effects can arise in specific $q^2$ regions. 
\end{abstract}

\section{Introduction}




Flavor-changing neutral current (FCNC) processes are forbidden at tree level in the Standard Model (SM) and first arise at the one-loop level through electroweak box and penguin diagrams. Their amplitudes are strongly suppressed by the 
Glashow–Iliopoulos–Maiani (GIM) mechanism and the hierarchical structure of the Cabibbo–Kobayashi–Maskawa (CKM) matrix, resulting in very small branching ratios. This suppression enhances the sensitivity of FCNC observables to virtual effects from heavy new particles, even when such states lie well beyond the kinematic reach of current collider experiments. In light of the discovery of a scalar resonance consistent with the SM Higgs boson, and the continued absence of direct signals of new physics (NP) at the LHC, rare FCNC processes in the $B$- and kaon sectors provide a crucial window into possible extensions of the SM. 

Current ongoing experiments such as LHCb and Belle-II are targeting clean observables in the $B$-system, whereas NA62 and KOTO look for precision physics in kaon modes. While the recent measurements of the lepton-flavor-universality ratios in $B \to K^{(*)} \ell^+ \ell^-$ are broadly consistent with the SM predictions, the individual branching fractions for $B^+ \to K^+ \mu^+ \mu^-$ and $B^0 \to K^{*0} \mu^+ \mu^-$ ($\ell=\mu,e$) continue to exhibit mild deviations from theoretical expectations~\cite{LHCb:2014cxe,LHCb:2016ykl,LHCb:2022vje,CMS:2024syx,LHCb:2025mqb}. In addition, the angular observable $P_5^\prime$ in $B \to K^* \mu^+\mu^-$ decays shows a persistent tension with the SM~\cite{LHCb:2020lmf,CMS:2024atz,LHCb:2025mqb}, providing a possible indication of NP effects.

Recently, the Belle~II collaboration reported the first evidence for the rare decay $B^+ \to K^+ \nu\bar{\nu}$~\cite{Belle-II:2023esi}, with a branching fraction exceeding the SM expectation. After accounting for the estimated $\sim 10\%$ contribution from the tree-level background process $B^+ \to \tau^+ \nu_\tau$ followed by $\tau^+ \to K^+ \bar{\nu}_\tau$, the measured rate exhibits an approximately $2.7\sigma$ tension with the SM prediction. 

Similar to the $B$-meson sector, the rare kaon decays $K^+ \to \pi^+ \nu \bar{\nu}$ and $K_L \to \pi^0 \nu \bar{\nu}$ provide theoretically clean probes of short-distance physics owing to strong GIM suppression and under control hadronic uncertainties~\cite{Buras:2004uu,Blanke:2009pq}. Consequently, these modes are highly sensitive to NP contributions, although their experimental study remains challenging due to the extremely small branching fractions. The NA62 collaboration has measured the branching fraction for the charged mode~\cite{NA62:2024pjp}, while the KOTO experiment has placed a $90\%$ C.L. upper bound on the neutral channel~\cite{KOTO:2024zbl}. Since the SM prediction for $K^+ \to \pi^+ \nu \bar{\nu}$ is in good agreement with the present data, these rare kaon decays provide stringent constraints on possible NP scenarios.

On the other hand, the rare charged lepton final states $K^\pm \to \pi^\pm \ell^+ \ell^-$ have been precisely measured, beginning with the first observation of $K^+ \to \pi^+ e^+ e^-$ at CERN~\cite{BLOCH1975201}. In the electron channel, the most extensive datasets are from BNL-E865~\cite{E865:1999ker} and NA48/2~\cite{NA482:2009pfe}, while in the muon channel, results have been reported by NA48/2~\cite{NA482:2010zrc} and, more recently, NA62~\cite{NA62:2022qes}. 
Due to the presence of a neutral pion in the final state, the rare decays of neutral kaon, $K_L \to \pi^0 \ell^+ \ell^-$ ($\ell = e, \mu$) have been relatively challenging and the most stringent upper limits are available from the Fermilab experiment~\cite{KTeV:2003sls,KTEV:2000ngj}. These bounds already approach the SM expectations, 
and the forthcoming KOTO-II experiment at J-PARC aims to achieve a precision of about $25\%$
in $\mathrm{BR}(K_L \to \pi^0 \ell^+ \ell^-)$ as mentioned in Ref.~\cite{KOTO:2018dsc}.

On the theory side, significant efforts produced precise form factor estimates for $B\!\to\!K^{(*)}$ within Lattice-QCD~\cite{Bailey:2015dka,Parrott:2022rgu} and light-cone sum rule~\cite{Buras:2014fpa,Bharucha:2015bzk} approaches, including the non-local contributions arising from the charm-loop effects~\cite{Khodjamirian:2010vf,Gubernari:2020eft,AlamKhan:2025mgb}, thus allowing consistent comparisons of data with theory. In parallel, 
in the kaon sector, chiral perturbation theory ($\chi$PT) analyses clarified the long-distance contributions to $K\!\to\!\pi\ell^+\ell^-$, identified the dominance of one-photon exchange, and established the standard form-factor parametrisation~\cite{DAmbrosio:1998gur,Chen:2003nz,Buras:2004uu}. These results are subsequently developed through detailed analysis of operator mixing, radiative corrections, and $\mathcal{CP}$-violating effects within the SM~\cite{Cirigliano:2011ny}.

In the absence of any direct evidence for states and/or interactions beyond the SM, the Standard Model Effective Field Theory (SMEFT) provides a particularly appealing and model-independent framework to investigate possible NP effects. Within SMEFT, the dilepton and dineutrino transitions are related through the underlying $SU(2)_L$ gauge symmetry, allowing unified analyses of FCNC processes that are sensitive to flavor structures, lepton-flavor-universality violation, and correlations among different semileptonic operators~\cite{Descotes-Genon:2020buf,Belle-II:2021rof,Browder:2021hbl,Aebischer:2018iyb,Mandal:2019gff}. 

In this work, we perform a dimension-six SMEFT analysis of semileptonic FCNC transitions, focusing primarily on rare $B$-meson decays. Allowing the relevant Wilson coefficients to be complex, we determine their preferred regions through a $\chi^2$ fit to a selected set of rare $B$-decay observables, including $B \to K^{(*)}\mu^+\mu^-$, $B \to K^{(*)}\nu\bar\nu$, and $B_s \to \mu^+\mu^-$. This also enables us to investigate possible new weak phases beyond the SM and their implications for $\mathcal{CP}$-violating observables. The rare kaon decay modes $K_{(L)} \to \pi \ell^+\ell^-$ and $K_{(L)} \to \pi \nu\bar\nu$ are subsequently used as correlated probes to study the implications of the preferred $b \to s$ solutions in the $s \to d$ sector. These modes provide particularly stringent tests of the flavor structure underlying the preferred NP scenarios. The interplay between the dilepton and dineutrino channels is governed by the $SU(2)_L$ structure of SMEFT, which correlates charged-lepton and neutrino interactions at the operator level. 

While the $b \to s$ and $s \to d$ transitions are initially treated independently in a model-independent setup, we further investigate the implications of flavor symmetries within the framework of Minimal Flavor Violation (MFV). In MFV, all sources of flavor and $\mathcal{CP}$ violation are aligned with the Yukawa structure of the SM, thereby naturally preserving the CKM hierarchies among different flavor-changing transitions. Motivated by this framework, we consider both $U(3)^5$ and $U(2)^5$ flavour symmetries, which relate the $b \to s$ and $s \to d$ sectors through characteristic CKM-like suppression. Within these scenarios, we explore how the preferred complex Wilson coefficients affect a broad set of observables and assess whether different NP scenarios can be distinguished through their correlated signatures across rare $B$-meson and kaon decays.

The paper is organized as follows. In Sec.~\ref{sec:framework}, we introduce the SMEFT setup at
dimension six and present the effective Hamiltonian relevant for the semileptonic
$b \to s(\ell^+\ell^-,\,\nu\bar\nu)$ and $s \to d(\ell^+\ell^-,\,\nu\bar\nu)$ transitions.  Section~\ref{chi2fit} provides a detailed description of our fit analysis.
In Secs.~\ref{subsec:BtoKmumu} and~\ref{subsec:BtoKnunu}\,, we provide the predictions for $B \to K^{(*)}\mu^+\mu^-$ and $B \to K^{(*)} \nu\bar{\nu}$ observables respectively from our fit results.
Section~\ref{subsec:Ktopinunu} is devoted to the implications of fit results for kaon decays, including a discussion on the $U(3)^5$ and $U(2)^5$ flavor symmetries within the framework of MFV. Section~\ref{subsec:ACP} contains a discussion on the impact of complex Wilson
coefficients on $\mathcal{CP}$-violating observables. Finally, in Sec.~\ref{sec:summary} we summarize the outcomes. Several appendices are provided at the end with relevant formulae, expressions, and input parameters employed in our analysis.

\section{Theoretical framework}
\label{sec:framework}

Within the SMEFT framework, the SM Lagrangian is systematically extended
by higher-dimensional operators that encode the low-energy effects of heavy NP states. These operators preserve the SM gauge symmetry group, $\mathrm{SU}(3)_C \times \mathrm{SU}(2)_L \times \mathrm{U}(1)_Y$, and their contributions are suppressed by the NP scale. The SMEFT therefore provides a well-defined operator basis that systematically parametrizes possible NP effects in low-energy observables through an expansion in inverse powers of the heavy scale. 
In the electroweak sector, the $\mathrm{SU}(2)_L$ gauge symmetry places left-handed charged leptons and neutrinos within the same weak-isospin doublet. This structure is naturally preserved in the SMEFT operator basis, leading to correlations between charged-lepton and neutrino processes. Consequently, SMEFT enables a unified treatment of the semileptonic flavor observables considered in this work.

\subsection{SMEFT operator basis}
\label{sec:theory}
 We start with the SMEFT Lagrangian that incorporates dimension-six operators, which is expressed as~\cite{Grzadkowski:2010es}
\begin{equation}
\label{eq:LSMEFT}
 \mathcal{L}^{(6)}=\sum_i \frac{c_i}{\Lambda^2}\, \mathcal{Q}_i\,,
\end{equation}
Here, the operators $\mathcal{Q}_i$ are constructed from SM fields that respect the full SM gauge symmetry group. The Wilson coefficients $c_i$ are dimensionless parameters that encode the details of the high-scale UV physics.
In this framework, both the quark level transitions $d_i \to d_j \ell_{\alpha}^+ \ell_{\beta}^-$ and $d_i \to d_j \nu_{\alpha} \bar{\nu}_{\beta}$ are induced by a restricted set of dimension-six operators that generate FCNCs via semileptonic four-fermion terms and modified electroweak interactions, given by

\begin{equation}
\begin{aligned}
[ \mathcal{Q}_{ql}^{(1)}]^{ij\alpha\beta} &= (\bar{Q}^j_L \gamma_{\mu} Q^i_L)(\bar{L}^\alpha_L \gamma^{\mu} L^\beta_L),
&  
[\mathcal{Q}_{Hq}^{(1)}]^{ij} &= i(\bar{Q}^j_L \gamma_{\mu} Q^i_L)\, H^\dagger \overleftrightarrow{D}^{\mu} H,
\\[6pt]
[\mathcal{Q}_{ql}^{(3)}]^{ij\alpha\beta} &= (\bar{Q}^j_L \gamma_{\mu} \tau^a Q^i_L)(\bar{L}^\alpha_L \gamma^{\mu} \tau_a L^\beta_L),
&  
[\mathcal{Q}_{Hq}^{(3)}]^{ij} &= i(\bar{Q}^j_L \gamma_{\mu} \tau^a Q^i_L)\, H^\dagger \overleftrightarrow{D}^{\mu} \tau_a H,
\\[6pt]
[\mathcal{Q}_{dl}]^{ij\alpha\beta} &= (\bar{d}^j_R \gamma_{\mu} d^i_R)(\bar{L}^\alpha_L \gamma^{\mu} L^\beta_L),
&  
[\mathcal{Q}_{Hd}]^{ij} &= i(\bar{d}^j_R \gamma_{\mu} d^i_R)\, H^\dagger \overleftrightarrow{D}^{\mu} H,
\end{aligned}
\label{eq:SMEFT_Ops}
\end{equation}
and the operators contributing only to $d_i\,\to\,d_j\,\ell_\alpha^+\,\ell_\beta^-$ decays are 
 \begin{eqnarray}\label{smeft-o2}
 [\mathcal{Q}_{de}]^{ij\alpha\beta}=(\bar{d}^j_R \gamma_{\mu} d^i_R)(\bar{e}^\alpha_R \gamma^{\mu} e^\beta_R), \qquad
 [\mathcal{Q}_{qe}]^{ij\alpha\beta}=(\bar{Q}^j_L \gamma_{\mu} Q^i_L)(\bar{e}^\alpha_R \gamma^{\mu} e^\beta_R)\,.
\end{eqnarray}
Here $Q_L (L_L)$ and $H$ denote the quark (lepton) and Higgs weak doublets and $d_R(e_R)$ are the corresponding fermion singlets. 

From the above operators, it is evident that the $SU(2)_L$ doublet structure imposes a powerful and model-independent correlation between processes involving charged leptons and those involving neutrinos. Consequently, any NP contribution generating a non-zero Wilson coefficient for an operator containing left-handed fermion doublets will simultaneously contribute to both $d_i \to d_j \nu_\alpha \bar{\nu}_\beta$ and $d_i \to d_j \ell_\alpha \bar{\ell}_\beta$ transitions, thereby inducing correlations among the corresponding meson decay observables.

In its most general form, however, SMEFT does not impose any relation between flavor transitions involving different quark generations. Such correlations can arise only after imposing an additional flavor principle, such as the MFV hypothesis, in which NP operators inherit the same CKM hierarchies that govern the corresponding SM transitions. We discuss these flavor-symmetric scenarios in Sec.~\ref{subsec:flavorUni}.
In the following subsections, we perform the matching of the above SMEFT operators onto the low-energy effective theory relevant at the corresponding meson mass scales.

\subsection{Effective Hamiltonian for 
  $d_i\to d_j\,\ell_\alpha\bar\ell_\beta$ 
}
\label{sec:Heff_didj_ll}

We consider the FCNC semileptonic transitions
$d_i \to d_j\,\ell_\alpha \bar\ell_\beta$, focusing in particular on the
$b \to s\,\ell_\alpha \bar\ell_\beta$ ($ij=32$) and
$s \to d\,\ell_\alpha \bar\ell_\beta$ ($ij=21$) sectors.
Defining the CKM combinations
$\lambda_t^{ij} \equiv V_{ti}V_{tj}^*$,
with
$\lambda_t^{32}=V_{tb}V_{ts}^*$ and
$\lambda_t^{21}=V_{ts}V_{td}^*$,
the corresponding $\Delta F=1$ effective Hamiltonian, evaluated at the scales $\mu \sim m_b$ for $b \to s$ transitions and $\mu \sim m_s$ for $s \to d$ transitions, can be written as~\cite{Altmannshofer:2009ma}

\begin{align}
 \mathcal{H}_{\rm eff}^{d_i\to d_j\ell_\alpha\ell_\beta} \!
 &= -\frac{4 G_F}{\sqrt{2}}\,
    \lambda_t^{ij}\,
    \frac{\alpha_{\rm em}}{4\pi}
    \sum_{\alpha,\beta}
    \Big[
        C_9^{ij,\alpha\beta}\,\mathcal{O}_9^{ij,\alpha\beta}
      \!+\! C_{10}^{ij,\alpha\beta}\,\mathcal{O}_{10}^{ij,\alpha\beta}
      \!+\! C_9^{\prime\,ij,\alpha\beta}\,\mathcal{O}_9^{\prime\,ij,\alpha\beta}
      \!+\! C_{10}^{\prime\,ij,\alpha\beta}\,\mathcal{O}_{10}^{\prime\,ij,\alpha\beta}
    \Big]
    + {\rm h.c.},
 \label{eq:Heff_didj_ll}
\end{align}
where $G_F$ is the Fermi constant and $\alpha_{\rm em}=e^2/(4\pi)$ is the fine structure constant. 
The Wilson coefficients $C_i$ describe short-distance physics after integrating out all heavy degrees of freedom above the scale $\mu$, and the relevant semileptonic operators specific to the particular transitions are the following.
\begin{align}
\mathcal{O}_9^{ij,\alpha\beta}
&= (\bar d_j \gamma_\mu P_L d_i)\,
   (\bar{\ell}^\alpha \gamma^\mu \ell^\beta)\,,
&
\mathcal{O}_{10}^{ij,\alpha\beta}
&= (\bar d_j \gamma_\mu P_L d_i)\,
   (\bar{\ell}^\alpha \gamma^\mu \gamma_5 \ell^\beta)\,,
\\[2pt]
\mathcal{O}_9^{\prime\,ij,\alpha\beta}
&= (\bar d_j \gamma_\mu P_R d_i)\,
   (\bar{\ell}^\alpha \gamma^\mu \ell^\beta)\,,
&
\mathcal{O}_{10}^{\prime\,ij,\alpha\beta}
&= (\bar d_j \gamma_\mu P_R d_i)\,
   (\bar{\ell}^\alpha \gamma^\mu \gamma_5 \ell^\beta)\,,
\label{eq:ops_didj_ll_unified}
\end{align}
where $P_{L,R}=(1\mp\gamma_5)/2$.
In this framework, the effective Hamiltonian is governed by the dominant
operators $\mathcal{O}_{9,10}^{ij,\alpha\beta}$ whereas their chirality-flipped
counterparts $\mathcal{O}_{9,10}^{\prime\,ij,\alpha\beta}$, with the
associated Wilson coefficients $C_{9,10}^{ij,\alpha\beta}$ and
$C_{9,10}^{\prime\,ij,\alpha\beta}$ are suppressed in the SM by the quark mass factors.
In the SM, the Wilson coefficients are flavor-universal in the lepton indices and are given by
\begin{align}
C_9^{ij,\alpha\beta}\big|_{\rm SM} &= C_{9,\,\rm SM}^{ij}\,\delta_{\alpha\beta}\,,
&
C_{10}^{ij,\alpha\beta}\big|_{\rm SM} &= C_{10,\,\rm SM }^{ij}\,\delta_{\alpha\beta}\,,
\\[2pt]
C_9^{\prime\,ij,\alpha\beta}\big|_{\rm SM} &= 0\,,\qquad
&
C_{10}^{\prime\,ij,\alpha\beta}\big|_{\rm SM} &= 0 .
\end{align}
In case of $b \to s\ell^+\ell^-$ decays, the SM Wilson coefficients $C^{32}_{9,\rm SM}(q^2)$ and $C^{32}_{7,\rm SM}(q^2)$, including the $\mathcal{O}(\alpha_s)$ matrix elements of the four-quark and chromomagnetic operators, are given in Eqs.~\eqref{eq:C9SM} and \eqref{eq:C7SM}, and the Wilson coefficients at $\mu_b$ are listed in Table~\ref{tab:new_SM_WCs} of Appendix~\ref{app:SM_WCs}.

In contrast to the $b\to s\ell\ell$ transitions, the $s\to d\ell\ell$ amplitudes receive sizable additional contributions from CKM-enhanced charm loops, particularly in the coefficient $C_9$. Consequently, the SM Wilson coefficients can be written as~\cite{Bobeth:2017ecx}
\begin{align}
C_{9,\,\rm SM}^{21}
&=
\frac{1}{2}\left[
\left(
\frac{Y(x_t)}{\sin^2\theta_W}
-4\,Z(x_t)
\right)
+
P_0
\right]\,,
\\[2pt]
C_{10,\,\rm SM}^{21}
&=
-\frac{1}{2}\frac{Y(x_t)}{\sin^2\theta_W}\,,
\end{align}
where $Y(x_t)$ and $Z(x_t)$ denote the loop functions with $x_t=m_t^2/m_W^2$, $\theta_W$ is the weak mixing angle, and $P_0$ parametrizes the dominant long-distance charm contribution to $C_9$.

The next step is to perform the matching between the SMEFT operators and the low-energy effective Hamiltonian, thereby establishing the explicit relations between the high-scale SMEFT Wilson coefficients $c_k$ in Eq.~\eqref{eq:LSMEFT} and the semileptonic Wilson coefficients $C_{9,10}^{ij,\alpha\beta}$ and $C_{9,10}^{\prime\,ij,\alpha\beta}$ appearing in Eq.~\eqref{eq:Heff_didj_ll}. For convenience, we introduce the following rescaled combinations of the SMEFT coefficients:
\begin{align}
\label{eq:ctilde_def}
[\widetilde c_k ]^{ij,\alpha \beta}
&= \frac{[c_k]^{ij,\alpha\beta}}{\Lambda^2}\,
   \frac{\pi}{\sqrt{2}\, G_F\, \alpha_{\rm em}\, \lambda_t^{ij}}\,, 
\end{align}
and
\begin{align}
\label{eq:cz_def}
[\widetilde{c}_{Z}]^{ij} 
&=\frac{1}{2}\Big(
   [\widetilde{c}_{Hq}^{(1)}]^{ij}
 + [\widetilde{c}_{Hq}^{(3)}]^{ij}\Big),\qquad
[\widetilde{c}_{Z}^{\,\prime}]^{ij}
=\frac{1}{2}[\widetilde{c}_{Hd}]^{ij}\,,
\end{align}
At the electroweak scale, the
modified vector and axial–vector Wilson coefficients take the form
\cite{Buras:2014fpa}
\begin{align}
\label{eq:C9b2sll}
  C_9^{ij,\alpha \beta} 
  &= C_9^{\rm SM} \delta_{\alpha \beta} 
   + [\widetilde{c}_{qe}]^{ij,\alpha \beta}
   + [\widetilde{c}_{ql}^{(1)}]^{ij,\alpha \beta} 
   + [\widetilde{c}_{ql}^{(3)}]^{ij,\alpha \beta} 
   - \zeta\, [\widetilde{c}_{Z}]^{ij},  \\
\label{eq:C10b2sll}
  C_{10}^{ij,\alpha \beta} 
  &= C_{10}^{\rm SM} \delta_{\alpha \beta} 
   + [\widetilde{c}_{qe}]^{ij,\alpha \beta} 
   - [\widetilde{c}_{ql}^{(1)}]^{ij,\alpha \beta} 
   - [\widetilde{c}_{ql}^{(3)}]^{ij,\alpha \beta} 
   + [\widetilde{c}_{Z} ]^{ij}, \\
  C_9^{\prime\,ij,\alpha \beta} 
  &= [\widetilde{c}_{de}]^{ij,\alpha \beta} 
   + [\widetilde{c}_{dl}]^{ij,\alpha \beta} 
   - \zeta\, [\widetilde{c}_{Z}^{\,\prime}]^{ij},  \\
\label{eq:C10pb2sll}
  C_{10}^{\prime\,ij,\alpha \beta} 
  &= [\widetilde{c}_{de}]^{ij,\alpha \beta} 
   - [\widetilde{c}_{dl}]^{ij,\alpha \beta} 
   + [\widetilde{c}_{Z}^{\,\prime}]^{ij}.
\end{align}
In addition to the semileptonic four-fermion operators, the electroweak operators $\mathcal{Q}_{Hx}$ defined in Eq.~\eqref{eq:SMEFT_Ops} also contribute to semileptonic transitions through effective flavor-changing $\bar d_j d_i Z$ couplings. Their contributions are governed by the factor
\[
\zeta = 1 - 4 \sin^2\theta_W \simeq 0.0754\,,
\]
which reflects the small vector coupling of the $Z$ boson to charged leptons. Nevertheless, we retain these electroweak operators in our analysis and find that the corresponding Wilson coefficients can play an important role in alleviating the tensions observed in $b \to s$ transitions, comparable to the effects of the semileptonic four-fermion operators.

A large number of observables have been measured in rare decays governed by the $b \to s \ell \ell$ transition. In this work, we focus on the processes $B \to K\mu^+\mu^-$, $B^0 \to K^{*0}\mu^+\mu^-$, and $B_s \to \mu^+\mu^-$. The observables included in the numerical analysis are summarized in Table~\ref{tab:exp_data1}. 
In particular, we consider the differential branching fractions
$d\mathcal{B}(B \to K\mu^+\mu^-)/dq^2$ and
$d\mathcal{B}(B^0 \to K^{*0}\mu^+\mu^-)/dq^2$
measured in several $q^2$ bins. For the vector mode $B^0 \to K^{*0}\mu^+\mu^-$, we additionally include the angular observables $F_L$, $P_5^\prime$, and $A_{FB}$, which are particularly sensitive to the helicity structure of the underlying short-distance interactions. Correlations among different observables, as well as among measurements of the same observable in different \(q^2\) bins, are included in our analysis wherever available. The detailed expressions for these observables are given in Appendices~\ref{app:obs_B2Kmumu} and~\ref{app:obs_B2Kstmumu}. 
In the case of the $s \to d \ell \ell$ transition, branching fraction measurements are available for $K^+ \to \pi^+\mu^+\mu^-$, whereas only an experimental upper bound currently exists for the neutral mode $K_L \to \pi^0\mu^+\mu^-$ as quoted in Table~\ref{tab:exp_data1}.

\begin{table}
\caption{Experimental measurements of the observables considered in the present analysis for the $d_i\to d_j\,\ell_\alpha\bar\ell_\beta$ transitions. Correlations among observables and across \(q^2\) bins are included wherever available.}
\centering
\setlength{\tabcolsep}{6pt}
\renewcommand{\arraystretch}{1.2}
\begin{tabular}{l c c}
\hline
Observable & $q^2$ bin [GeV$^2$] & Experimental measurement \\
\hline
\multirow{8}{*}{$\mathcal{B}(B^+ \to K^+ \mu^+ \mu^-) \times 10^{-8}$\cite{CMS:2024syx}} 
 & [0.1,\,0.98] & $(2.91 \pm 0.24) \times 10^{-8}$ \\
 & [1.1,\,2.0]  & $(1.93 \pm 0.20) \times 10^{-8}$ \\
 & [2.0,\,3.0]  & $(3.06 \pm 0.25) \times 10^{-8}$ \\
 & [3.0,\,4.0]  & $(2.54 \pm 0.23) \times 10^{-8}$ \\
 & [4.0,\,5.0]  & $(2.47 \pm 0.24) \times 10^{-8}$ \\
 & [5.0,\,6.0]  & $(2.53 \pm 0.27) \times 10^{-8}$ \\
 & [6.0,\,7.0]  & $(2.50 \pm 0.23) \times 10^{-8}$ \\
 & [7.0,\,8.0]  & $(2.34 \pm 0.25) \times 10^{-8}$ \\
\hline
\multirow{5}{*}{
$\displaystyle
\frac{d\mathcal{B}}{dq^2}(B^0\to K^{*0}\mu^+\mu^-) 
\times 10^{-7}$~\cite{LHCb:2025mqb}}
 & [0.06,\,0.98] & $1.042 \pm 0.060$ \\
 & [1.1,\,2.5]   & $0.339 \pm 0.024$ \\
 & [2.5,\,4.0]   & $0.272 \pm 0.022$ \\
 & [4.0,\,6.0]   & $0.336 \pm 0.020$ \\
 & [6.0,\,8.0]   & $0.376 \pm 0.022$ \\
\hline
\multirow{5}{*}{$F_L(B^0 \to K^{*0}\mu^+\mu^-)$~\cite{LHCb:2025mqb}} 
 & [0.06,\,0.98] & $0.294 \pm 0.088$ \\
 & [1.1,\,2.5]   & $0.742 \pm 0.094$ \\
 & [2.5,\,4.0]   & $0.862 \pm 0.090$ \\
 & [4.0,\,6.0]   & $0.627 \pm 0.055$ \\
 & [6.0,\,8.0]   & $0.643 \pm 0.046$ \\
\hline
\multirow{5}{*}{$P'_5(B^0 \to K^{*0}\mu^+\mu^-)$~\cite{LHCb:2025mqb}}
 & [0.06,\,0.98] & $0.452 \pm 0.155$ \\
 & [1.1,\,2.5]   & $0.441 \pm 0.075$ \\
 & [2.5,\,4.0]   & $-0.209 \pm 0.092$ \\
 & [4.0,\,6.0]   & $-0.479 \pm 0.073$ \\
 & [6.0,\,8.0]   &  $0.586\pm 0.056$  \\
\hline
\multirow{5}{*}{$A_{FB}(B^0 \to K^{*0}\mu^+\mu^-)$~\cite{LHCb:2025mqb}}
 & [0.06,\,0.98] & $-0.053 \pm 0.029$ \\
 & [1.1,\,2.5]   & $-0.196 \pm 0.029$ \\
 & [2.5,\,4.0]   & $-0.118 \pm 0.027$ \\
 & [4.0,\,6.0]   & $ 0.034 \pm 0.020$ \\
 & [6.0,\,8.0]   & $ 0.144 \pm 0.018$ \\
\hline
$\mathcal{B}(K^+ \to \pi^+ \mu^+ \mu^-)$~\cite{NA62:2022qes} & -- & $(9.16\pm 0.06) \times 10^{-8}$ \\
$\mathcal{B}(K_L \to \pi^0 \mu^+ \mu^-)$~\cite{KTEV:2000ngj} & -- & $< 38 \times 10^{-11}$ @ 90\% C.L. \\
$\mathcal{B}(B_s\to \mu^+\mu^-)$~\cite{CMS:2022mgd} & -- & $(3.52 \pm 0.32)\times 10^{-9}$ \\
\hline 
\end{tabular}
\label{tab:exp_data1}
\end{table}

\subsection{Effective Hamiltonian for 
  $d_i\to d_j\,\nu_\alpha\bar\nu_\beta$ 
}
\label{sec:Heff_didj_nunu}

We next consider the FCNC dineutrino transitions
$d_i \to d_j\,\nu_\alpha \bar\nu_\beta$, which provide theoretically clean probes of short-distance physics and are closely correlated with the corresponding dilepton modes within the SMEFT framework. Here $(d_1,d_2,d_3)=(d,s,b)$, and we focus in particular on the transitions $ij=32$, corresponding to $b\to s\,\nu_\alpha \bar\nu_\beta$, and $ij=21$, corresponding to $s\to d\,\nu_\alpha \bar\nu_\beta$. Using the CKM combinations $\lambda_t^{ij}=V_{ti}V_{tj}^*$ defined above, the corresponding $\Delta F=1$ low-energy effective Hamiltonian, evaluated at the scales $\mu \sim m_b$ for $b \to s$ transitions and $\mu \sim m_s$ for $s \to d$ transitions, can be written as
\begin{align}
 \mathcal{H}_{\rm eff}^{d_i\to d_j\nu \bar\nu}
 &= -\frac{4 G_F}{\sqrt{2}}\,
    \lambda_t^{ij}\,
    \frac{\alpha_{\rm em}}{4\pi}\,
    \sum_{\alpha,\beta}
    \Big[
        C_L^{ij,\alpha\beta}\,\mathcal{O}_L^{ij,\alpha\beta}
      + C_R^{ij,\alpha\beta}\,\mathcal{O}_R^{ij,\alpha\beta}
    \Big]
    + {\rm h.c.}\,,
 \label{eq:Heff_didj_nunu}
\end{align}
with the semileptonic operators
\begin{align}
\mathcal{O}_L^{ij,\alpha\beta}
&= (\bar d_j \gamma_\mu P_L d_i)\,
   (\bar{\nu}^\alpha \gamma^\mu (1-\gamma_5)\nu^\beta)\,,
&
\mathcal{O}_R^{ij,\alpha\beta}
&= (\bar d_j \gamma_\mu P_R d_i)\,
   (\bar{\nu}^\alpha \gamma^\mu (1-\gamma_5)\nu^\beta)\,.
\end{align}

In the SM, the decay is generated purely from left-handed interactions with the same neutrino flavor, and   thus the Wilson coefficients are given by
\begin{align}
C_R^{ij,\alpha\beta}\big|_{\rm SM} &= 0\,,\\
C_L^{ij,\alpha\beta}\big|_{\rm SM} &= C^{ij}_{L,\,{\rm SM}}\,\delta_{\alpha\beta}\,,
\end{align}
The SM contribution arises from $Z$-penguin and box diagrams. The corresponding short-distance functions are conventionally defined with $\alpha_{\rm em}(M_{\rm Z})$ and the $\overline{MS}$ weak mixing angle, the scheme in which their higher-order QCD and electroweak corrections are evaluated. In the normalisation of Eq.~\eqref{eq:Heff_didj_nunu}, where the product $\alpha_{\rm em}\,C_L$ is fixed, this gives 
\begin{align}
C^{ij}_{L,\,\rm SM} = -\,\frac{\alpha_{\rm em}(M_Z)}{\alpha_{\rm em}}\,\frac{X^{ij}}{s_W^2}\,,
\label{eq:CLSM_norm}
\end{align}
with $\alpha_{\rm em}(M_Z)=1/127.95$ and $s_W^2\equiv\sin^2\theta_W=0.23116$. In the case of $b\to s\nu\bar\nu$, the charm contribution is CKM suppressed and negligible, so that $X^{32}=X_t$ and
\begin{align}
C_{L,\, \rm SM}^{32} = -\,\frac{\alpha_{\rm em}(M_Z)}{\alpha_{\rm em}}\,\frac{X_t}{s_W^2} = -6.59 \pm 0.04\,,
\label{eq:CL32SM}
\end{align}
where the dominant contribution arises from top-quark loops in the box and $Z$-penguin diagrams~\cite{Buras:2014fpa}. Here $X_t\equiv X(x_t)$, with $x_t=m_t^2/M_W^2$, is the top-quark function including NLO QCD and two-loop electroweak corrections, for which we use $X_t = 1.475(8)$~\cite{Blanke:2026rdi}. In the $\alpha_{\rm em}(M_Z)$ normalisation scheme, Eq.~\eqref{eq:CL32SM} corresponds to $-X_t/s_W^2=-6.38$.


Beyond the SM, the dimension-six SMEFT operators induce additional contributions to the Wilson coefficients $C_{L,R}^{ij,\alpha\beta}$. Using the same rescaled SMEFT combinations $[\widetilde c_k]^{ij,\alpha\beta}$, $[\widetilde c_Z]^{ij}$, and $[\widetilde c_Z']^{ij}$ introduced in Eqs.~\eqref{eq:ctilde_def} and~\eqref{eq:cz_def} for the dilepton case, the Wilson coefficients relevant for the dineutrino transitions at the electroweak scale can be matched onto the SMEFT coefficients as
\begin{align}
\label{eq:CLb2snunu}
  C_L^{ij,\alpha \beta} 
  &= C^{ij}_{L,\,\rm SM}\,\delta_{\alpha \beta} 
   + [\widetilde{c}_{ql}^{(1)}]^{ij,\alpha \beta} 
   - [\widetilde{c}_{ql}^{(3)}]^{ij,\alpha \beta} 
   + [\widetilde{c}_{Z}]^{ij}, \\
\label{eq:CRb2snunu}
  C_R^{ij,\alpha \beta} 
  &= [\widetilde{c}_{dl}]^{ij,\alpha \beta} 
   + [\widetilde{c}_{Z}^{\,\prime}]^{ij}.
\end{align}

Similarly, for the rare kaon decays $K^+ \to \pi^+ \nu\bar{\nu}$ the SM contribution to the left-handed Wilson coefficient is given by~\cite{Buchalla:1995vs}
\begin{align}
\label{eq:CL21SM}
C_{L,\,\rm SM}^{21}= -\,\frac{\alpha_{\rm em}(M_Z)}{\alpha_{\rm em}}\,\frac{1}{s_W^{2}} \left( X_t + \frac{\lambda_c^{21}}{\lambda_t^{21}}\, \big(X_c + \delta X_{c,u}\big)\right),
\qquad
\lambda_c^{21} = V_{cs}V_{cd}^*\,.
\end{align}
 Here $X_c\equiv\lambda^4P_c^{\rm SD}$, with $\lambda\equiv|V_{us}|$, denotes the charm-quark contribution at NNLO in QCD including electroweak corrections, $P_c^{\rm SD}=0.365(12)$~\cite{Buras:2015qea}, and $\delta X_{c,u}\equiv\lambda^4\,\delta P_{c,u}$, with $\delta P_{c,u}=0.04(2)$, accounts for dimension-eight charm and long-distance up-quark contributions~\cite{Isidori:2005xm}. Numerically,  we obtain, $C_{L,\,\rm SM}^{21} = -9.22 + 1.16\,i\,.$

For $K_L \to \pi^0 \nu\bar{\nu}$, the top contribution dominates in the amplitude as only the imaginary part is relevant there, and hence the second term in $C_{L,\,\rm SM}^{21}$ in Eq.~\eqref{eq:CL21SM} can be neglected.

At present, only branching-fraction measurements or upper bounds at $90\%$ confidence level (C.L.) are available for these dineutrino channels. The corresponding experimental observables included in our analysis are summarized in Table~\ref{tab:exp_data2}, while the explicit theoretical expressions for the branching ratios are collected in Appendix~\ref{app:obs_Knunu}.

\begin{table}
\caption{Experimental measurements of the observables considered in the present analysis for the $d_i\to d_j\,\nu_\alpha\bar\nu_\beta$ transitions.}
\centering
\setlength{\tabcolsep}{8pt}
\renewcommand{\arraystretch}{1.1}
\begin{tabular}{l c}
\hline
Observable & Experimental measurement \\
\hline
$\mathcal{B}(B^+ \to K^+ \nu \bar{\nu})$ & $(2.3\pm 0.7) \times 10^{-5}$~\cite{Belle-II:2023esi} \\
$\mathcal{B}(B \to K^* \nu \bar{\nu})$ & $< 2.7 \times 10^{-5} \quad \rm{(90\% \,C.L.)}$~\cite{Belle:2017oht} \\
$\mathcal{B}(K^+ \to \pi^+ \nu \nu)$ & $(13.0^{+3.3}_{-3.0}) \times 10^{-11}$~\cite{NA62:2024pjp} \\
$\mathcal{B}(K_L \to \pi \nu \nu)$ & $< 2.2 \times 10^{-9} \rm{(90\% \,C.L.)}$~\cite{KOTO:2024zbl} \\
\hline 
\end{tabular}
\label{tab:exp_data2}
\end{table}

\section{Results}
\label{sec:results}

We perform a $\chi^2$ analysis of the $B$-meson observables summarized in Tables~\ref{tab:exp_data1} and~\ref{tab:exp_data2} in order to determine the preferred regions of the NP Wilson coefficients. The fit includes observables from $b \to s$ transitions in both the dilepton and dineutrino channels. Using the resulting best-fit values, we provide predictions for a wide range of observables in both the $b \to s$ and $s \to d$ sectors, including currently measured quantities as well as additional observables that can further probe rare meson decays. All numerical inputs used in our analysis are listed in Table~\ref{tab:WC_inputs} in Appendix~\ref{input-para}.

\subsection{Numerical $\chi^2$-analysis }
\label{chi2fit}

\begin{table}
\caption{Best fit values and correlation of SMEFT coefficients at $68\%$ CL.}
\centering
\setlength{\tabcolsep}{8pt}
\renewcommand{\arraystretch}{1.8}
\resizebox{\textwidth}{!}{
\begin{tabular}{l c c c}
\hline
SMEFT couplings & Best fit $(x_i, y_i)$ & $\chi^2_{\rm min}/{\rm d.o.f}$ & Correlation \\
\hline
SM & $-$ & 2.970 & $-$ \\
\hline
\cqlonebsT   & $[-0.744\pm0.154,\ 0.006\pm0.945]$  & 1.864 & $0.010$ \\
\cqlthreebsT & $[-0.739\pm0.188,\ 0.020\pm1.001]$  & 1.881 & $-0.084$ \\
\czbsT       & $[1.377\pm0.278,\ 2.127\pm0.351]$    & 2.195 & $0.684$ \\
\cdlbsT      & $[-0.481\pm0.100,\ 0.025\pm0.569]$   & 2.316 & $-0.001$ \\
\czpbsT      & $[0.529\pm0.140,\ 0.001\pm0.455]$    & 2.667 & $0.000$ \\
\cqebsT      & $[-0.108\pm0.178,\ 0.042\pm0.514]$   & 3.146 & $-0.068$ \\
\cdebsT      & $[-0.010\pm0.166,\ 0.001\pm0.247]$   & 3.182 & $0.000$ \\
\hline
(\cqlonebsT, \czbsT)   & $[(-1.128\pm0.354\,,~1.181\pm0.729\,),~(-0.810\pm0.546\,,~1.402\pm1.149\,)]$ & 1.564 &  Eq.~\eqref{eq:corr_cql1_cz} \\
(\cqlthreebsT, \czbsT) & $[(-1.161\pm0.367\,,~0.218\pm0.995\,),~(-0.622\pm0.575\,,~1.656\pm0.962\,)]$ & 1.562 &  Eq.~\eqref{eq:corr_cql3_cz} \\
\hline
\end{tabular}}
\label{tab_bestfit_extended}
\end{table}
As discussed in the previous sections, our primary objective is to employ a model-independent SMEFT framework to investigate the anomalies observed in rare $B$-decays.  Starting from the SMEFT Wilson coefficients contributing to $b \to s \ell \bar{\ell}$ (given in Eqs.~\eqref{eq:C9b2sll} -- \eqref{eq:C10pb2sll}) and $b \to s \nu \bar{\nu}$ (given in Eqs.~\eqref{eq:CLb2snunu} and \eqref{eq:CRb2snunu}) channels, we first consider one-dimensional (1D) NP scenarios in which only a single NP operator is considered at a time. 
Based on the results of the 1D fits, we then identify several preferred NP scenarios involving simultaneous contributions from two distinct NP operators (2D scenarios). Within the SMEFT framework, these operators generate correlated effects both in dineutrino and dilepton channels. In the present analysis, we restrict ourselves to flavor-conserving muonic operators, corresponding to $\alpha=\beta=2$, motivated by the current tensions observed primarily in the $b \to s \mu^+\mu^-$ sector. Contributions involving electrons, taus, or lepton-flavor-violating interactions are assumed to vanish. Consequently, all NP Wilson coefficients considered in the numerical analysis correspond to the muon sector only.
We construct the following $\chi^2$ function, including a total of 30 observables. 
\begin{equation}
\chi^2= \sum_{i,j} \Big ({\cal O}_i^{\rm th} -{\cal O}_i^{\rm exp} \Big )^T {\mathcal{C}ov}^{-1}_{ij} \Big ({\cal O}_j^{\rm th} -{\cal O}_j^{\rm exp} \Big )\,, 
\end{equation}
where ${\cal O}^{\rm th}$ and ${\cal O}^{\rm exp}$ denote the theoretical predictions and experimental measurements of each observable, respectively. The uncertainties associated with theory and data, including the correlations, are included in the covariance matrix $\mathcal{C}ov$ = $\mathcal{C}ov_{\rm exp}$+ $\mathcal{C}ov_{\rm th}$. $\mathcal{C}ov_{\rm exp}$ contains the experimental uncertainties and correlations between $P_5'$, $F_L$ and $A_{\rm FB}$ within each $q^2$ bin, taken from Ref.~\cite{LHCb:2025mqb}. $\mathcal{C}ov_{\rm th}$ is obtained from the form-factor samples at the SM point and includes all correlations between $q^2$ bins and between observables; the $B\to K$ and $B\to K^*$ form factors are treated as independent. $\mathcal{B}(B_s\to\mu\mu)$ and $\mathcal{B}(B^+\to K^+\nu\bar\nu)$ enter as
uncorrelated terms.

The best-fit values of the SMEFT coefficients for the various 1D and 2D scenarios are presented in Table~\ref{tab_bestfit_extended}, together with their corresponding $1\sigma$ uncertainties and correlations. In order to quantify the statistical preference of a given scenario over the SM, we report $\chi^2_{\rm min}/{\rm d.o.f.}$ for each case. It should be emphasized that, throughout our analysis, the NP Wilson coefficients are treated as complex parameters.

The SM fit is obtained by setting all NP Wilson coefficients to zero, giving
\begin{equation}
\chi^{2}_{\text{SM}}/{\rm d.o.f.} = 2.97,
\end{equation}
with 30 degrees of freedom (d.o.f.). 
This value serves as the reference against which improvements from NP contributions are evaluated.
For the 1D fits, two real parameters (corresponding to the real and imaginary parts of a single complex Wilson coefficient) are varied, leading to 
${\rm d.o.f.} = 30 - 2 = 28$. 

We can see from Table~\ref{tab_bestfit_extended} that
among the 1D scenarios, the lowest  $\chi^2_{\rm min}/{\rm d.o.f.}$ values are obtained for \cqlonebsT and \cqlthreebsT, corresponding to the four-fermion operators given in Eq.~\eqref{eq:SMEFT_Ops}.  Interestingly, the electroweak operator involving the Higgs field also yields a competitive fit through the Wilson coefficient \czbsT. 
The scenarios with \cdlbsT and, to a lesser extent, \czpbsT also improve the fit relative to the SM, whereas \cqebsT and \cdebsT do not lead to any improvement over the SM hypothesis.
These results indicate a clear preference for left-handed SM-like operators over right-handed structures, as can be understood from the matching relations in Eqs.~\eqref{eq:C9b2sll} -- \eqref{eq:C10pb2sll}, \eqref{eq:CLb2snunu}, and \eqref{eq:CRb2snunu}. 
We also observe that the imaginary parts of the preferred Wilson coefficients are consistent with zero within the $1\sigma$ uncertainty except \czbsT where the best-fit central value is sizable. This may lead interesting signatures in observables sensitive to new weak phase. 
\begin{figure}
\centering
\includegraphics[width=0.23\textwidth]{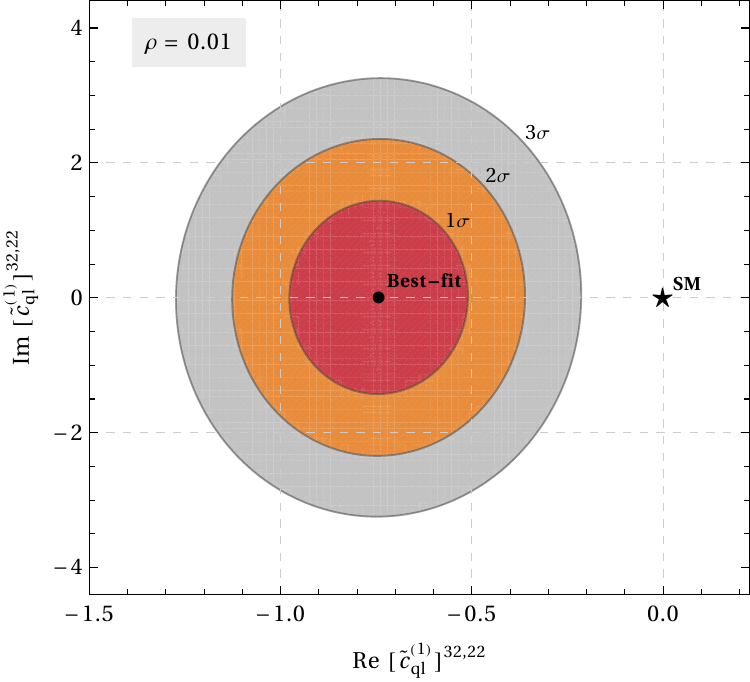}~
\includegraphics[width=0.23\textwidth]{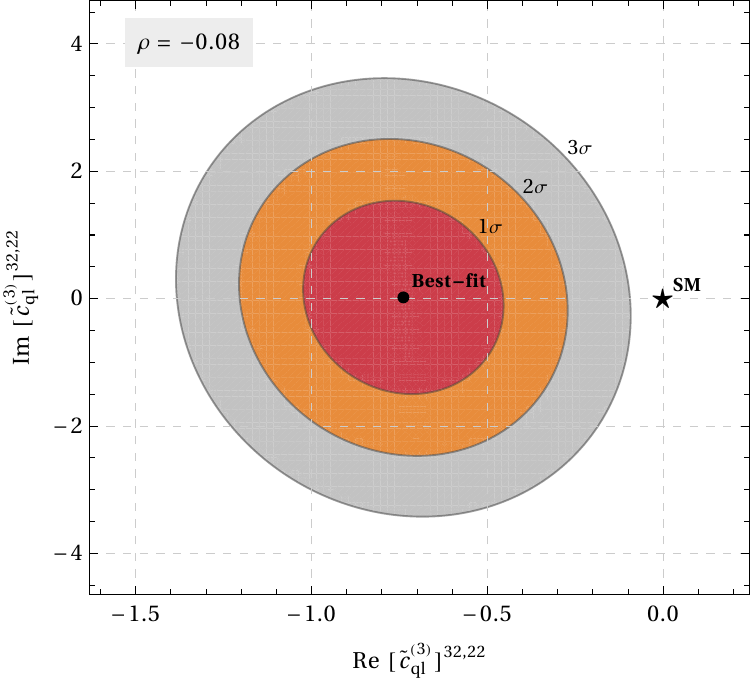}~
\includegraphics[width=0.23\textwidth]{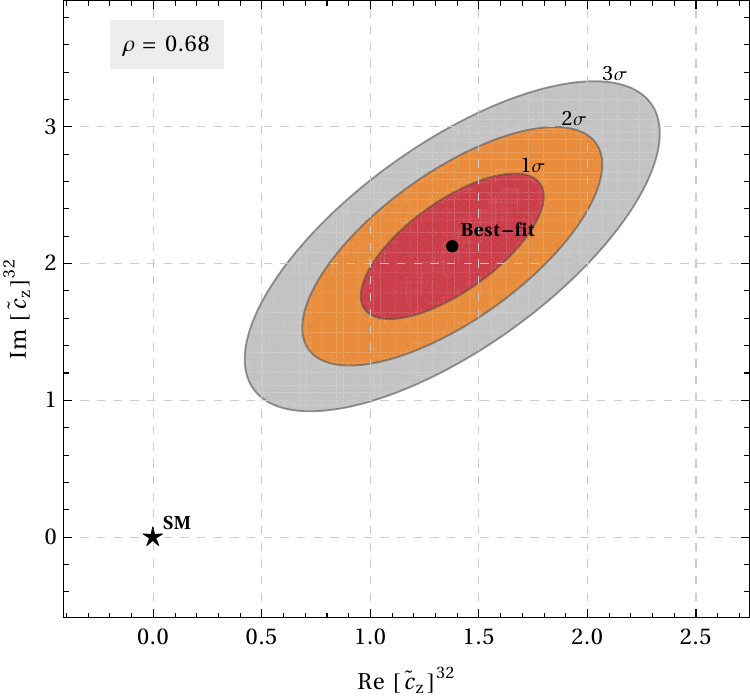}\\
\includegraphics[width=0.23\textwidth]{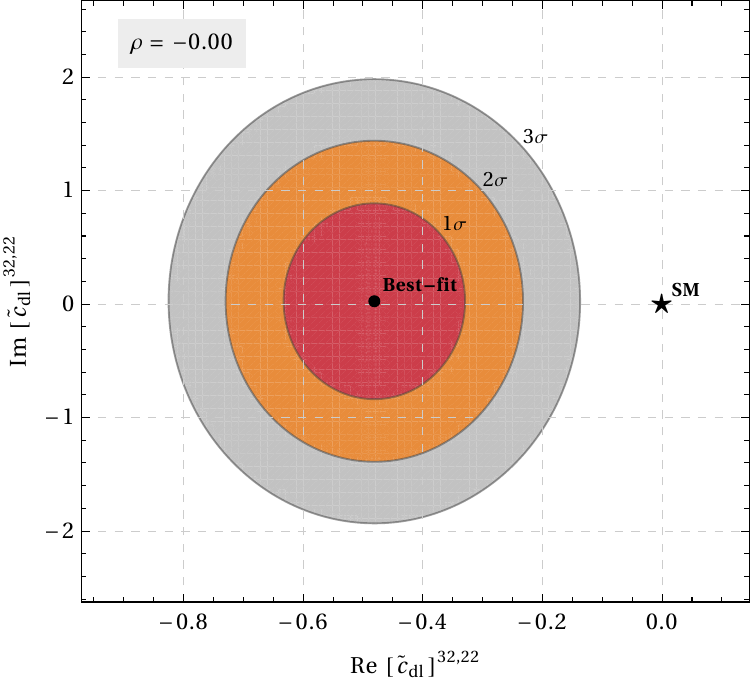}~
\includegraphics[width=0.23\textwidth]{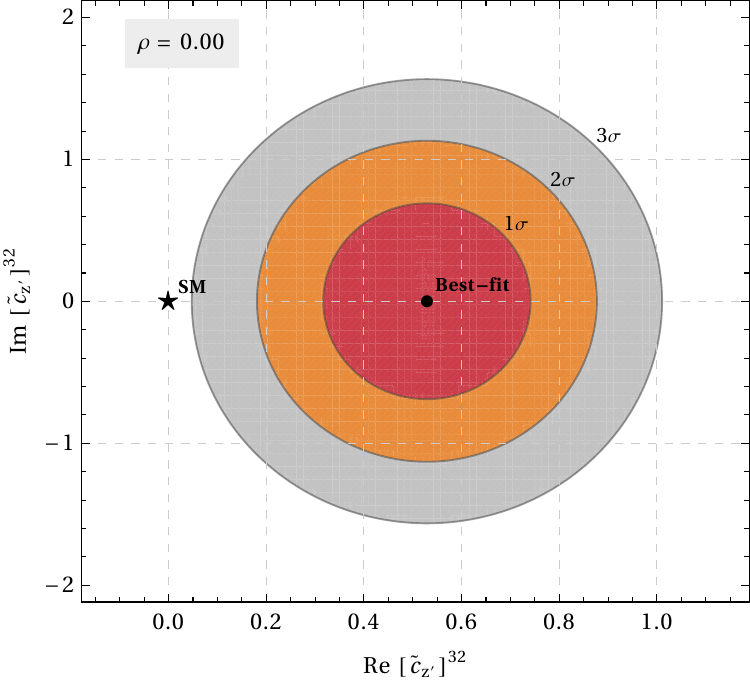}~
\includegraphics[width=0.23\textwidth]{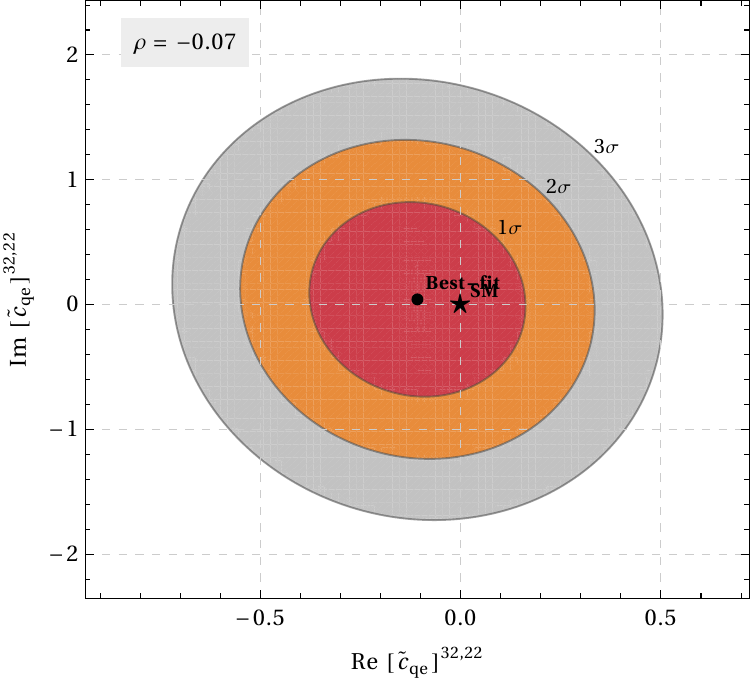}~
\includegraphics[width=0.23\textwidth]{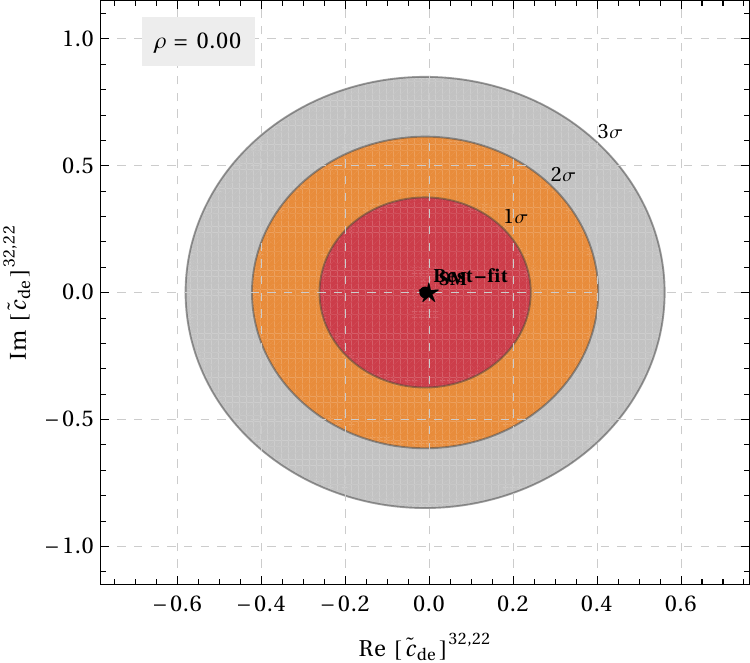}
\caption{Confidence level contours in the complex plane $(\Re[\tilde{c}],\,\Im[\tilde{c}])$ for the NP Wilson coefficients considered in this work. The inner (red), middle (orange), and outer (gray) regions correspond to the $1\sigma$, $2\sigma$, and $3\sigma$ allowed regions, respectively. The black dot indicates the best-fit point, while the SM point is shown by a star at the origin. The correlation coefficient $\rho$ between the real and imaginary parts is displayed in each panel.}
\label{fig:contours}
\end{figure}
In Fig.~\ref{fig:contours}, we present the $1\sigma$, $2\sigma$, and $3\sigma$ confidence level contours in the complex Wilson coefficient plane $(\Re[\tilde{c}],\,\Im[\tilde{c}])$ for the operator scenarios listed in Table~\ref{tab_bestfit_extended}. The SM point is located at the origin, denoted by a `star', and the correlation coefficient $\rho$ between $\Re[\tilde{c}]$ and $\Im[\tilde{c}]$ is indicated in each panel.
For \cqlonebsT and \cqlthreebsT (top row), the contours are nearly circular with negligible correlations. In both cases, the SM point lies well outside the $3\sigma$ regions, indicating a clear preference for NP.
For \czbsT (top right), the contour is strongly elongated with a large positive correlation ($\rho \simeq 0.68$), and the SM point is excluded at more than $3\sigma$.
In the remaining panels (\cdlbsT, \czpbsT, \cqebsT, and \cdebsT), the contours are nearly circular with negligible correlations, indicating little interplay between the real and imaginary parts. For \cdlbsT and \czpbsT, the SM point is displaced from the best-fit regions along the real axis and lies outside the $3\sigma$ contour, only marginally so in the case of \czpbsT. In contrast, for \cqebsT and \cdebsT the SM point lies within the $1\sigma$ region.
Overall, while strong NP preference is observed for \cqlonebsT, \cqlthreebsT, and \czbsT, followed by \cdlbsT and \czpbsT, the coefficients \cqebsT and \cdebsT remain fully consistent with the SM.

We next perform selected 2D fits, pairing operators guided by the preferred 1D results. In this case, varying four real parameters gives ${\rm d.o.f.} = 26$. The combinations (\cqlonebsT, \czbsT) and (\cqlthreebsT, \czbsT) yield the most significant improvement in $\chi^2_{\rm min}/{\rm d.o.f.}$ with values slightly lower than those obtained in the corresponding 1D fit scenarios. However, the improvement comes at the cost of larger uncertainties in the extracted best-fit parameters in the 2D scenarios. 
To validate our fit results, we also perform an independent fit to the same set of observables using the \texttt{flavio} package. The results, presented in Appendix~\ref{app:corr}, show very good agreement between the two fits, despite slight differences in the theoretical inputs used to predict the observables. We now discuss the impact of these preferred scenarios on observables in the $b \to s$ and $s \to d$ semileptonic decay  \footnote{
The SMEFT operators studied here also induce rare top-quark processes such as $t\to c\mu^+\mu^-$, $t\to c\nu\bar\nu$, and, in the case of $\mathcal O_{lq}^{(3)}$, $t\to s\mu^+\nu$. The best-fit values of \cqlonebsT and \cqlthreebsT can lead to enhancements of order two to three of these rare top-quark decay rates relative to their highly suppressed SM predictions. Nevertheless, the current experimental limits on top-FCNC processes remain substantially weaker than the flavor constraints considered in this work. Future measurements of rare top decays may offer complementary probes of the SMEFT scenarios favoured by the present analysis. See Ref.~\cite{Kala:2026xzo} for a recent study of these channels within the SMEFT framework.}.

\subsection{Predictions for $B \to K^{(*)}\mu^+\mu^-$ observables}
\label{subsec:BtoKmumu}

In this subsection, we investigate the impact of the 1D and 2D fit results in  $b \to s \mu^+ \mu^-$ observables. We start with 1D fit where a single NP coupling is varied at a time while the remaining couplings are set to zero. In particular, we study how these best-fit values affect the differential branching fractions $d\mathcal{B}/dq^2(B \to K \mu^+\mu^-)$ and $d\mathcal{B}/dq^2(B \to K^* \mu^+\mu^-)$ in different dimuon invariant mass square $(q^2)$ bins. Moreover, for $B \to K^* \mu^+\mu^-$ channel, we provide predictions for the angular observables $F_L$, $P_5'$ and $A_{FB}$, which serve as complementary probes of possible NP effects.
The predictions are given in Tables~\ref{tab:clean_final} and \ref{tab:FL_transposed}  for all observables across different $q^2$ regions. For comparison, the corresponding SM predictions and current experimental measurements of
are also presented. Below we discuss each case in detail. 

\begin{table}
\caption{Predictions for the differential branching fractions of $B^+ \to K^+ \mu^+\mu^-$ and $B^0 \to K^{*0}\mu^+\mu^-$
in different $q^2$ bins from the 1D and 2D fits, together with the SM predictions and the experimental measurements.
Central values correspond to the best-fit point; the uncertainties are $68\%$ intervals combining the fit uncertainty of the
coefficients and the form-factor uncertainty.}
\centering
\small
\setlength{\tabcolsep}{5pt}
\renewcommand{\arraystretch}{1.55}
\resizebox{\textwidth}{!}{%
\begin{tabular}{c cccccccc}
\hline
\multicolumn{9}{c}{$d\mathcal{B}/dq^2(B^+ \to K^+ \mu^+\mu^-)\times 10^{-8}~[\mathrm{GeV}^{-2}]$} \\
\hline
\textbf{$q^2\,[\mathrm{GeV}^2]$}
& $[0.1,\,0.98]$ & $[1.1,\,2]$ & $[2,\,3]$ & $[3,\,4]$
& $[4,\,5]$ & $[5,\,6]$ & $[6,\,7]$ & $[7,\,8]$ \\
\hline
Exp.~\cite{CMS:2024syx}
& $2.91\pm0.24$ & $1.93\pm0.20$ & $3.06\pm0.25$ & $2.54\pm0.23$
& $2.47\pm0.24$ & $2.53\pm0.27$ & $2.50\pm0.23$ & $2.34\pm0.25$ \\
SM
& $3.681 \pm 0.210$ & $3.696 \pm 0.201$ & $3.692 \pm 0.195$ & $3.686 \pm 0.191$
& $3.677 \pm 0.189$ & $3.663 \pm 0.189$ & $3.644 \pm 0.190$ & $3.622 \pm 0.191$ \\
\hline
\textbf{\cqlonebsT}
& $2.465^{+0.514}_{-0.177}$ & $2.474^{+0.522}_{-0.170}$ & $2.472^{+0.525}_{-0.162}$ & $2.470^{+0.537}_{-0.160}$
& $2.466^{+0.564}_{-0.159}$ & $2.459^{+0.542}_{-0.157}$ & $2.450^{+0.519}_{-0.156}$ & $2.439^{+0.498}_{-0.157}$ \\
\textbf{\cqlthreebsT}
& $2.472^{+0.541}_{-0.200}$ & $2.482^{+0.561}_{-0.211}$ & $2.480^{+0.597}_{-0.199}$ & $2.477^{+0.599}_{-0.199}$
& $2.473^{+0.609}_{-0.202}$ & $2.466^{+0.586}_{-0.218}$ & $2.458^{+0.569}_{-0.213}$ & $2.447^{+0.568}_{-0.208}$ \\
\textbf{\czbsT}
& $3.001^{+0.215}_{-0.224}$ & $3.012^{+0.210}_{-0.218}$ & $3.010^{+0.215}_{-0.220}$ & $3.007^{+0.224}_{-0.206}$
& $3.002^{+0.220}_{-0.196}$ & $2.993^{+0.218}_{-0.185}$ & $2.982^{+0.224}_{-0.177}$ & $2.968^{+0.223}_{-0.177}$ \\
\textbf{\cdlbsT}
& $2.867^{+0.304}_{-0.204}$ & $2.878^{+0.314}_{-0.197}$ & $2.876^{+0.318}_{-0.178}$ & $2.872^{+0.329}_{-0.183}$
& $2.866^{+0.329}_{-0.176}$ & $2.857^{+0.328}_{-0.185}$ & $2.845^{+0.320}_{-0.180}$ & $2.831^{+0.310}_{-0.185}$ \\
\textbf{\czpbsT}
& $3.177^{+0.222}_{-0.229}$ & $3.190^{+0.222}_{-0.222}$ & $3.187^{+0.220}_{-0.207}$ & $3.183^{+0.227}_{-0.193}$
& $3.177^{+0.239}_{-0.196}$ & $3.167^{+0.239}_{-0.207}$ & $3.155^{+0.234}_{-0.207}$ & $3.139^{+0.252}_{-0.214}$ \\
\textbf{\cqebsT}
& $3.694^{+0.283}_{-0.166}$ & $3.710^{+0.275}_{-0.169}$ & $3.705^{+0.267}_{-0.149}$ & $3.699^{+0.256}_{-0.149}$
& $3.689^{+0.255}_{-0.163}$ & $3.674^{+0.268}_{-0.176}$ & $3.655^{+0.283}_{-0.185}$ & $3.632^{+0.277}_{-0.189}$ \\
\textbf{\cdebsT}
& $3.682^{+0.240}_{-0.187}$ & $3.697^{+0.234}_{-0.166}$ & $3.693^{+0.240}_{-0.162}$ & $3.687^{+0.230}_{-0.179}$
& $3.677^{+0.218}_{-0.178}$ & $3.664^{+0.215}_{-0.190}$ & $3.645^{+0.225}_{-0.189}$ & $3.623^{+0.227}_{-0.187}$ \\
\hline
\textbf{(\cqlonebsT,\,\czbsT)}
& $2.738^{+0.458}_{-0.086}$ & $2.750^{+0.458}_{-0.079}$ & $2.747^{+0.438}_{-0.092}$ & $2.743^{+0.433}_{-0.082}$
& $2.736^{+0.417}_{-0.057}$ & $2.725^{+0.390}_{-0.068}$ & $2.712^{+0.367}_{-0.069}$ & $2.695^{+0.369}_{-0.064}$ \\
\textbf{(\cqlthreebsT,\,\czbsT)}
& $2.619^{+0.452}_{-0.124}$ & $2.630^{+0.473}_{-0.122}$ & $2.627^{+0.464}_{-0.100}$ & $2.623^{+0.464}_{-0.095}$
& $2.616^{+0.454}_{-0.091}$ & $2.606^{+0.425}_{-0.069}$ & $2.593^{+0.423}_{-0.069}$ & $2.578^{+0.406}_{-0.071}$ \\
\hline
\end{tabular}}

\vspace{0.45cm}

\resizebox{\textwidth}{!}{%
\begin{tabular}{c ccccc}
\hline
\multicolumn{6}{c}{$d\mathcal{B}/dq^2(B^0 \to K^{*0}\mu^+\mu^-)\times 10^{-7}~[\mathrm{GeV}^{-2}]$} \\
\hline
\textbf{$q^2\,[\mathrm{GeV}^2]$}
& $[0.06,\,0.98]$ & $[1.1,\,2.5]$ & $[2.5,\,4.0]$
& $[4.0,\,6.0]$ & $[6.0,\,8.0]$ \\
\hline
Exp.~\cite{LHCb:2025mqb}
& $1.042\pm0.060$ & $0.339\pm0.024$ & $0.272\pm0.022$
& $0.336\pm0.020$ & $0.376\pm0.022$ \\
SM
& $0.921 \pm 0.126$ & $0.440 \pm 0.090$ & $0.436 \pm 0.089$ & $0.480 \pm 0.089$ & $0.544 \pm 0.091$ \\
\hline
\textbf{\cqlonebsT}
& $0.816^{+0.144}_{-0.096}$ & $0.317^{+0.104}_{-0.069}$ & $0.294^{+0.114}_{-0.060}$ & $0.317^{+0.112}_{-0.055}$ & $0.358^{+0.120}_{-0.063}$ \\
\textbf{\cqlthreebsT}
& $0.816^{+0.149}_{-0.096}$ & $0.318^{+0.110}_{-0.066}$ & $0.295^{+0.124}_{-0.061}$ & $0.318^{+0.125}_{-0.062}$ & $0.359^{+0.128}_{-0.070}$ \\
\textbf{\czbsT}
& $0.852^{+0.128}_{-0.127}$ & $0.361^{+0.087}_{-0.077}$ & $0.347^{+0.090}_{-0.077}$ & $0.381^{+0.090}_{-0.078}$ & $0.433^{+0.092}_{-0.082}$ \\
\textbf{\cdlbsT}
& $1.014^{+0.163}_{-0.141}$ & $0.535^{+0.126}_{-0.111}$ & $0.532^{+0.126}_{-0.109}$ & $0.578^{+0.124}_{-0.105}$ & $0.644^{+0.132}_{-0.094}$ \\
\textbf{\czpbsT}
& $0.979^{+0.136}_{-0.135}$ & $0.500^{+0.120}_{-0.125}$ & $0.496^{+0.120}_{-0.116}$ & $0.541^{+0.114}_{-0.125}$ & $0.606^{+0.113}_{-0.118}$ \\
\textbf{\cqebsT}
& $0.926^{+0.125}_{-0.129}$ & $0.445^{+0.102}_{-0.090}$ & $0.441^{+0.100}_{-0.084}$ & $0.485^{+0.100}_{-0.086}$ & $0.548^{+0.096}_{-0.074}$ \\
\textbf{\cdebsT}
& $0.921^{+0.125}_{-0.121}$ & $0.440^{+0.100}_{-0.109}$ & $0.436^{+0.094}_{-0.097}$ & $0.480^{+0.094}_{-0.096}$ & $0.544^{+0.094}_{-0.087}$ \\
\hline
\textbf{(\cqlonebsT,\,\czbsT)}
& $0.856^{+0.145}_{-0.102}$ & $0.361^{+0.115}_{-0.055}$ & $0.341^{+0.120}_{-0.054}$ & $0.367^{+0.114}_{-0.051}$ & $0.411^{+0.126}_{-0.059}$ \\
\textbf{(\cqlthreebsT,\,\czbsT)}
& $0.844^{+0.129}_{-0.103}$ & $0.348^{+0.093}_{-0.051}$ & $0.326^{+0.096}_{-0.056}$ & $0.350^{+0.109}_{-0.049}$ & $0.391^{+0.110}_{-0.049}$ \\
\hline
\end{tabular}}
\label{tab:clean_final}
\end{table}

\begin{table}
\caption{Predictions for the angular observables of $B \to K^{*} \mu^{+}\mu^{-}$ in different $q^2$ bins, obtained from the 1D and 2D fits. The corresponding SM predictions and experimental data for each bin are also shown.}
\centering
\small
\setlength{\tabcolsep}{10pt}
\renewcommand{\arraystretch}{1.4}
\resizebox{\textwidth}{!}{
\begin{tabular}{c ccccc}
\hline
\multicolumn{6}{c}{$F_L$} \\
\hline
$q^2\,[\rm{GeV^{2}}]$
& [0.06,\,\,0.98] & [1.1,\,\,2.5] & [2.5,\,\,4.0] & [4.0,\,\,6.0] & [6.0,\,\,8.0] \\
\hline
Exp.~\cite{LHCb:2025mqb}
& $0.294 \pm 0.088$ & $0.742 \pm 0.094$ & $0.862 \pm 0.090$ & $0.627 \pm 0.055$ & $0.643 \pm 0.046$ \\
SM
& $0.329 \pm 0.061$ & $0.799 \pm 0.042$ & $0.808 \pm 0.044$ & $0.722 \pm 0.061$ & $0.621 \pm 0.072$ \\
\hline
\cqlonebsT
& $0.249^{+0.076}_{-0.051}$ & $0.739^{+0.062}_{-0.055}$ & $0.798^{+0.040}_{-0.062}$ & $0.730^{+0.054}_{-0.076}$ & $0.632^{+0.066}_{-0.096}$ \\
\cqlthreebsT
& $0.250^{+0.081}_{-0.053}$ & $0.740^{+0.062}_{-0.056}$ & $0.798^{+0.039}_{-0.062}$ & $0.730^{+0.053}_{-0.076}$ & $0.632^{+0.065}_{-0.096}$ \\
\czbsT
& $0.290^{+0.062}_{-0.056}$ & $0.790^{+0.042}_{-0.043}$ & $0.824^{+0.041}_{-0.049}$ & $0.740^{+0.061}_{-0.066}$ & $0.635^{+0.072}_{-0.082}$ \\
\cdlbsT
& $0.374^{+0.064}_{-0.058}$ & $0.824^{+0.034}_{-0.032}$ & $0.830^{+0.036}_{-0.046}$ & $0.751^{+0.051}_{-0.062}$ & $0.656^{+0.066}_{-0.071}$ \\
\czpbsT
& $0.358^{+0.066}_{-0.068}$ & $0.815^{+0.038}_{-0.037}$ & $0.822^{+0.039}_{-0.045}$ & $0.740^{+0.059}_{-0.065}$ & $0.643^{+0.071}_{-0.080}$ \\
\cqebsT
& $0.329^{+0.065}_{-0.058}$ & $0.793^{+0.044}_{-0.042}$ & $0.802^{+0.046}_{-0.053}$ & $0.717^{+0.064}_{-0.070}$ & $0.618^{+0.074}_{-0.087}$ \\
\cdebsT
& $0.329^{+0.061}_{-0.059}$ & $0.799^{+0.039}_{-0.040}$ & $0.808^{+0.040}_{-0.050}$ & $0.722^{+0.064}_{-0.064}$ & $0.621^{+0.071}_{-0.082}$ \\
\hline
(\cqlonebsT, \czbsT)
& $0.265^{+0.080}_{-0.050}$ & $0.725^{+0.064}_{-0.062}$ & $0.768^{+0.048}_{-0.059}$ & $0.702^{+0.057}_{-0.077}$ & $0.612^{+0.066}_{-0.083}$ \\
(\cqlthreebsT, \czbsT)
& $0.257^{+0.073}_{-0.048}$ & $0.719^{+0.067}_{-0.063}$ & $0.768^{+0.048}_{-0.058}$ & $0.704^{+0.058}_{-0.078}$ & $0.614^{+0.067}_{-0.084}$ \\
\hline
\multicolumn{6}{c}{$P_5^{'}$} \\
\hline
$q^2\,[\rm{GeV^{2}}]$
& [0.06,\,\,0.98] & [1.1,\,\,2.5] & [2.5,\,\,4.0] & [4.0,\,\,6.0] & [6.0,\,\,8.0] \\
\hline
Exp.~\cite{LHCb:2025mqb}
& $0.452 \pm 0.155$ & $0.441 \pm 0.075$ & $-0.209 \pm 0.092$ & $-0.479 \pm 0.073$ & $-0.586 \pm 0.056$ \\
SM
& $0.643 \pm 0.073$ & $0.031 \pm 0.149$ & $-0.567 \pm 0.135$ & $-0.780 \pm 0.100$ & $-0.845 \pm 0.084$ \\
\hline
\cqlonebsT
& $0.681^{+0.045}_{-0.120}$ & $0.217^{+0.098}_{-0.222}$ & $-0.395^{+0.115}_{-0.203}$ & $-0.705^{+0.113}_{-0.132}$ & $-0.812^{+0.099}_{-0.089}$ \\
\cqlthreebsT
& $0.681^{+0.045}_{-0.127}$ & $0.215^{+0.096}_{-0.213}$ & $-0.397^{+0.111}_{-0.210}$ & $-0.706^{+0.112}_{-0.133}$ & $-0.813^{+0.098}_{-0.090}$ \\
\czbsT
& $0.486^{+0.074}_{-0.075}$ & $0.022^{+0.114}_{-0.124}$ & $-0.508^{+0.144}_{-0.137}$ & $-0.690^{+0.120}_{-0.096}$ & $-0.731^{+0.110}_{-0.088}$ \\
\cdlbsT
& $0.663^{+0.076}_{-0.076}$ & $0.132^{+0.139}_{-0.143}$ & $-0.416^{+0.166}_{-0.102}$ & $-0.628^{+0.150}_{-0.065}$ & $-0.700^{+0.116}_{-0.054}$ \\
\czpbsT
& $0.695^{+0.077}_{-0.076}$ & $0.136^{+0.144}_{-0.150}$ & $-0.443^{+0.151}_{-0.135}$ & $-0.669^{+0.138}_{-0.109}$ & $-0.746^{+0.105}_{-0.085}$ \\
\cqebsT
& $0.661^{+0.081}_{-0.085}$ & $0.058^{+0.149}_{-0.149}$ & $-0.531^{+0.169}_{-0.106}$ & $-0.754^{+0.146}_{-0.064}$ & $-0.827^{+0.117}_{-0.055}$ \\
\cdebsT
& $0.641^{+0.073}_{-0.072}$ & $0.029^{+0.148}_{-0.161}$ & $-0.568^{+0.154}_{-0.154}$ & $-0.781^{+0.130}_{-0.103}$ & $-0.845^{+0.093}_{-0.079}$ \\
\hline
(\cqlonebsT, \czbsT)
& $0.735^{+0.054}_{-0.109}$ & $0.301^{+0.119}_{-0.151}$ & $-0.225^{+0.153}_{-0.150}$ & $-0.523^{+0.155}_{-0.106}$ & $-0.654^{+0.151}_{-0.078}$ \\
(\cqlthreebsT, \czbsT)
& $0.711^{+0.086}_{-0.140}$ & $0.299^{+0.097}_{-0.173}$ & $-0.215^{+0.158}_{-0.166}$ & $-0.511^{+0.126}_{-0.124}$ & $-0.641^{+0.118}_{-0.092}$ \\
\hline
\multicolumn{6}{c}{$A_{FB}$} \\
\hline
$q^2\,[\rm{GeV^{2}}]$
& [0.06,\,\,0.98] & [1.1,\,\,2.5] & [2.5,\,\,4.0] & [4.0,\,\,6.0] & [6.0,\,\,8.0] \\
\hline
Exp.~\cite{LHCb:2025mqb}
& $-0.053 \pm 0.029$ & $-0.196 \pm 0.029$ & $-0.118 \pm 0.027$ & $0.034 \pm 0.020$ & $0.144 \pm 0.018$ \\
SM
& $-0.084 \pm 0.010$ & $-0.106 \pm 0.026$ & $0.005 \pm 0.027$ & $0.128 \pm 0.042$ & $0.234 \pm 0.051$ \\
\hline
\cqlonebsT
& $-0.081^{+0.011}_{-0.007}$ & $-0.148^{+0.048}_{-0.025}$ & $-0.046^{+0.048}_{-0.018}$ & $0.087^{+0.058}_{-0.031}$ & $0.206^{+0.060}_{-0.042}$ \\
\cqlthreebsT
& $-0.081^{+0.011}_{-0.006}$ & $-0.148^{+0.050}_{-0.025}$ & $-0.046^{+0.050}_{-0.015}$ & $0.087^{+0.057}_{-0.028}$ & $0.206^{+0.059}_{-0.043}$ \\
\czbsT
& $-0.061^{+0.009}_{-0.010}$ & $-0.088^{+0.024}_{-0.027}$ & $0.005^{+0.022}_{-0.021}$ & $0.111^{+0.038}_{-0.040}$ & $0.201^{+0.053}_{-0.052}$ \\
\cdlbsT
& $-0.077^{+0.009}_{-0.009}$ & $-0.090^{+0.023}_{-0.019}$ & $0.000^{+0.020}_{-0.026}$ & $0.102^{+0.026}_{-0.038}$ & $0.193^{+0.038}_{-0.052}$ \\
\czpbsT
& $-0.079^{+0.010}_{-0.010}$ & $-0.095^{+0.023}_{-0.025}$ & $0.003^{+0.023}_{-0.021}$ & $0.112^{+0.037}_{-0.042}$ & $0.209^{+0.048}_{-0.053}$ \\
\cqebsT
& $-0.086^{+0.012}_{-0.011}$ & $-0.111^{+0.028}_{-0.032}$ & $-0.001^{+0.025}_{-0.029}$ & $0.121^{+0.029}_{-0.046}$ & $0.228^{+0.042}_{-0.056}$ \\
\cdebsT
& $-0.084^{+0.010}_{-0.010}$ & $-0.106^{+0.027}_{-0.025}$ & $0.005^{+0.027}_{-0.024}$ & $0.128^{+0.047}_{-0.041}$ & $0.234^{+0.059}_{-0.051}$ \\
\hline
(\cqlonebsT, \czbsT)
& $-0.088^{+0.013}_{-0.010}$ & $-0.159^{+0.048}_{-0.032}$ & $-0.071^{+0.037}_{-0.028}$ & $0.047^{+0.041}_{-0.033}$ & $0.156^{+0.045}_{-0.051}$ \\
(\cqlthreebsT, \czbsT)
& $-0.084^{+0.012}_{-0.008}$ & $-0.157^{+0.047}_{-0.027}$ & $-0.071^{+0.048}_{-0.030}$ & $0.044^{+0.049}_{-0.036}$ & $0.152^{+0.056}_{-0.048}$ \\
\hline
\end{tabular}}
\label{tab:FL_transposed}
\end{table}

\begin{itemize}
\item {\bf  $d\mathcal{B}/dq^2(B \to K^{(*)} \mu^+\mu^- )$ :} For $B \to K \mu^+\mu^-$, the experimental measurements lie consistently below the SM predictions, particularly in the low-$q^2$ region $[0.1,\,8]~\text{GeV}^2$. 
We find that, for the best-fit values of \cqlonebsT or \cqlthreebsT, the predicted differential distributions $d\mathcal{B}/dq^2(B \to K \mu^+\mu^-)$ are brought into agreement with the data at the level of $1\sigma$ in all the considered $q^2$ bins except $[1.1,\,2]~\text{GeV}^2$, with the NP contributions lowering the theory predictions toward the experimental measurements.
The scenarios with \czbsT and \cdlbsT also alleviate the tension between the theoretical prediction and the measurements, with deviations reduced to $\lesssim 2\sigma$ in most bins, and \czpbsT leads to a milder reduction, whereas the remaining 1D fits involving \cqebsT or \cdebsT do not lead to any noticeable improvement.
A similar pattern is observed for $d\mathcal{B}/dq^2(B \to K^{*} \mu^+\mu^-)$. The 1D fits with \cqlonebsT, \cqlthreebsT, or \czbsT lower the theoretical predictions relative to the SM and thus make them consistent with the experimental results within $1\sigma$ uncertainties in the $q^2$ region $[1.1,\,8]~\text{GeV}^2$. In contrast, the right-handed scenarios \cdlbsT and \czpbsT enhance the predictions relative to the SM and therefore increase the tension with the data.

 In the 2D analysis, both preferred combinations (\cqlonebsT, \czbsT) and (\cqlthreebsT, \czbsT) slightly increase $d\mathcal{B}/dq^2(B \to K \mu^+\mu^-)$ relative to the individual 1D fits with \cqlonebsT and \cqlthreebsT, giving predictions $\sim (2.6 -2.7) \times  10^{-8}~\text{GeV}^{-2}$ across $q^2 \in [0.1,\,8]~\text{GeV}^2$, and hence in agreement with the data within $\sim 1\sigma$ in most bins. Same conclusion can be drawn for $d\mathcal{B}/dq^2(B \to K^{*} \mu^+\mu^-)$,
where both the 2D scenarios increase the theoretical predictions
relative to the corresponding 1D fits across $q^2 \in [1.1,\,8]~\text{GeV}^2$, and provide
good agreement with the experimental data. Interestingly, the tension in the lowest bin $[0.06,\,0.98]~\text{GeV}^2$ of $B \to K^{*} \mu^+\mu^-$, where the data lie above the SM prediction, is not alleviated by any of the scenarios considered here, and likewise the deviation in the $[1.1,\,2]~\text{GeV}^2$ bin of $B \to K \mu^+\mu^-$ persists at the $\gtrsim 2\sigma$ level in all scenarios.

\item {\bf  $F_L$:} The measurements of the longitudinal polarization fraction $F_L$ in $B \to K^*\mu^+\mu^-$ are slightly lower than the SM predictions in most $q^2$ bins, although they remain consistent within the experimental and theoretical uncertainties. In the low $q^2$ region $[1.1,\,2.5]~\text{GeV}^2$, we find that the 1D fits with \cqlonebsT or \cqlthreebsT individually lead to a reduction in the predicted values of $F_L$, bringing them into excellent agreement with the experimental data. In contrast, the remaining 1D scenarios yield predictions that stay close to the SM expectation.

In the 2D analysis, the combinations (\cqlonebsT,\,\czbsT) and (\cqlthreebsT,\,\czbsT) yield predictions for $F_L$ similar to those of the corresponding 1D fit scenarios. In particular, in the $[1.1,\,2.5]~\GeV^2$ bin, the predicted values remain fully consistent with the measured result, while a mild improvement is observed in the $[4.0,\,6.0]~\GeV^2$ bin. In all $q^2$ bins the predictions lie within $1\sigma$ of the data.

\item {\bf $P_5'$:} The angular observable $P_5^\prime$ exhibits a well-known tension between the experimental measurements and the SM predictions, particularly in the $q^2$ region $[4,\,8]~\text{GeV}^2$. In the 1D analysis, we find that the scenarios with \cqlonebsT or \cqlthreebsT individually provide an improved agreement with the data in the region $[1.1,\,8]~\text{GeV}^2$, with the corresponding deviations reduced to about $(1.3-2)\sigma$.
The 1D scenarios with \czbsT, \cdlbsT, or \czpbsT also improve the agreement in the region $[2.5,\,8]~\text{GeV}^2$ by shifting the predicted values of $P_5'$ closer to the experimental central values, with \cdlbsT yielding deviations at the $1\sigma$ level.
Among all the scenarios considered, the best overall agreement with the data is obtained in the 2D fits, with both combinations (\cqlonebsT,\czbsT) and (\cqlthreebsT,\czbsT) yielding nearly identical predictions. Both combinations shift the predictions closer to the experimental measurements in the $[1.1,\,8.0]~\GeV^2$ region, with the predictions lying within $0.5\sigma$ of the data in $[2.5,\,8.0]~\GeV^2$ and especially good agreement observed in the $[2.5,\,4.0]~\GeV^2$ bin
where the deviation reduces to $\lesssim 0.1\sigma$.

\item {\bf $A_{FB}$:} The forward-backward lepton asymmetry $A_{FB}$ in $B \to K^* \mu^+ \mu^-$ exhibits tension with the SM predictions in the $q^2$ region $[1.1,\,8]~\text{GeV}^2$, most pronounced in the $[2.5,\,4.0]~\text{GeV}^2$ bin, correlated with the behavior observed in $P_5'$. In the 1D scenarios, we find that the cases with \cqlonebsT or \cqlthreebsT reduce the tension, bringing the predicted values of $A_{FB}$ to within about $1.5\sigma$ of the experimental measurements in the $q^2$ regions $[1.1,\,2.5]~\text{GeV}^2$ and $[4,\,8]~\text{GeV}^2$.
In the 2D analysis, both (\cqlonebsT,\,\czbsT) and
(\cqlthreebsT,\,\czbsT) further improve the agreement with data
relative to the 1D fits, with the predictions in $[4.0,\,6.0]~\text{GeV}^2$ and $[6.0,\,8.0]~\text{GeV}^2$
in agreement with the measurements well within $1\sigma$ uncertainty, while the tension in
$[1.1,\,4.0]~\text{GeV}^2$ is reduced to about $1\sigma$ level.

\item {\bf $\overline{\mathcal{B}}(B_s\to \mu^+\mu^-)$:} Being sensitive only to $C_{10}^{32,22}-C_{10}^{\prime \, 32,22}$, this mode separates the scenarios by quark chirality. The left-handed scenarios lower the branching fraction relative to the SM, while the right-handed \cdlbsT\ and \czpbsT\ raise it, and \cqebsT\ and \cdebsT\ remain SM-like. Relative to the experimental measurement~\cite{CMS:2022mgd}, the 1D left-handed scenarios lie $2.0$--$2.7\sigma$ below and the right-handed ones about $1.4\sigma$ above, whereas the 2D fits are consistent within $1\sigma$.

\end{itemize}

\subsection{Predictions for $B \to K^{(*)} \nu\bar{\nu}$ observables}
\label{subsec:BtoKnunu}

\begin{table}
\caption{The SM predictions, experimental measurements and NP predictions for $B \to K^{(*)}\nu\bar{\nu}$ branching fractions in the 1D and 2D fit scenarios.}
\centering
\renewcommand{\arraystretch}{1.4}
\begin{tabular}{c c c}
\hline
Couplings &
$\mathcal{B}(B \to K \nu\bar{\nu})$ [$\times 10^{-6}$] &
$\mathcal{B}(B \to K^* \nu\bar{\nu})$ [$\times 10^{-6}$] \\
\hline
Exp. & $23\pm 7.0$~\cite{Belle-II:2023esi} & $<27$~\cite{Belle:2017oht} \\
SM & $4.879 \pm 0.223$ & $8.449 \pm 1.218$ \\
\hline
\cqlonebsT   & $5.267^{+0.320}_{-0.238}$ & $9.121^{+1.369}_{-1.095}$ \\
\cqlthreebsT & $4.534^{+0.299}_{-0.214}$ & $7.852^{+1.185}_{-1.071}$ \\
\czbsT       & $4.439^{+0.213}_{-0.198}$ & $7.688^{+1.178}_{-1.131}$ \\
\cdlbsT      & $5.125^{+0.237}_{-0.230}$ & $8.188^{+1.116}_{-1.139}$ \\
\czpbsT      & $4.628^{+0.212}_{-0.220}$ & $8.770^{+1.254}_{-1.295}$ \\
\hline
(\cqlonebsT, \czbsT)   & $6.226^{+0.673}_{-0.538}$ & $10.782^{+1.951}_{-1.573}$ \\
(\cqlthreebsT, \czbsT) & $4.701^{+0.268}_{-0.176}$ & $8.141^{+1.119}_{-1.043}$ \\
\hline
\end{tabular}
\label{tab:nunudecays_trimmed}
\end{table}

We now turn to the implications of the fit results for semileptonic decay modes, with particular emphasis on channels involving dineutrino in the final state. These include 
$B \to K \nu\bar{\nu}$ and
$B \to K^* \nu\bar{\nu}$.
The dineutrino modes are free from photon-pole contributions and long-distance  effects, which makes their theoretical predictions particularly robust and sensitive to short-distance physics. We show the estimates for branching fractions for both the modes in Table~\ref{tab:nunudecays_trimmed} and below discuss the cases in detail.
\begin{itemize}
\item {\bf $\mathcal{B}(B \to K \nu\bar{\nu})$:} The measured branching fraction is reported to lie approximately $2.7\sigma$ above the SM prediction. In the 1D scenarios where either \cqlonebsT or  \cdlbsT is varied individually, the predicted value is shifted toward the experimental central value, partially reducing the tension. Nevertheless, even in these cases, the predicted branching ratio remains more than $2\sigma$ away from the experimental result. Interestingly, although the NP coefficient \cqlthreebsT is as well preferred as \cqlonebsT in $b \to s \mu^+ \mu^- $ observables, it leads to a reduction of the predicted branching ratios in the $b \to s \nu \bar\nu$  channel, thereby increasing the tension with data. This behavior originates from the different sign entering the matching between SMEFT and the low-energy Hamiltonian (see Eqs.~\eqref{eq:C9b2sll}, \eqref{eq:C10b2sll}, and \eqref{eq:CLb2snunu}).  

In the 2D fit scenarios, the combination (\cqlonebsT,\czbsT) leads to an enhancement of the predicted branching fraction relative to the corresponding 1D cases, thereby shifting the prediction closer to the experimental central value. Nevertheless, the prediction still remains more than $2\sigma$ away from the measured result. 
In contrast, the (\cqlthreebsT,\czbsT) combination yields a branching fraction smaller than the SM prediction and therefore increases the tension with the data. This behavior originates from the opposite sign contributions of \cqlonebsT and \cqlthreebsT to the effective Wilson coefficients, as discussed above, which also govern the pattern observed in the 2D fit results.

\item {\bf $\mathcal{B}(B \to K^* \nu\bar{\nu})$:} As shown in Table~\ref{tab:nunudecays_trimmed}, this decay mode currently has only an experimental upper bound at $90\%$ C.L. The NP scenarios with left-handed quark currents (\cqlonebsT, \cqlthreebsT, and \czbsT) exhibit trends similar to those observed in $\mathcal{B}(B \to K \nu\bar{\nu})$, whereas the right-handed scenarios \cdlbsT and \czpbsT shift the branching fraction in the opposite direction, since $B \to K \nu\bar{\nu}$ is sensitive to $(C_L^{32,22} + C_R^{32,22})$ while $B \to K^* \nu\bar{\nu}$ is dominated by $(C_L^{32,22} - C_R^{32,22})$ (see Appendix~\ref{app:BtoKnunuexp} for the expressions). In all cases, the predicted branching fractions remain well within the experimentally allowed range.

In the 2D fit scenarios, the combination (\cqlonebsT,\,\czbsT) yields an enhanced branching fraction relative to the SM prediction, whereas the (\cqlthreebsT,\, \czbsT) combination leads to a suppression of the branching ratio. In both cases, however, the predictions remain below the current experimental upper bound.

\end{itemize}
We next discuss the impact of our fit results on the differential distributions of the dineutrino modes, which are expected to be accessible at experiments such as Belle~II. 

Experimentally, the momentum transfer to the invisible neutrino pair cannot be reconstructed directly due to the missing energy. Instead, one introduces the recoil variable $q^{2}_{\mathrm{rec}}$~\cite{Belle-II:2023esi}, defined using the kinematics of the visible hadron in the $B\bar{B}$ rest frame. While the true momentum transfer is given by $q^{2} = (p_{B} - p_{K^{(*)}})^{2}$, the quantity $q^{2}_{\mathrm{rec}}$ serves as an experimentally accessible proxy. Importantly, the corresponding differential distribution $d\Gamma/dq^{2}_{\mathrm{rec}}$ retains full sensitivity to the short-distance Wilson coefficients governing the $b \to s\nu\bar{\nu}$ transition, as well as to the hadronic form factors describing the $B \to K^{(*)}$ matrix elements. Consequently, measurements of the $q^{2}_{\mathrm{rec}}$ spectrum at Belle~II will play a crucial role in testing the SM description of these transitions and in probing possible NP effects.

Following Ref.~\cite{Berezhnoy:2025tiw}, the relation between the true momentum transfer and the reconstructed variable can be approximated as
\begin{align}
\label{eq:qrec_def}
    q^2 \approx q^2_{\mathrm{rec}} + \cos{\theta} \, v \, \lambda^{1/2}(m_B^2, m_{K^*}^{2}, q^2_{\mathrm{rec}}),
\end{align}
where $\theta$ is the angle between the direction of the $K^{(*)}$ meson in the $B$-meson rest frame, and $v$ denotes the velocity of the $B$ meson in the $B\bar{B}$ center-of-mass frame. 

The differential decay distribution in terms of $q^2_{\mathrm{rec}}$ is then obtained as
\begin{align}
    \frac{d \Gamma(B \to K^{(*)} \nu \bar{\nu})}{d q^2_{\mathrm{rec}}}
    =
    \frac{1}{2}
    \int_{-1}^{1}
    \frac{d \Gamma(B \to K^{(*)} \nu \bar{\nu})}{d q^2}
    ~
    \frac{d q^2}{d q^2_{\mathrm{rec}}}
    ~ d \cos \theta \, ,
\end{align}
where the expression for $d \Gamma(B \to K^{(*)} \nu \bar{\nu})/dq^2$ is given in Appendix~\ref{app:BtoKnunuexp}.

\begin{figure}[t]
\centering
\includegraphics[width=0.45\textwidth]{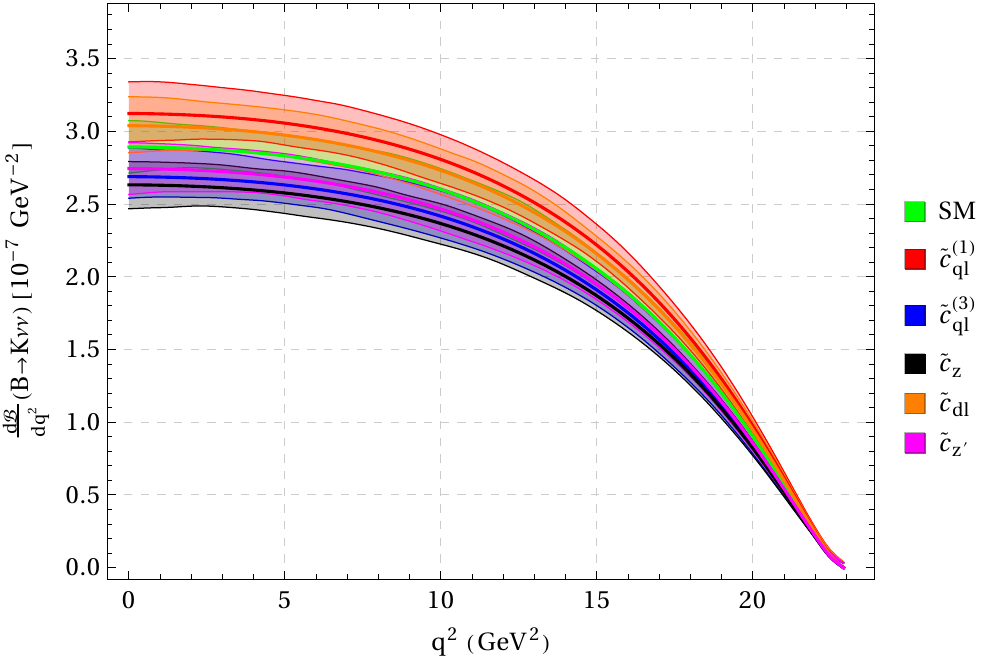}~~
\includegraphics[width=0.45\textwidth]{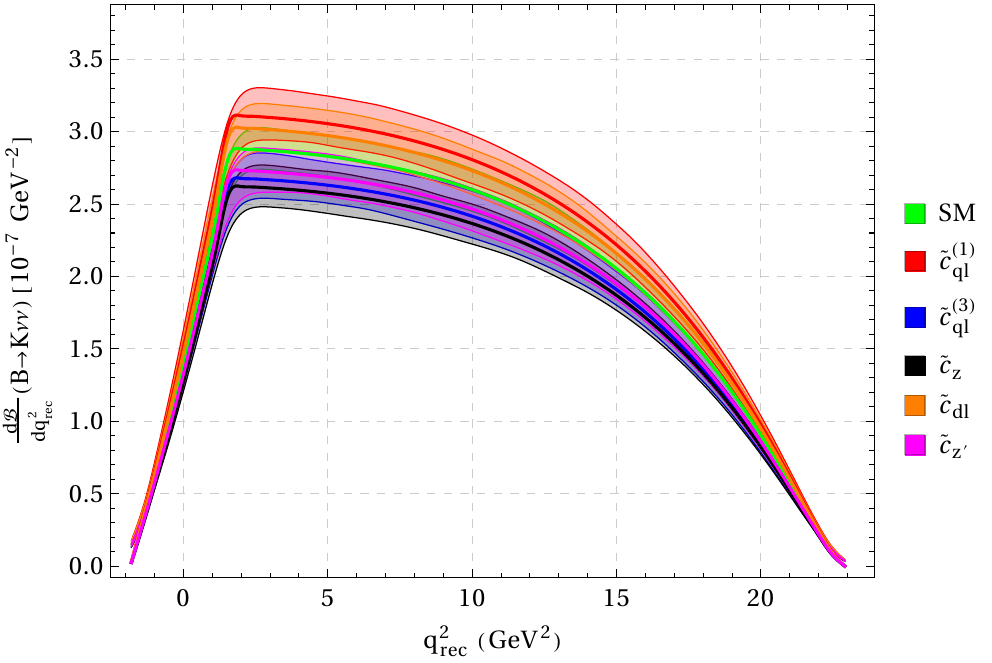}\\
\includegraphics[width=0.45\textwidth]{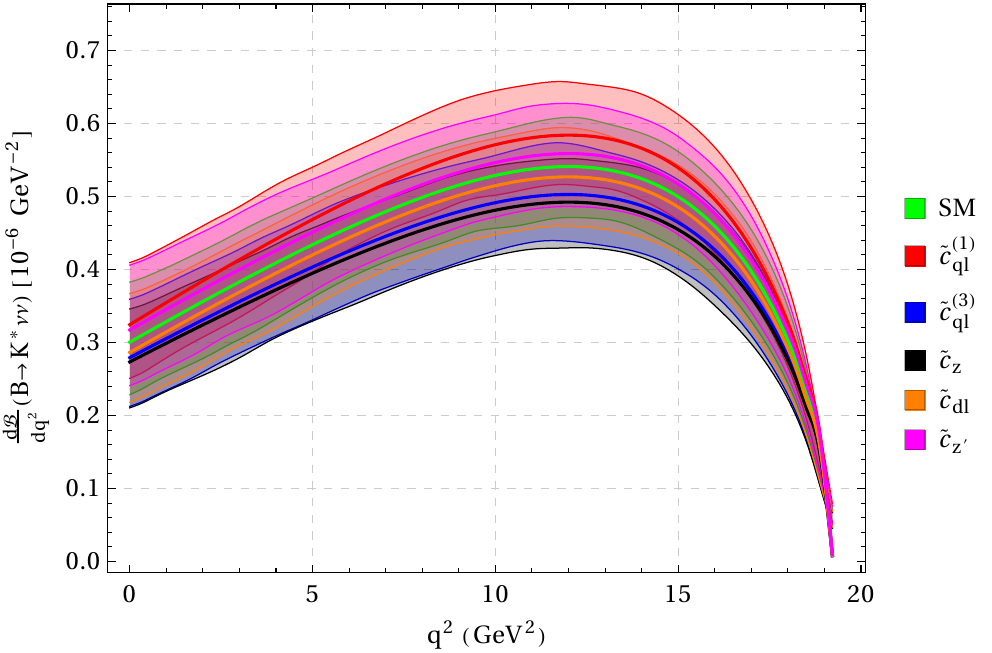}~~
\includegraphics[width=0.45\textwidth]{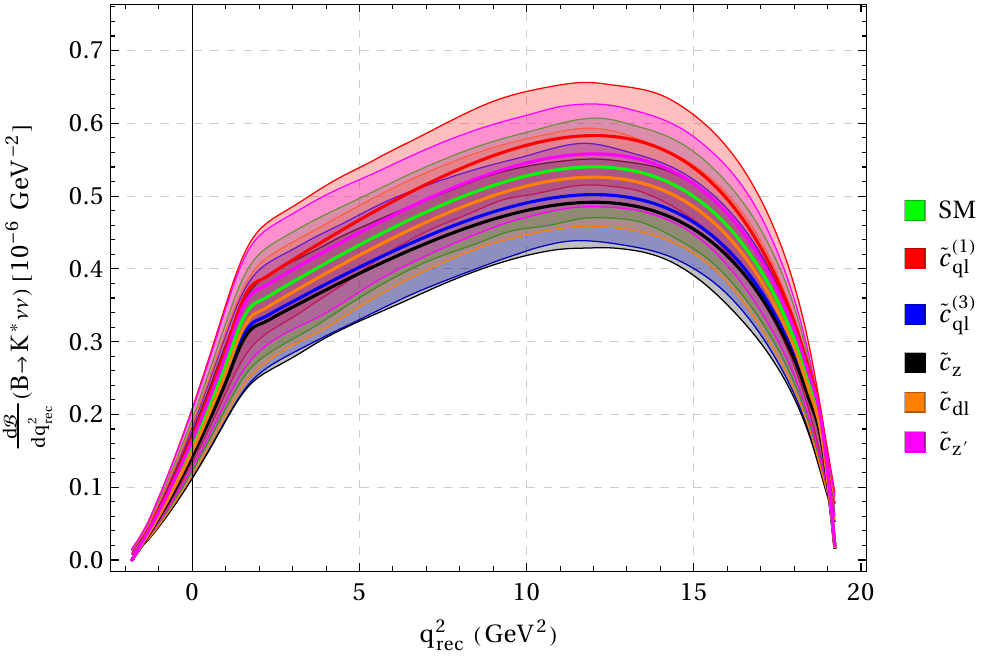}
\caption{Differential distributions for $B \to K \nu\bar{\nu}$ (top panels) and $B \to K^* \nu\bar{\nu}$ (bottom panels), shown as functions of $q^2$ (left) and $q^2_{\mathrm{rec}}$ (right) for different NP Wilson coefficients.}
\label{plotsforB2K}
\end{figure}

In Fig.~\ref{plotsforB2K}, we show the differential distributions as functions of $q^2$ (in left panels) and $q^2_{\rm rec}$ (in right panels) for $B \to K^{(*)} \nu \bar{\nu}$ decays, both within the SM and for the best-fit results of the various 1D NP scenarios considered in this work. 
The SM prediction is shown by the green curve, with the corresponding $1\sigma$ uncertainty indicated by the shaded band. 
The red, blue, black, yellow, and magenta curves, together with their respective $1\sigma$ uncertainty bands obtained from best-fit results by including correlations among form factors and NP Wilson coefficients, correspond to \cqlonebsT, \cqlthreebsT, \czbsT, \cdlbsT, and \czpbsT, respectively. In the plot legends, these scenarios are denoted by the shorthand labels
$\tilde c_{ql}^{(1)}$, $\tilde c_{ql}^{(3)}$, $\tilde c_Z$,
$\tilde c_{dl}$, and $\tilde c_{Z'}$, where the flavor indices have been suppressed for brevity.

We observe that for both $B\!\to\!K\nu\bar{\nu}$ and $B\!\to\!K^{*}\nu\bar{\nu}$ decays, the differential distributions in $q^2$ and the reconstructed variable $q^2_{\mathrm{rec}}$ exhibit the same shape as in the SM, decreasing smoothly with increasing $q^2$. 
This indicates that the NP scenarios considered here modify the short-distance Wilson coefficients $C_L^{32,22}$ and $C_R^{32,22}$ (see Eqs.~\eqref{eq:CLb2snunu} and \eqref{eq:CRb2snunu}), while the $q^2$ dependence is governed by the hadronic form factors. 
As a result, NP effects mainly rescale the overall normalization of the distributions, leaving their kinematic dependence essentially unchanged. Also note that by definition, the variable $q^2_{\mathrm{rec}}$ (given in eq.~\eqref{eq:qrec_def}) can take negative values depending on the kinamatics. 

The distributions corresponding to \cqlonebsT lie above the SM prediction, as this operator enhances the left-handed contribution through $C_L^{32,22}$ and increases the $b\!\to\!s\nu\bar{\nu}$ rate across the entire $q^2$ range.
In contrast, \cqlthreebsT and \czbsT shift $C_L^{32,22}$ with an opposite sign, leading to a partial cancellation of the SM contribution and resulting in spectra that lie below the SM curve.
This pattern reflects the $SU(2)_L$ structure of the SMEFT operators, which enforces different combinations of \cqlonebsT and \cqlthreebsT in the $b \to s \ell \ell$ and $b \to s \nu \bar{\nu}$ transitions.
Finally, the distributions obtained with \cdlbsT and \czpbsT remain closer to the SM expectation. Since these coefficients enter through $C_R^{32,22}$, they shift the $B \to K\nu\bar{\nu}$ and $B \to K^*\nu\bar{\nu}$ spectra in opposite directions.

It is interesting to note that, for several NP Wilson coefficients, the predicted uncertainty bands are clearly separated from the SM expectation.
Moreover, the hierarchy among the left-handed NP scenarios is preserved across both decay modes, indicating that they probe the same underlying short-distance dynamics.
In realistic experimental environments such as Belle~II, detector and reconstruction effects may reduce the overall normalization and slightly distort the high-$q^2$ region. However, these effects are not expected to alter the relative pattern among the different NP scenarios. 
This implies that the $q^2_{\mathrm{rec}}$ distributions remain theoretically clean and provide a powerful probe of the underlying short-distance Wilson coefficients.

\subsection{Implications for rare kaon decays}
\label{subsec:Ktopinunu}
\begin{table}
\caption{Experimental measurements, SM predictions, and NP predictions for kaon decay observables in the 1D and 2D fit scenarios. For entries given as intervals, the best-fit point lies outside the 68\% interval (see text), and the 68\% interval is quoted. 
}
\centering
\setlength{\tabcolsep}{8pt}
\renewcommand{\arraystretch}{1.5}
\resizebox{\textwidth}{!}{%
\begin{tabular}{l c c c c}
\hline
Coupling
& $\mathcal{B}(K \to \pi \nu\bar{\nu})$
& $\mathcal{B}(K_L \to \pi \nu\bar{\nu})$
& $\mathcal{B}(K \to \pi \mu^+\mu^-)$
& $\mathcal{B}(K_L \to \pi \mu^+\mu^-)$ \\
\hline
Exp.
& $\left(13.0^{+3.3}_{-3.0}\right)\times 10^{-11}$~\cite{NA62:2024pjp}
& $<2.2 \times 10^{-9}$~\cite{KOTO:2024zbl}
& $(9.15 \pm 0.08)\times 10^{-8}$~\cite{NA62:2022qes}
& $<3.8 \times 10^{-10}$~\cite{KTEV:2000ngj} \\
SM
& $\left(8.856^{+0.429}_{-0.410}\right)\times 10^{-11}$
& $\left(2.950^{+0.036}_{-0.035}\right)\times 10^{-11}$
& $\left(8.902^{+0.061}_{-0.067}\right)\times 10^{-8}$
& $\left(0.905^{+0.224}_{-0.206}\right)\times 10^{-11}$ \\
\hline
\multicolumn{5}{c}{Flavor-universal scenario} \\
\hline
\cqlonebsT
& $\left(3.018^{+8.180}_{-0.052}\right)\times 10^{-9}$
& $[0.070,\,3.405]\times 10^{-8}$
& $\left(8.022^{+0.386}_{-0.086}\right)\times 10^{-8}$
& $[0.032,\,1.378]\times 10^{-8}$ \\
\cqlthreebsT
& $\left(1.975^{+9.193}_{-0.000}\right)\times 10^{-9}$
& $[0.072,\,3.927]\times 10^{-8}$
& $\left(8.028^{+0.453}_{-0.109}\right)\times 10^{-8}$
& $[0.029,\,1.580]\times 10^{-8}$ \\
\czbsT
& $\left(8.011^{+2.897}_{-2.453}\right)\times 10^{-8}$
& $\left(2.450^{+0.898}_{-0.752}\right)\times 10^{-7}$
& $\left(9.500^{+0.248}_{-0.201}\right)\times 10^{-8}$
& $\left(2.349^{+0.885}_{-0.730}\right)\times 10^{-8}$ \\
\cdlbsT
& $\left(1.428^{+3.006}_{-0.053}\right)\times 10^{-9}$
& $[0.025,\,1.238]\times 10^{-8}$
& $\left(8.313^{+0.204}_{-0.079}\right)\times 10^{-8}$
& $[0.103,\,5.243]\times 10^{-9}$ \\
\czpbsT
& $\left(2.669^{+6.195}_{-0.362}\right)\times 10^{-9}$
& $[0.048,\,2.426]\times 10^{-8}$
& $\left(8.882^{+0.088}_{-0.039}\right)\times 10^{-8}$
& $[0.044,\,2.228]\times 10^{-9}$ \\
\cqebsT
& $-$ & $-$
& $\left(8.764^{+0.306}_{-0.183}\right)\times 10^{-8}$
& $[0.094,\,4.274]\times 10^{-9}$ \\
\cdebsT
& $-$ & $-$
& $\left(8.889^{+0.251}_{-0.196}\right)\times 10^{-8}$
& $[0.261,\,9.272]\times 10^{-10}$ \\
(\cqlonebsT,\,\czbsT)
& $\left(6.990^{+7.616}_{-2.082}\right)\times 10^{-8}$
& $\left(2.009^{+3.644}_{-1.611}\right)\times 10^{-7}$
& $\left(7.648^{+0.600}_{-0.281}\right)\times 10^{-8}$
& $\left(3.295^{+15.168}_{-0.918}\right)\times 10^{-9}$ \\
(\cqlthreebsT,\,\czbsT)
& $\left(3.724^{+4.420}_{-1.353}\right)\times 10^{-8}$
& $\left(1.396^{+1.929}_{-0.955}\right)\times 10^{-7}$
& $\left(7.780^{+0.653}_{-0.340}\right)\times 10^{-8}$
& $\left(1.083^{+1.852}_{-0.640}\right)\times 10^{-8}$ \\
\hline
\multicolumn{5}{c}{$U(3)^5$ symmetry scenario} \\
\hline
\cqlonesdT
& $\left(9.329^{+0.478}_{-0.422}\right)\times 10^{-11}$
& $\left(3.180^{+0.840}_{-0.621}\right)\times 10^{-11}$
& $\left(8.895^{+0.061}_{-0.066}\right)\times 10^{-8}$
& $\left(0.813^{+0.365}_{-0.259}\right)\times 10^{-11}$ \\
\cqlthreesdT
& $\left(8.420^{+0.474}_{-0.391}\right)\times 10^{-11}$
& $\left(2.754^{+0.730}_{-0.488}\right)\times 10^{-11}$
& $\left(8.895^{+0.061}_{-0.066}\right)\times 10^{-8}$
& $\left(0.818^{+0.387}_{-0.261}\right)\times 10^{-11}$ \\
\czsdT
& $\left(7.475^{+0.541}_{-0.486}\right)\times 10^{-11}$
& $\left(0.010^{+0.147}_{-0.007}\right)\times 10^{-11}$
& $\left(8.902^{+0.061}_{-0.067}\right)\times 10^{-8}$
& $\left(0.787^{+0.221}_{-0.196}\right)\times 10^{-11}$ \\
(\cqlonesdT,\,\czsdT)
& $\left(11.981^{+1.288}_{-0.866}\right)\times 10^{-11}$
& $\left(0.965^{+2.429}_{-0.712}\right)\times 10^{-11}$
& $\left(8.886^{+0.060}_{-0.065}\right)\times 10^{-8}$
& $\left(0.978^{+0.350}_{-0.237}\right)\times 10^{-11}$ \\
(\cqlthreesdT,\,\czsdT)
& $\left(9.956^{+0.834}_{-0.610}\right)\times 10^{-11}$
& $\left(0.714^{+1.419}_{-0.521}\right)\times 10^{-11}$
& $\left(8.890^{+0.061}_{-0.066}\right)\times 10^{-8}$
& $\left(0.727^{+0.318}_{-0.207}\right)\times 10^{-11}$ \\
\hline
\multicolumn{5}{c}{$U(2)^5$ symmetry scenario} \\
\hline
\cqlonesdT
& $\left(8.418^{+0.463}_{-0.393}\right)\times 10^{-11}$
& $\left(2.744^{+0.686}_{-0.469}\right)\times 10^{-11}$
& $\left(8.910^{+0.060}_{-0.065}\right)\times 10^{-8}$
& $\left(1.005^{+0.453}_{-0.326}\right)\times 10^{-11}$ \\
\cqlthreesdT
& $\left(9.327^{+0.494}_{-0.427}\right)\times 10^{-11}$
& $\left(3.168^{+0.894}_{-0.627}\right)\times 10^{-11}$
& $\left(8.910^{+0.060}_{-0.065}\right)\times 10^{-8}$
& $\left(1.000^{+0.472}_{-0.326}\right)\times 10^{-11}$ \\
\czsdT
& $\left(11.479^{+0.801}_{-0.728}\right)\times 10^{-11}$
& $\left(11.131^{+1.805}_{-1.712}\right)\times 10^{-11}$
& $\left(8.903^{+0.061}_{-0.067}\right)\times 10^{-8}$
& $\left(1.521^{+0.276}_{-0.235}\right)\times 10^{-11}$ \\
(\cqlonesdT,\,\czsdT)
& $\left(6.742^{+1.387}_{-0.979}\right)\times 10^{-11}$
& $\left(6.146^{+5.093}_{-3.561}\right)\times 10^{-11}$
& $\left(8.918^{+0.061}_{-0.067}\right)\times 10^{-8}$
& $\left(0.850^{+0.374}_{-0.233}\right)\times 10^{-11}$ \\
(\cqlthreesdT,\,\czsdT)
& $\left(8.312^{+1.081}_{-0.849}\right)\times 10^{-11}$
& $\left(6.736^{+3.442}_{-2.590}\right)\times 10^{-11}$
& $\left(8.914^{+0.061}_{-0.068}\right)\times 10^{-8}$
& $\left(1.273^{+0.472}_{-0.369}\right)\times 10^{-11}$ \\
\hline
\end{tabular}}
\label{tab:kaondecays}
\end{table}

In the present analysis, the $\chi^2$ fit is performed exclusively using rare $b \to s$ observables, and the corresponding SMEFT Wilson coefficients are thus determined solely from the $b \to s$ sector. In this section, we explore the implications for rare kaon decays using these best-fit coefficients.

\subsubsection{Flavor-universal SMEFT scenario}
\label{subsec:flavorUni}

As a simplifying, flavor-agnostic assumption, we first take the SMEFT coupling strengths in the $b \to s$ and $s \to d$ sectors to be identical, i.e.,
\begin{align}
[c_{X}]^{32,\alpha \alpha}= [c_{X}]^{21,\alpha \alpha}\,,
\end{align}
so that any differences in the corresponding low-energy coefficients $[\widetilde{c}_X]^{ij,\alpha}$ arise solely from the CKM factors entering their definition in Eq.~\eqref{eq:ctilde_def}. 
This setup allows us to test whether a flavor-universal NP structure, as preferred by the $b \to s$ data, remains compatible with rare kaon observables before introducing additional flavor-dependent effects. We note that, although the SMEFT coupling strengths are assumed to be identical at the electroweak scale, the corresponding effective Wilson coefficients relevant for the $B$-meson and kaon sectors can in principle differ due to renormalization-group (RG) evolution down to the respective low-energy scales. However, these running effects are expected to be negligible in the present analysis. In particular, electroweak running effects are numerically small, while the QCD running of vector quark currents vanishes at leading order owing to their zero anomalous dimensions. For this reason, RG running effects are not included in our numerical analysis, since our primary aim is to demonstrate the qualitative implications of the preferred $b \to s$ fit results for the kaon sector.


The results for the flavor-universal scenario are shown in first few rows of Table~\ref{tab:kaondecays}. We observe the following.
\begin{itemize}

\item {\bf $\mathcal{B}(K_{(L)} \to \pi \nu\bar{\nu})$:} For the $K^+ \to \pi^+ \nu\bar{\nu}$ decay channel, the predicted branching fractions show a significant enhancement compared to the SM value in the presence of all the NP couplings. In particular, the predictions are enhanced by one to three orders of magnitude, thereby exceeding the current experimental measurement.

A similar trend is observed for \(K_L \to \pi^0 \nu\bar{\nu}\). Owing to the sizable uncertainties in the imaginary parts of the NP Wilson coefficients, we quote the \(68\%\) C.L. interval for the branching fraction. The upper end of this interval can be enhanced by approximately three orders of magnitude relative to the SM prediction. Despite the relatively weak experimental constraint on this mode, primarily due to the presence of a \(\pi^0\) and two undetected neutrinos in the final state, all the NP predictions considered exceed the current experimental upper limit.

\item {\bf $\mathcal{B}(K_{(L)} \to \pi \mu^+\mu^-)$ :} 
For $K^+ \to \pi^+ \mu^+\mu^-$ decay channel, we find that in the presence of the coupling \czsdT, the predicted central values slightly exceed the experimental measurement, while still remaining consistent within the $2\sigma$ uncertainty. In contrast, for the remaining NP couplings, the predicted branching fractions are fully consistent with the experimental data.

For $K_L \to \pi^0 \mu^+\mu^-$ decay channel, a significant enhancement in the predicted branching fraction is observed, reaching up to three orders of magnitude above the SM expectation. Although the current experimental upper bound is about one order of magnitude above the SM prediction, the NP-enhanced branching ratios exceed this bound and are therefore excluded.
\end{itemize}

The results show that fitting the SMEFT Wilson coefficients particularly $[c_{ql}^{(1)}]^{32,22}$, $[c_{ql}^{(3)}]^{32,22}$, and $[c_{z}]^{32}$ to $b \to s \mu^+\mu^-$ data, and subsequently applying the same numerical values to $s \to d$ transitions (up to the corresponding matching factors), leads to excessively large predictions for rare kaon decays. 
In particular, the dimuon channel $K_L \to \pi^0 \mu^+\mu^-$ and the dineutrino mode $K \to \pi \nu\bar{\nu}$ are strongly enhanced and exceed the experimental bounds reported in Table~\ref{tab:kaondecays}. The decay $K_L \to \pi^0 \nu\bar{\nu}$ also exhibits a significant enhancement, although it remains within the current experimental upper limit.
This behaviour clearly indicates that a flavor-universal assignment of SMEFT coefficients to both $b \to s$ and $s \to d$ transitions is incompatible with the kaon data. Such a tension is consistent with the broader experimental pattern, where no significant deviations from the SM have been observed in transitions between the second and first quark generations.

The pronounced sensitivity of rare kaon decays to short-distance FCNC processes implies that any viable NP scenario addressing the $b \to s$ anomalies must exhibit a non-trivial flavor structure. In particular, the absence of CKM suppression in the equal-coupling assumption effectively removes the hierarchical separation between $b \to s$ and $s \to d$ amplitudes, thereby generating unacceptably large contributions in the kaon sector.
This observation motivates the consideration of  MFV, in which all sources of flavor and $\mathcal{CP}$ violation are aligned with the SM Yukawa couplings. Within MFV, flavor-changing transitions are governed by the CKM matrix, and the strong suppression of $s \to d$ processes relative to $b \to s$ transitions arises naturally from the hierarchy $|V_{td}| \ll |V_{tb}|$. The lack of such a CKM hierarchy in the flavor-universal scenario therefore renders it incompatible with the observed suppression of $s \to d$ transitions.

This situation strongly motivates the implementation of MFV-like flavor alignment and highlights the critical role of rare kaon decays as powerful probes of the flavor structure of NP. As emphasized in Refs.~\cite{Buras:2003jf,Buras:2005gr}, experimental bounds on rare $K$ decays provide some of the most stringent constraints on extensions of the SM in the flavor sector. In the following subsections, we explore explicit realizations of MFV based on $U(3)^5$ and $U(2)^5$ flavor symmetries, and assess their implications for rare kaon observables and the associated $s \to d$ constraints.

\subsubsection{$U(3)^5$ Minimal Flavor Violation }
\label{sec:MFV-U35}

In the framework of MFV, based on the full $U(3)^5$ flavor symmetry acting on the five fermion multiplets $Q_L$, $u_R$, $d_R$, $L_L$, and $e_R$, all sources of flavor violation are aligned with the Yukawa structure of the SM~\cite{Buras:2003jf,Isidori:2012ts,DAmbrosio:2002vsn}. In this setup, flavor violation arises solely through insertions of the Yukawa matrices, and NP operators are constructed as invariants under the $U(3)^5$ symmetry with Yukawa spurion insertions. 

For operators involving left-handed quark currents, the leading flavor-breaking structure is proportional to $(Y_u Y_u^\dagger)_{ij}$. Retaining only the dominant top-quark contribution, one obtains
\begin{align}
[c_X]^{ij,\alpha\alpha} = C_X\, y_t^2\, V_{ti}^* V_{tj}\,,
\end{align}
where $y_t \sim \mathcal{O}(1)$ denotes the top-quark Yukawa coupling. This structure enforces NP contributions to down-type FCNC transitions to inherit the same CKM hierarchy as in the SM. Consequently, the relative scaling between $s \to d$ and $b \to s$ amplitudes is given by
\begin{equation}
\frac{\mathcal{A}(s \to d)}{\mathcal{A}(b \to s)}
\propto
\frac{[c_X]^{21,\alpha\alpha}}{[c_X]^{32,\alpha\alpha}}
\sim
\left| \frac{V_{td}}{V_{tb}} \right|
\simeq 8.5 \times 10^{-3} .
\label{eqn:U35scaling}
\end{equation}

Having established the CKM-aligned structure of NP contributions in the $U(3)^5$ framework, we now translate the NP coefficients extracted from the $b \to s \mu^{+} \mu^{-}$ fit to the $s \to d$ sector using the scaling relation in Eq.~\eqref{eqn:U35scaling}. After implementing this rescaling, the predicted branching ratios return to values close to the SM expectations, as shown in the middle rows of Table~\ref{tab:kaondecays}. 

In particular, the dineutrino modes, as well as the $\mathcal{CP}$-violating decay $K_L\to \pi^0\mu^+\mu^-$, remain at the level of $\mathcal{O}(10^{-11})$, while the dimuon channel $K^+ \to \pi^+ \mu^+ \mu^-$ remains essentially unchanged from its SM prediction. No observable exhibits a significant enhancement, and all predictions are consistent with the experimental data within uncertainties. In contrast to the flavor-universal scenario discussed previously, the $U(3)^5$ MFV framework naturally restores consistency with kaon observables through the CKM suppression of $s \to d$ transitions.

\subsubsection{$U(2)^5$ Minimal Flavor Violation }
\label{sec:MFV-U25}

We now consider the $U(2)^5$ flavor symmetry framework, in which the first two generations transform as doublets under $U(2)$, while the third generation is treated as a singlet. In this setup, NP predominantly couples to third-generation fermions, and flavor violation arises through the breaking of the $U(2)^5$ symmetry by spurion fields that connect the light generations to the third one~\cite{Faroughy:2020ina}. Working in the down-quark mass basis, where $Q_L^3$ contains the physical $b_L$ field, the leading spurion structure can be parameterized as
\begin{equation}
\label{eq:U2spurion-final}
\widetilde{V} \equiv \begin{pmatrix}
\widetilde{V}_1 \\[1.0ex]
\widetilde{V}_2
\end{pmatrix}
\;=\;
-\,\kappa\,V_{ts}
\begin{pmatrix}
\delta\, \dfrac{V_{td}}{V_{ts}} \\[1.0ex]
1
\end{pmatrix},
\qquad
Q_L^i \;\to\; Q_L^i + \widetilde{V}_i\, Q_L^3, 
\quad i=1,2,
\end{equation}
where $\kappa = \mathcal{O}(1)$ controls the overall size of flavor breaking, while $\delta$ parameterizes possible deviations from the minimal structure, with $\kappa = 1$ corresponding to the minimal limit.

Now, for a semileptonic NP interaction constructed from third-generation fields of the following form,
\begin{equation}
(\bar Q_L^3 \Gamma_X Q_L^3)
(\bar L_L^3 \Gamma_X L_L^3)
\label{eq:u2op}
\end{equation}
the flavor-changing transitions are generated through insertions of the spurion $\widetilde{V}_i$. Expanding the operator in Eq.~\eqref{eq:u2op} in powers of the spurion, one obtains
\begin{equation}
\mathcal A(b \to s)
\sim
C_X\, \widetilde{V}_2^*
(\bar s_L \gamma_\mu b_L),
\qquad
\mathcal A(s \to d)
\sim
C_X\, \widetilde{V}_1^* \widetilde{V}_2
(\bar d_L \gamma_\mu s_L).
\end{equation}
Using Eq.~\eqref{eq:U2spurion-final}, the relative scaling between the two transitions is therefore given by
\begin{equation}
\frac{\mathcal A(s \to d)}
{\mathcal A(b \to s)}
\propto
\frac{[c_X]^{21,\alpha\alpha}}{[c_X]^{32,\alpha\alpha}}
\sim
\delta \,\kappa\,V_{td}.
\label{eqn:U25scaling}
\end{equation}

Unlike the $U(3)^5$ MFV scenario, where the suppression of flavor-changing transitions is entirely dictated by the CKM hierarchy, the $U(2)^5$ framework allows for additional flexibility through the parameters $\kappa$ and $\delta$. Nevertheless, the suppression of $s \to d$ transitions remains dominantly controlled by the small CKM element $V_{td}$, thereby naturally alleviating the tension with rare kaon observables. 
Implementing the scaling relation in Eq.~\eqref{eqn:U25scaling} with minimal set up (i.e., for $\kappa=\delta=1$) in our numerical analysis, we show in Table~\ref{tab:kaondecays} lower rows that the predicted branching fractions remain numerically close to their SM expectations and comfortably satisfy the current experimental bounds.

Therefore, the flavor symmetries $ {U(3)}^5$ and ${ U(2)}^5$ provide two complementary symmetry frameworks to describe the flavor structure of physics beyond the SM within the SMEFT. 
The $U(3)^5$ MFV limit fully aligns with the SM Yukawa structures, yields CKM-like suppressions, and is consistent with the stringent constraints from rare kaon decays.  
On the other hand, the minimally broken $U(2)^5$ framework allows significant NP effects in third-generation transitions, while flavor violation among the light families remains controlled by CKM-like suppression. Taken together, these symmetry patterns provide coherent and predictive descriptions of flavor and $\mathcal{CP}$ violation, with the observed hierarchies and possible NP effects emerging directly from the underlying symmetry structure of the effective field theory. Our analysis therefore illustrates how rare kaon decays provide a highly sensitive discriminator of the underlying flavor structure of NP scenarios addressing the $b \to s$ anomalies.

\subsection{ $\mathcal{CP}$ asymmetries in $B \to K^{(*)}\mu^+\mu^-$ from complex Wilson coefficients}
\label{subsec:ACP}

The presence of complex NP Wilson coefficients obtained in Table~\ref{tab_bestfit_extended} naturally motivates an investigation of possible $\mathcal{CP}$-violating observables in rare semileptonic B decays. Such observables provide sensitive probes of new weak phases and therefore offer complementary information to branching ratios and angular distributions. In particular, complex phases in semileptonic operators may induce direct $\mathcal{CP}$ asymmetries at a level potentially accessible to current or future measurements at LHCb and Belle~II. Experimentally, direct $\mathcal{CP}$ asymmetries have been measured in both charged and neutral $B \to K^{(*)}\mu^+\mu^-$ decay modes, including the dielectron channels. The Belle collaboration, using a data sample of $657\times10^6$ $B\bar B$ pairs, reported~\cite{Belle:2009zue}
\[
A_{\mathcal{CP}}(B^+\!\to K^+\mu^+\mu^-)=0.04\pm0.10, \qquad
A_{\mathcal{CP}}(B^0\!\to K^{*0}\mu^+\mu^-)=-0.10\pm0.10
\,.\]
The BaBar collaboration, with a dataset of $471\times10^6$ $B\bar B$ events, obtained~\cite{BaBar:2012mrf}
\[
A_{\mathcal{CP}}(B^+\!\to K^+\mu^+\mu^-)=-0.03\pm0.14\,, \qquad
A_{\mathcal{CP}}(B^0\!\to K^{*0}\mu^+\mu^-)=0.03\pm0.13\,.
\]
 More recently, the LHCb collaboration~\cite{LHCb:2014mit}, using $3\,\mathrm{fb}^{-1}$ of data collected at $\sqrt{s}=7$ and $8~\mathrm{TeV}$, measured
\[
A_{\mathcal{CP}}(B^+\!\to K^+\mu^+\mu^-)=0.012\pm0.017\,, \qquad
A_{\mathcal{CP}}(B^0\!\to K^{*0}\mu^+\mu^-)=-0.035\pm0.024\,,
\]
in the full-$q^2$ region. All current measurements are consistent with the SM expectations within experimental uncertainties, although asymmetries at the percent level remain experimentally allowed.

In this section, we explore the possible $\mathcal{CP}$ asymmetries arising in our analysis. For the decay $B^+\!\to K^+\mu^+\mu^-$,  the direct $\mathcal{CP}$ asymmetry  is given by
\begin{equation}
\mathcal{A}_{\mathcal{CP}}\equiv
\frac{\Gamma(\bar B\to \bar K^{(*)}\mu^+\mu^-)-\Gamma(B\to K^{(*)}\mu^+\mu^-)}
{\Gamma(\bar B\to \bar K^{(*)}\mu^+\mu^-)+\Gamma(B\to K^{(*)}\mu^+\mu^-)}\,,
\label{eq:ACP}
\end{equation}
where $\bar B\to\bar K^{(*)}$ denotes $B^-\to K^-$ ($\bar B^0\to\bar K^{*0}$) and $B\to K^{(*)}$ denotes $B^+\to K^+$ ($B^0\to K^{*0}$) for the charged (neutral) mode.
where $\Gamma(B^\pm)$ denote the respective partial width integrated over a given $q^2$ bin.
In the SM, the dominant short-distance amplitude is proportional to $V_{tb}^*V_{ts}$,
while the subleading contribution proportional to $V_{ub}^*V_{us}$ carries a sizable weak phase
but is strongly CKM suppressed.
After applying unitarity to eliminate the $V_{cb}^*V_{cs}$ term,
the remaining amplitudes are nearly aligned in weak phase.
Strong phases arise mainly from long-distance rescattering,
in particular from intermediate $c\bar c$ loops entering $C_9^{\rm eff}$,
but their overall effect remains small far from the $c\bar c$ resonance region.
As a result, the expected asymmetry within the SM is tiny,
$|\mathcal{A}_{\mathcal{CP}}|\!\lesssim\!10^{-3}$.
Any effect at the percent level would therefore signal the presence of new complex phases
in the semileptonic operators, such as $\mathcal{O}_9$, $\mathcal{O}_{10}$,
or their chirality-flipped counterparts $\mathcal{O}_9'$ and $\mathcal{O}_{10}'$.

\begin{table}
\caption{Experimental measurements, SM predictions, and NP predictions for the direct $\mathcal{CP}$ asymmetry
$A_{\mathcal{CP}}$ in $B\to K\mu^+\mu^-$ and $B\to K^*\mu^+\mu^-$ across different $q^2$ bins for the 1D and 2D fit scenarios.}
\centering
\small
\setlength{\tabcolsep}{6pt}
\renewcommand{\arraystretch}{1.6}
\resizebox{\textwidth}{!}{%
\begin{tabular}{c cccccccc}
\hline
\multicolumn{9}{c}{$A_{\mathcal{CP}}(B \to K \mu^+\mu^-)\times 10^{3}$} \\
\hline
$q^2\,[\mathrm{GeV}^2]$
& [0.1,\,0.98] & [1.1,\,2] & [2,\,3] & [3,\,4]
& [4,\,5] & [5,\,6] & [6,\,7] & [7,\,8] \\
\hline
Exp.~\cite{LHCb:2014mit}
& $88\pm57\pm1$ & $-4\pm68\pm2$ & $42\pm59\pm1$
& $-34\pm63\pm1$ & $-21\pm64\pm1$
& $31\pm62\pm2$ & $26\pm60\pm1$ & $41\pm59\pm2$ \\
SM
& $0.168 \pm 0.021$ & $0.679 \pm 0.021$ & $0.904 \pm 0.021$ & $1.069 \pm 0.020$
& $1.198 \pm 0.019$ & $1.306 \pm 0.019$ & $1.402 \pm 0.018$ & $1.491 \pm 0.018$ \\
\hline
\textbf{\cqlonebsT}
& $0.233^{+5.868}_{-5.886}$ & $0.851^{+5.470}_{-5.645}$ & $1.123^{+5.243}_{-5.487}$ & $1.322^{+5.037}_{-5.332}$
& $1.478^{+4.836}_{-5.172}$ & $1.609^{+4.622}_{-4.992}$ & $1.725^{+4.378}_{-4.777}$ & $1.832^{+4.076}_{-4.502}$ \\
\textbf{\cqlthreebsT}
& $0.331^{+5.875}_{-6.319}$ & $0.943^{+5.569}_{-6.043}$ & $1.210^{+5.381}_{-5.870}$ & $1.406^{+5.202}_{-5.702}$
& $1.559^{+5.023}_{-5.530}$ & $1.686^{+4.827}_{-5.338}$ & $1.798^{+4.598}_{-5.111}$ & $1.900^{+4.311}_{-4.815}$ \\
\textbf{\czbsT}
& $-0.724^{+0.161}_{-0.145}$ & $-0.061^{+0.159}_{-0.150}$ & $0.238^{+0.155}_{-0.134}$ & $0.464^{+0.151}_{-0.131}$
& $0.646^{+0.150}_{-0.132}$ & $0.805^{+0.143}_{-0.135}$ & $0.955^{+0.138}_{-0.136}$ & $1.104^{+0.131}_{-0.137}$ \\
\textbf{\cdlbsT}
& $0.333^{+3.272}_{-3.240}$ & $0.902^{+3.060}_{-3.073}$ & $1.150^{+2.938}_{-2.976}$ & $1.332^{+2.826}_{-2.885}$
& $1.473^{+2.716}_{-2.792}$ & $1.591^{+2.599}_{-2.690}$ & $1.693^{+2.464}_{-2.571}$ & $1.787^{+2.297}_{-2.419}$ \\
\textbf{\czpbsT}
& $0.192^{+0.169}_{-0.189}$ & $0.778^{+0.142}_{-0.191}$ & $1.036^{+0.144}_{-0.186}$ & $1.224^{+0.142}_{-0.176}$
& $1.371^{+0.126}_{-0.174}$ & $1.495^{+0.130}_{-0.177}$ & $1.603^{+0.118}_{-0.162}$ & $1.703^{+0.116}_{-0.143}$ \\
\textbf{\cqebsT}
& $0.358^{+2.316}_{-2.421}$ & $0.842^{+2.203}_{-2.315}$ & $1.054^{+2.133}_{-2.239}$ & $1.208^{+2.055}_{-2.159}$
& $1.327^{+1.976}_{-2.082}$ & $1.426^{+1.891}_{-1.998}$ & $1.512^{+1.792}_{-1.902}$ & $1.590^{+1.670}_{-1.782}$ \\
\textbf{\cdebsT}
& $0.172^{+1.141}_{-1.153}$ & $0.682^{+1.119}_{-1.056}$ & $0.905^{+1.099}_{-1.018}$ & $1.070^{+1.067}_{-0.990}$
& $1.198^{+1.025}_{-0.960}$ & $1.306^{+0.980}_{-0.927}$ & $1.402^{+0.929}_{-0.887}$ & $1.491^{+0.866}_{-0.836}$ \\
\hline
\textbf{(\cqlonebsT,\,\czbsT)}
& $6.933^{+3.592}_{-4.687}$ & $7.051^{+3.400}_{-4.424}$ & $7.046^{+3.238}_{-4.267}$ & $7.000^{+3.150}_{-4.119}$
& $6.921^{+3.060}_{-3.972}$ & $6.805^{+2.961}_{-3.813}$ & $6.641^{+2.842}_{-3.629}$ & $6.405^{+2.664}_{-3.399}$ \\
\textbf{(\cqlthreebsT,\,\czbsT)}
& $0.767^{+4.885}_{-6.450}$ & $1.249^{+4.587}_{-6.124}$ & $1.456^{+4.451}_{-5.927}$ & $1.606^{+4.316}_{-5.740}$
& $1.720^{+4.177}_{-5.552}$ & $1.813^{+3.999}_{-5.345}$ & $1.891^{+3.794}_{-5.102}$ & $1.958^{+3.542}_{-4.794}$ \\
\hline
\multicolumn{9}{c}{$A_{\mathcal{CP}}(B \to K^* \mu^+\mu^-)\times 10^{3}$} \\
\hline
$q^2\,[\mathrm{GeV}^2]$
& [0.1,\,0.98] & [1.1,\,2] & [2,\,3] & [3,\,4]
& [4,\,5] & [5,\,6] & [6,\,7] & [7,\,8] \\
\hline
Exp.~\cite{LHCb:2014mit}
& $-87 \pm 60 \pm 6$
& $-176 \pm 106 \pm 9$
& $-146 \pm 102 \pm 8$
& $-13 \pm 113 \pm 14$
& $-76 \pm 106 \pm 12$
& $-30 \pm 97 \pm 9$
& $20 \pm 95 \pm 8$
& $99 \pm 87 \pm 6$ \\
SM
& $3.341 \pm 0.377$ & $1.433 \pm 0.228$ & $0.933 \pm 0.135$ & $0.770 \pm 0.131$
& $0.758 \pm 0.133$ & $0.813 \pm 0.129$ & $0.898 \pm 0.122$ & $0.999 \pm 0.114$ \\
\hline
\textbf{\cqlonebsT}
& $3.897^{+0.731}_{-0.967}$ & $2.018^{+1.909}_{-2.059}$ & $1.347^{+2.325}_{-2.619}$ & $1.069^{+2.554}_{-2.598}$
& $1.000^{+2.629}_{-2.627}$ & $1.034^{+2.620}_{-2.769}$ & $1.118^{+2.529}_{-2.791}$ & $1.229^{+2.340}_{-2.644}$ \\
\textbf{\cqlthreebsT}
& $3.908^{+0.779}_{-1.024}$ & $2.046^{+2.022}_{-1.935}$ & $1.384^{+2.465}_{-2.716}$ & $1.110^{+2.719}_{-2.968}$
& $1.044^{+2.818}_{-2.888}$ & $1.078^{+2.757}_{-2.923}$ & $1.160^{+2.622}_{-2.864}$ & $1.268^{+2.416}_{-2.699}$ \\
\textbf{\czbsT}
& $3.495^{+0.420}_{-0.544}$ & $1.409^{+0.371}_{-0.406}$ & $0.765^{+0.222}_{-0.222}$ & $0.530^{+0.134}_{-0.159}$
& $0.498^{+0.130}_{-0.154}$ & $0.562^{+0.104}_{-0.144}$ & $0.677^{+0.102}_{-0.136}$ & $0.824^{+0.091}_{-0.117}$ \\
\textbf{\cdlbsT}
& $2.975^{+0.902}_{-1.104}$ & $1.164^{+1.469}_{-1.597}$ & $0.764^{+1.404}_{-1.672}$ & $0.649^{+1.309}_{-1.616}$
& $0.657^{+1.240}_{-1.520}$ & $0.717^{+1.072}_{-1.412}$ & $0.803^{+0.940}_{-1.266}$ & $0.902^{+0.812}_{-1.106}$ \\
\textbf{\czpbsT}
& $3.129^{+0.365}_{-0.422}$ & $1.271^{+0.174}_{-0.227}$ & $0.825^{+0.157}_{-0.195}$ & $0.685^{+0.133}_{-0.155}$
& $0.679^{+0.133}_{-0.160}$ & $0.732^{+0.118}_{-0.153}$ & $0.813^{+0.117}_{-0.142}$ & $0.909^{+0.111}_{-0.128}$ \\
\textbf{\cqebsT}
& $3.368^{+0.535}_{-0.586}$ & $1.493^{+0.774}_{-0.797}$ & $1.002^{+0.980}_{-1.019}$ & $0.840^{+0.993}_{-1.031}$
& $0.825^{+0.978}_{-1.081}$ & $0.874^{+0.942}_{-1.086}$ & $0.953^{+0.888}_{-1.055}$ & $1.044^{+0.835}_{-0.984}$ \\
\textbf{\cdebsT}
& $3.340^{+0.597}_{-0.580}$ & $1.431^{+0.728}_{-0.774}$ & $0.932^{+0.780}_{-0.928}$ & $0.769^{+0.753}_{-0.846}$
& $0.758^{+0.692}_{-0.768}$ & $0.813^{+0.589}_{-0.697}$ & $0.899^{+0.513}_{-0.631}$ & $1.000^{+0.458}_{-0.556}$ \\
\hline
\textbf{(\cqlonebsT,\,\czbsT)}
& $4.754^{+0.844}_{-0.832}$ & $4.007^{+1.679}_{-1.741}$ & $3.816^{+1.907}_{-2.056}$ & $3.796^{+1.946}_{-2.098}$
& $3.829^{+1.837}_{-2.149}$ & $3.853^{+1.771}_{-2.142}$ & $3.832^{+1.683}_{-2.051}$ & $3.735^{+1.574}_{-1.890}$ \\
\textbf{(\cqlthreebsT,\,\czbsT)}
& $3.872^{+0.717}_{-0.954}$ & $2.055^{+1.718}_{-2.168}$ & $1.433^{+2.157}_{-2.513}$ & $1.168^{+2.463}_{-2.863}$
& $1.089^{+2.587}_{-2.950}$ & $1.098^{+2.616}_{-2.867}$ & $1.151^{+2.559}_{-2.861}$ & $1.223^{+2.423}_{-2.733}$ \\
\hline
\end{tabular}}
\label{tab:Acp_para}
\end{table}

The results of our analysis are summarised in Table~\ref{tab:Acp_para}, which compares the SM expectations, current measurements, and predictions obtained in respective fit scenarios. 
Within the SM, $\mathcal{A}_{\mathcal{CP}}$ remains strongly suppressed in both $B\!\to K\mu^+\mu^-$ and $B\!\to K^{*}\mu^+\mu^-$, at the level of $\mathcal{O}(10^{-4})$, consistent with the alignment of weak phases and the smallness of hadronic rescattering. 

In the presence of complex Wilson coefficients, the central values are only mildly modified, while the $1\sigma$ ranges broaden considerably, reflecting the weakly constrained imaginary parts of the fitted coefficients. Among the 1D scenarios, \cqlonebsT and \cqlthreebsT allow the largest asymmetries, with $|\mathcal{A}_{\mathcal{CP}}|$ up to $\simeq6\times10^{-3}$ in $B\!\to K\mu^+\mu^-$ and $\simeq(4$--$5)\times10^{-3}$ in $B\!\to K^{*}\mu^+\mu^-$ at $1\sigma$. They are followed by \cdlbsT, \cqebsT{} and \cdebsT, whereas \czbsT{} and \czpbsT{} remain at the SM level.

In the 2D fit scenarios, (\cqlonebsT,\,\czbsT) yields the largest effects among all scenarios. It gives $\mathcal{A}_{\mathcal{CP}}(B\!\to K\mu^+\mu^-)\simeq(6$--$7)\times10^{-3}$, nearly independent of $q^2$ and approaching $1\%$ at the upper end of the $1\sigma$ range. For $B\!\to K^{*}\mu^+\mu^-$ it gives $\mathcal{A}_{\mathcal{CP}}\simeq(4$--$5)\times10^{-3}$, about four times the SM value for $q^2\gtrsim2\,\mathrm{GeV}^2$.
The scenario (\cqlthreebsT,\,\czbsT) instead leads to values similar to those of the corresponding 1D scenario.
In all cases, the predicted asymmetries remain well below the present experimental sensitivity. This is expected, since a direct $\mathcal{CP}$ asymmetry requires both weak- and strong-phase differences, and in our framework the latter arise only from perturbative quark loops.

\begin{table}
\caption{SM and NP predictions for the CP-averaged angular observables $S_{7,8,9}$ of $B^0\to K^{*0}\mu^+\mu^-$ in the 1D and 2D fit scenarios}
\centering
\small
\setlength{\tabcolsep}{6pt}
\renewcommand{\arraystretch}{1.6}
\resizebox{\textwidth}{!}{%
\begin{tabular}{c cccccc}
\hline
$q^2\,[\mathrm{GeV}^2]$
& [0.06,\,0.98] & [1.1,\,2.5] & [2.5,\,4.0] & [4.0,\,6.0] & [1.1,\,6.0] & [6.0,\,8.0] \\
\hline
\multicolumn{7}{c}{$S_7\times10^{3}$} \\
\hline
Exp.~\cite{LHCb:2025mqb}
& $-7 \pm 28 \pm 2$ & $-112 \pm 42 \pm 2$ & $-107 \pm 44 \pm 3$ & $-87 \pm 33 \pm 2$ & -- & $-76 \pm 30 \pm 2$ \\
\hline
SM
& $-10.078 \pm 0.911$ & $-11.788 \pm 1.613$ & $-8.603 \pm 1.350$ & $-5.996 \pm 1.058$ & $-8.351 \pm 1.272$ & $-4.137 \pm 0.898$ \\
\hline
\textbf{\cqlonebsT}
& $-9.417^{+1.232}_{-0.762}$ & $-13.492^{+2.667}_{-1.103}$ & $-10.531^{+2.407}_{-1.083}$ & $-7.510^{+1.864}_{-0.900}$ & $-10.127^{+2.193}_{-1.088}$ & $-5.203^{+1.426}_{-0.780}$ \\
\textbf{\cqlthreebsT}
& $-9.420^{+1.298}_{-0.794}$ & $-13.475^{+2.951}_{-1.135}$ & $-10.512^{+2.729}_{-1.033}$ & $-7.495^{+1.928}_{-0.857}$ & $-10.109^{+2.509}_{-0.950}$ & $-5.192^{+1.538}_{-0.770}$ \\
\textbf{\czbsT}
& $-7.874^{+0.774}_{-0.857}$ & $-10.398^{+1.721}_{-1.661}$ & $-7.824^{+1.334}_{-1.333}$ & $-5.469^{+0.946}_{-1.056}$ & $-7.541^{+1.258}_{-1.205}$ & $-3.743^{+0.833}_{-0.865}$ \\
\textbf{\cdlbsT}
& $-10.188^{+0.939}_{-0.732}$ & $-10.809^{+1.738}_{-1.576}$ & $-7.855^{+1.414}_{-1.239}$ & $-5.553^{+1.071}_{-0.792}$ & $-7.682^{+1.344}_{-1.094}$ & $-3.895^{+0.844}_{-0.691}$ \\
\textbf{\czpbsT}
& $-10.648^{+1.042}_{-0.767}$ & $-11.671^{+1.740}_{-1.570}$ & $-8.495^{+1.421}_{-1.376}$ & $-5.980^{+0.982}_{-0.961}$ & $-8.289^{+1.323}_{-1.293}$ & $-4.174^{+0.820}_{-0.813}$ \\
\textbf{\cqebsT}
& $-10.284^{+1.040}_{-0.859}$ & $-11.950^{+1.787}_{-1.506}$ & $-8.721^{+1.511}_{-1.365}$ & $-6.088^{+1.124}_{-0.890}$ & $-8.472^{+1.387}_{-1.221}$ & $-4.208^{+0.898}_{-0.784}$ \\
\textbf{\cdebsT}
& $-10.055^{+1.071}_{-0.903}$ & $-11.764^{+1.893}_{-1.800}$ & $-8.585^{+1.487}_{-1.434}$ & $-5.984^{+1.064}_{-1.014}$ & $-8.334^{+1.446}_{-1.288}$ & $-4.128^{+0.849}_{-0.888}$ \\
\hline
\textbf{(\cqlonebsT,\,\czbsT)}
& $-10.096^{+1.405}_{-0.865}$ & $-13.314^{+2.912}_{-1.569}$ & $-10.194^{+2.594}_{-1.363}$ & $-7.283^{+2.020}_{-1.121}$ & $-9.867^{+2.530}_{-1.387}$ & $-5.093^{+1.502}_{-1.006}$ \\
\textbf{(\cqlthreebsT,\,\czbsT)}
& $-9.939^{+1.424}_{-0.909}$ & $-13.426^{+2.697}_{-0.857}$ & $-10.360^{+2.473}_{-0.721}$ & $-7.416^{+1.821}_{-0.662}$ & $-10.014^{+2.296}_{-0.698}$ & $-5.181^{+1.385}_{-0.717}$ \\
\hline
\multicolumn{7}{c}{$S_8\times10^{3}$} \\
\hline
Exp.~\cite{LHCb:2025mqb}
& $-36 \pm 33 \pm 2$ & $-41 \pm 45 \pm 4$ & $13 \pm 48 \pm 2$ & $-12 \pm 35 \pm 1$ & -- & $-69 \pm 30 \pm 3$ \\
\hline
SM
& $-1.591 \pm 0.147$ & $-2.369 \pm 0.282$ & $-1.966 \pm 0.260$ & $-1.552 \pm 0.221$ & $-1.897 \pm 0.241$ & $-1.226 \pm 0.208$ \\
\hline
\textbf{\cqlonebsT}
& $-0.887^{+0.245}_{-0.171}$ & $-1.879^{+0.375}_{-0.134}$ & $-1.810^{+0.346}_{-0.145}$ & $-1.563^{+0.319}_{-0.134}$ & $-1.726^{+0.333}_{-0.115}$ & $-1.318^{+0.329}_{-0.143}$ \\
\textbf{\cqlthreebsT}
& $-0.891^{+0.294}_{-0.195}$ & $-1.882^{+0.464}_{-0.147}$ & $-1.811^{+0.398}_{-0.133}$ & $-1.562^{+0.360}_{-0.131}$ & $-1.726^{+0.384}_{-0.120}$ & $-1.317^{+0.336}_{-0.145}$ \\
\textbf{\czbsT}
& $-1.616^{+0.177}_{-0.166}$ & $-2.740^{+0.295}_{-0.365}$ & $-2.360^{+0.289}_{-0.373}$ & $-1.880^{+0.288}_{-0.223}$ & $-2.261^{+0.267}_{-0.305}$ & $-1.486^{+0.300}_{-0.236}$ \\
\textbf{\cdlbsT}
& $-1.848^{+0.202}_{-0.184}$ & $-2.216^{+0.296}_{-0.293}$ & $-1.653^{+0.276}_{-0.259}$ & $-1.201^{+0.247}_{-0.216}$ & $-1.614^{+0.263}_{-0.261}$ & $-0.894^{+0.236}_{-0.179}$ \\
\textbf{\czpbsT}
& $-1.531^{+0.159}_{-0.145}$ & $-2.112^{+0.222}_{-0.251}$ & $-1.732^{+0.214}_{-0.245}$ & $-1.369^{+0.187}_{-0.218}$ & $-1.680^{+0.192}_{-0.225}$ & $-1.088^{+0.207}_{-0.208}$ \\
\textbf{\cqebsT}
& $-1.462^{+0.249}_{-0.237}$ & $-2.191^{+0.380}_{-0.284}$ & $-1.833^{+0.355}_{-0.248}$ & $-1.458^{+0.250}_{-0.185}$ & $-1.769^{+0.341}_{-0.205}$ & $-1.162^{+0.248}_{-0.167}$ \\
\textbf{\cdebsT}
& $-1.600^{+0.210}_{-0.221}$ & $-2.377^{+0.280}_{-0.294}$ & $-1.968^{+0.261}_{-0.255}$ & $-1.550^{+0.224}_{-0.234}$ & $-1.899^{+0.246}_{-0.235}$ & $-1.223^{+0.249}_{-0.200}$ \\
\hline
\textbf{(\cqlonebsT,\,\czbsT)}
& $-0.359^{+0.325}_{-0.322}$ & $-0.957^{+0.510}_{-0.412}$ & $-1.026^{+0.410}_{-0.316}$ & $-0.969^{+0.331}_{-0.287}$ & $-0.981^{+0.386}_{-0.303}$ & $-0.885^{+0.272}_{-0.233}$ \\
\textbf{(\cqlthreebsT,\,\czbsT)}
& $-0.409^{+0.275}_{-0.267}$ & $-1.062^{+0.454}_{-0.400}$ & $-1.126^{+0.386}_{-0.269}$ & $-1.051^{+0.355}_{-0.239}$ & $-1.075^{+0.391}_{-0.271}$ & $-0.951^{+0.336}_{-0.187}$ \\
\hline
\multicolumn{7}{c}{$S_9\times10^{3}$} \\
\hline
Exp.~\cite{LHCb:2025mqb}
& $-38 \pm 27 \pm 1$ & $-30 \pm 30 \pm 2$ & $-7 \pm 36 \pm 4$ & $-40 \pm 27 \pm 2$ & -- & $48 \pm 26 \pm 3$ \\
\hline
SM
& $-0.041 \pm 0.081$ & $-0.127 \pm 0.247$ & $-0.158 \pm 0.288$ & $-0.175 \pm 0.296$ & $-0.157 \pm 0.279$ & $-0.189 \pm 0.297$ \\
\hline
\textbf{\cqlonebsT}
& $-0.023^{+0.043}_{-0.046}$ & $-0.102^{+0.183}_{-0.202}$ & $-0.146^{+0.234}_{-0.272}$ & $-0.176^{+0.250}_{-0.311}$ & $-0.146^{+0.225}_{-0.268}$ & $-0.203^{+0.282}_{-0.335}$ \\
\textbf{\cqlthreebsT}
& $-0.023^{+0.042}_{-0.046}$ & $-0.102^{+0.179}_{-0.192}$ & $-0.146^{+0.234}_{-0.261}$ & $-0.176^{+0.251}_{-0.308}$ & $-0.146^{+0.225}_{-0.257}$ & $-0.203^{+0.280}_{-0.329}$ \\
\textbf{\czbsT}
& $-0.042^{+0.079}_{-0.097}$ & $-0.147^{+0.278}_{-0.330}$ & $-0.189^{+0.314}_{-0.381}$ & $-0.211^{+0.315}_{-0.389}$ & $-0.187^{+0.299}_{-0.368}$ & $-0.229^{+0.321}_{-0.415}$ \\
\textbf{\cdlbsT}
& $-0.226^{+0.081}_{-0.084}$ & $-0.447^{+0.209}_{-0.233}$ & $-0.373^{+0.235}_{-0.270}$ & $-0.264^{+0.249}_{-0.258}$ & $-0.346^{+0.220}_{-0.253}$ & $-0.180^{+0.247}_{-0.279}$ \\
\textbf{\czpbsT}
& $-0.055^{+0.075}_{-0.092}$ & $-0.142^{+0.204}_{-0.244}$ & $-0.160^{+0.234}_{-0.261}$ & $-0.165^{+0.238}_{-0.272}$ & $-0.157^{+0.226}_{-0.263}$ & $-0.171^{+0.243}_{-0.298}$ \\
\textbf{\cqebsT}
& $-0.038^{+0.069}_{-0.079}$ & $-0.118^{+0.217}_{-0.237}$ & $-0.147^{+0.240}_{-0.281}$ & $-0.164^{+0.249}_{-0.304}$ & $-0.147^{+0.231}_{-0.278}$ & $-0.179^{+0.248}_{-0.323}$ \\
\textbf{\cdebsT}
& $-0.045^{+0.101}_{-0.120}$ & $-0.136^{+0.238}_{-0.338}$ & $-0.164^{+0.262}_{-0.337}$ & $-0.178^{+0.275}_{-0.336}$ & $-0.162^{+0.266}_{-0.316}$ & $-0.190^{+0.269}_{-0.331}$ \\
\hline
\textbf{(\cqlonebsT,\,\czbsT)}
& $-0.010^{+0.018}_{-0.025}$ & $-0.052^{+0.086}_{-0.110}$ & $-0.083^{+0.129}_{-0.154}$ & $-0.110^{+0.160}_{-0.163}$ & $-0.085^{+0.131}_{-0.143}$ & $-0.137^{+0.194}_{-0.221}$ \\
\textbf{(\cqlthreebsT,\,\czbsT)}
& $-0.011^{+0.020}_{-0.025}$ & $-0.058^{+0.097}_{-0.104}$ & $-0.091^{+0.143}_{-0.146}$ & $-0.119^{+0.173}_{-0.192}$ & $-0.093^{+0.140}_{-0.148}$ & $-0.147^{+0.207}_{-0.204}$ \\
\hline
\end{tabular}}
\label{tab:S789}
\end{table}

\begin{table}
\caption{SM and NP predictions for the angular CP asymmetries $A_{7,8,9}$ of $B^0\to K^{*0}\mu^+\mu^-$ in the 1D and 2D fit scenarios, in units of $10^{-3}$.}
\centering
\small
\setlength{\tabcolsep}{6pt}
\renewcommand{\arraystretch}{1.6}
\resizebox{\textwidth}{!}{%
\begin{tabular}{c cccccc}
\hline
$q^2\,[\mathrm{GeV}^2]$
& [0.06,\,0.98] & [1.1,\,2.5] & [2.5,\,4.0] & [4.0,\,6.0] & [1.1,\,6.0] & [6.0,\,8.0] \\
\hline
\multicolumn{7}{c}{$A_7\times10^{3}$} \\
\hline
SM
& $0.952 \pm 0.086$ & $0.944 \pm 0.129$ & $0.580 \pm 0.091$ & $0.331 \pm 0.058$ & $0.573 \pm 0.086$ & $0.181 \pm 0.039$ \\
\hline
\textbf{\cqlonebsT}
& $1.374^{+74.305}_{-76.624}$ & $1.740^{+91.239}_{-97.175}$ & $1.198^{+64.165}_{-72.039}$ & $0.739^{+42.644}_{-47.111}$ & $1.164^{+61.957}_{-67.967}$ & $0.435^{+27.614}_{-31.677}$ \\
\textbf{\cqlthreebsT}
& $2.505^{+75.519}_{-77.152}$ & $3.272^{+91.161}_{-94.828}$ & $2.328^{+66.623}_{-68.990}$ & $1.493^{+41.961}_{-44.648}$ & $2.254^{+63.119}_{-65.141}$ & $0.917^{+27.604}_{-31.545}$ \\
\textbf{\czbsT}
& $-163.790^{+30.940}_{-32.103}$ & $-205.029^{+45.979}_{-44.447}$ & $-146.083^{+34.357}_{-30.377}$ & $-95.731^{+24.711}_{-21.789}$ & $-141.165^{+32.934}_{-28.113}$ & $-60.615^{+16.977}_{-17.073}$ \\
\textbf{\cdlbsT}
& $-0.662^{+35.409}_{-37.956}$ & $-0.770^{+34.105}_{-40.112}$ & $-0.595^{+23.424}_{-28.317}$ & $-0.437^{+16.009}_{-18.422}$ & $-0.575^{+22.983}_{-27.555}$ & $-0.311^{+10.810}_{-11.501}$ \\
\textbf{\czpbsT}
& $1.073^{+30.482}_{-30.452}$ & $1.005^{+29.862}_{-30.477}$ & $0.621^{+20.797}_{-20.159}$ & $0.362^{+13.792}_{-13.291}$ & $0.616^{+20.549}_{-19.765}$ & $0.203^{+8.948}_{-9.195}$ \\
\textbf{\cqebsT}
& $-2.016^{+33.494}_{-36.262}$ & $-2.337^{+36.884}_{-37.424}$ & $-1.686^{+24.108}_{-26.078}$ & $-1.149^{+15.736}_{-17.431}$ & $-1.633^{+23.069}_{-25.710}$ & $-0.763^{+10.423}_{-10.887}$ \\
\textbf{\cdebsT}
& $1.021^{+18.076}_{-17.626}$ & $1.022^{+18.244}_{-18.562}$ & $0.634^{+12.803}_{-12.911}$ & $0.366^{+8.653}_{-8.471}$ & $0.625^{+12.616}_{-12.568}$ & $0.203^{+5.604}_{-5.581}$ \\
\hline
\textbf{(\cqlonebsT,\,\czbsT)}
& $-16.063^{+87.656}_{-82.456}$ & $-20.225^{+95.616}_{-101.985}$ & $-14.739^{+70.872}_{-67.285}$ & $-9.909^{+49.487}_{-43.085}$ & $-14.287^{+69.646}_{-65.960}$ & $-6.427^{+30.892}_{-26.826}$ \\
\textbf{(\cqlthreebsT,\,\czbsT)}
& $-111.306^{+70.800}_{-66.137}$ & $-142.681^{+97.410}_{-79.096}$ & $-104.286^{+70.514}_{-55.809}$ & $-69.955^{+48.192}_{-38.703}$ & $-101.031^{+68.190}_{-51.673}$ & $-45.172^{+30.657}_{-28.529}$ \\
\hline
\multicolumn{7}{c}{$A_8\times10^{3}$} \\
\hline
SM
& $0.127 \pm 0.012$ & $0.206 \pm 0.025$ & $0.156 \pm 0.021$ & $0.110 \pm 0.016$ & $0.150 \pm 0.019$ & $0.079 \pm 0.013$ \\
\hline
\textbf{\cqlonebsT}
& $0.281^{+34.702}_{-32.065}$ & $0.500^{+44.180}_{-44.735}$ & $0.400^{+32.587}_{-31.175}$ & $0.286^{+21.845}_{-20.116}$ & $0.381^{+31.903}_{-30.192}$ & $0.200^{+12.600}_{-13.891}$ \\
\textbf{\cqlthreebsT}
& $0.803^{+34.554}_{-33.322}$ & $1.261^{+44.682}_{-44.761}$ & $0.965^{+32.162}_{-31.362}$ & $0.663^{+21.491}_{-20.768}$ & $0.924^{+30.454}_{-30.821}$ & $0.441^{+14.447}_{-13.581}$ \\
\textbf{\czbsT}
& $-5.594^{+1.149}_{-0.959}$ & $-7.445^{+1.342}_{-1.055}$ & $-5.320^{+0.892}_{-0.844}$ & $-3.485^{+0.666}_{-0.653}$ & $-5.134^{+0.834}_{-0.868}$ & $-2.197^{+0.598}_{-0.469}$ \\
\textbf{\cdlbsT}
& $-0.188^{+6.291}_{-7.985}$ & $1.062^{+16.726}_{-16.659}$ & $1.852^{+34.120}_{-35.479}$ & $2.264^{+43.157}_{-47.421}$ & $1.809^{+33.123}_{-34.769}$ & $2.433^{+50.762}_{-52.174}$ \\
\textbf{\czpbsT}
& $0.115^{+2.777}_{-2.868}$ & $0.146^{+15.409}_{-15.318}$ & $0.085^{+21.792}_{-21.774}$ & $0.038^{+25.268}_{-26.086}$ & $0.081^{+21.293}_{-21.577}$ & $0.010^{+26.676}_{-28.252}$ \\
\textbf{\cqebsT}
& $1.493^{+15.891}_{-15.154}$ & $1.824^{+18.021}_{-17.539}$ & $1.282^{+12.925}_{-11.844}$ & $0.848^{+8.541}_{-7.606}$ & $1.244^{+12.711}_{-11.337}$ & $0.550^{+4.831}_{-5.061}$ \\
\textbf{\cdebsT}
& $0.093^{+7.987}_{-8.981}$ & $0.162^{+9.520}_{-10.572}$ & $0.122^{+7.136}_{-8.102}$ & $0.085^{+5.221}_{-6.171}$ & $0.117^{+6.896}_{-7.937}$ & $0.061^{+3.835}_{-4.576}$ \\
\hline
\textbf{(\cqlonebsT,\,\czbsT)}
& $38.220^{+25.718}_{-26.378}$ & $51.444^{+29.711}_{-35.369}$ & $37.548^{+20.255}_{-25.325}$ & $25.190^{+14.097}_{-17.176}$ & $36.353^{+20.394}_{-24.721}$ & $16.284^{+9.515}_{-11.362}$ \\
\textbf{(\cqlthreebsT,\,\czbsT)}
& $3.371^{+30.846}_{-35.673}$ & $4.724^{+38.034}_{-47.297}$ & $3.502^{+27.353}_{-34.298}$ & $2.373^{+18.627}_{-22.308}$ & $3.384^{+26.694}_{-32.824}$ & $1.551^{+13.185}_{-14.244}$ \\
\hline
\multicolumn{7}{c}{$A_9\times10^{3}$} \\
\hline
SM
& $0.003 \pm 0.007$ & $0.011 \pm 0.022$ & $0.013 \pm 0.023$ & $0.012 \pm 0.021$ & $0.012 \pm 0.022$ & $0.012 \pm 0.019$ \\
\hline
\textbf{\cqlonebsT}
& $0.007^{+1.054}_{-1.010}$ & $0.027^{+3.248}_{-2.884}$ & $0.032^{+3.608}_{-3.317}$ & $0.032^{+3.390}_{-2.579}$ & $0.031^{+3.412}_{-3.113}$ & $0.031^{+2.991}_{-2.036}$ \\
\textbf{\cqlthreebsT}
& $0.021^{+1.311}_{-1.038}$ & $0.067^{+4.329}_{-2.838}$ & $0.077^{+4.627}_{-3.098}$ & $0.074^{+4.283}_{-2.863}$ & $0.073^{+4.396}_{-2.909}$ & $0.068^{+3.579}_{-2.151}$ \\
\textbf{\czbsT}
& $-0.141^{+0.263}_{-0.305}$ & $-0.397^{+0.738}_{-0.786}$ & $-0.425^{+0.733}_{-0.819}$ & $-0.389^{+0.608}_{-0.674}$ & $-0.402^{+0.686}_{-0.767}$ & $-0.336^{+0.476}_{-0.549}$ \\
\textbf{\cdlbsT}
& $-0.196^{+4.256}_{-4.656}$ & $0.134^{+1.823}_{-1.937}$ & $0.834^{+16.126}_{-18.107}$ & $1.533^{+28.817}_{-33.824}$ & $0.941^{+18.605}_{-20.189}$ & $2.133^{+39.679}_{-45.859}$ \\
\textbf{\czpbsT}
& $0.003^{+0.631}_{-0.626}$ & $-0.005^{+7.429}_{-6.851}$ & $-0.021^{+14.320}_{-12.631}$ & $-0.036^{+20.106}_{-17.660}$ & $-0.023^{+14.939}_{-13.052}$ & $-0.047^{+24.479}_{-23.100}$ \\
\textbf{\cqebsT}
& $0.038^{+0.683}_{-0.481}$ & $0.097^{+1.775}_{-1.158}$ & $0.103^{+1.786}_{-1.111}$ & $0.095^{+1.558}_{-1.037}$ & $0.098^{+1.711}_{-1.016}$ & $0.084^{+1.175}_{-0.732}$ \\
\textbf{\cdebsT}
& $-0.011^{+3.108}_{-3.236}$ & $-0.025^{+8.365}_{-8.153}$ & $-0.025^{+8.498}_{-8.213}$ & $-0.020^{+7.246}_{-7.152}$ & $-0.023^{+7.924}_{-7.750}$ & $-0.014^{+5.938}_{-5.959}$ \\
\hline
\textbf{(\cqlonebsT,\,\czbsT)}
& $0.968^{+2.312}_{-1.627}$ & $2.756^{+6.097}_{-4.608}$ & $3.011^{+6.199}_{-4.686}$ & $2.825^{+5.817}_{-3.818}$ & $2.860^{+5.890}_{-4.350}$ & $2.500^{+4.905}_{-3.372}$ \\
\textbf{(\cqlthreebsT,\,\czbsT)}
& $0.086^{+1.716}_{-1.308}$ & $0.253^{+4.956}_{-3.431}$ & $0.281^{+4.943}_{-3.485}$ & $0.266^{+4.527}_{-2.755}$ & $0.267^{+4.438}_{-3.305}$ & $0.238^{+3.744}_{-1.843}$ \\
\hline
\end{tabular}}
\label{tab:A789}
\end{table}

Beyond the direct $\mathcal{CP}$ asymmetry, the angular distribution of $B^0\to K^{*0}\mu^+\mu^-$ gives access to the T-odd observables $S_7$, $S_8$, and $S_9$ and to their $\mathcal{CP}$-asymmetric counterparts $A_7$, $A_8$, and $A_9$, defined in Appendix~\ref{app:obs_B2Kstmumu}. The corresponding angular coefficients $I_{7,8,9}$ are proportional to the imaginary parts of interference terms between transversity amplitudes. For an interference term with relative weak phase $\Delta\phi$ and relative strong phase $\Delta\delta$, one finds $S_{7,8,9}\propto\cos\Delta\phi\,\sin\Delta\delta$ and $A_{7,8,9}\propto\sin\Delta\phi\,\cos\Delta\delta$~\cite{Altmannshofer:2008dz}.
Hence $S_{7,8,9}$ vanishes in the absence of strong phases, even for complex Wilson coefficients. In contrast, $A_{7,8,9}$ directly probe new weak phases and are not suppressed by small strong phases.

The predictions for $S_{7,8,9}$ are summarised in Table~\ref{tab:S789}. Within the SM, these observables are generated solely by the perturbative strong phases, giving $S_7\simeq-(0.4$--$1.2)\times10^{-2}$, $S_8\simeq-(1.2$--$2.4)\times10^{-3}$, and $S_9=\mathcal{O}(10^{-4})$.
In all NP scenarios, the predictions remain close to the SM. For $S_7$, the scenarios \cqlonebsT, \cqlthreebsT, and both 2D scenarios enhance $|S_7|$ by about $15$--$25\%$ for $q^2>1.1\,\mathrm{GeV}^2$, while \czbsT{} reduces it by about $10\%$. In absolute terms, these shifts do not exceed $2\times10^{-3}$, comparable to the SM uncertainties.

For $S_8$, the 1D scenarios \cqlonebsT{} and \cqlthreebsT{} reduce $|S_8|$ mainly in the lowest bin. Both 2D scenarios reduce it by up to a factor of four at low $q^2$, with absolute shifts below $1.5\times10^{-3}$.
Finally, $S_9$ remains at the level of $\mathcal{O}(10^{-4})$ in all considered scenarios, compatible with the SM within uncertainties.
All these effects lie far below the present experimental precision of $(3$--$5)\times10^{-2}$~\cite{LHCb:2025mqb}.
We note that the measured values of $S_7$ lie systematically below the SM prediction, by about $2\sigma$ in each bin above $1.1\,\mathrm{GeV}^2$. None of the considered scenarios reproduces this pattern. Since $S_7$ vanishes identically in the absence of strong phases, values of this size would require strong phases considerably larger than the perturbative ones included here, for instance from nonlocal hadronic contributions, rather than new weak phases.

The $\mathcal{CP}$ asymmetries $A_{7,8,9}$, summarised in Table~\ref{tab:A789}, exhibit the strongest sensitivity to the complex Wilson coefficients\footnote{The only available measurements for these cases are of \(A_9\), with values of approximately \(10\%\) at $68\%$ C.L., reported in different \(q^2\) bins~\cite{LHCb:2013zuf}; hence, they are not included in Table~\ref{tab:A789}.}. In the SM, they are suppressed by the small weak phase of the $\lambda_u^{32}$ term, giving $A_7\lesssim10^{-3}$, $A_8=\mathcal{O}(10^{-4})$, and $A_9=\mathcal{O}(10^{-5})$.
The largest effect arises in $A_7$ for the \czbsT{} scenario. There $|A_7|\simeq(16$--$21)\%$ for $q^2<2.5\,\mathrm{GeV}^2$, decreasing to $\simeq6\%$ in the $[6,8]\,\mathrm{GeV}^2$ bin, and it differs from zero by $3.5$--$5\sigma$ in all bins. This directly reflects the large imaginary part of $C_{10}$ preferred by this scenario, which interferes with $C_7$.
A similar but less significant pattern is found for (\cqlthreebsT,\,\czbsT), with $|A_7|\simeq(4.5$--$14)\%$ at about $1.5$--$2\sigma$ from zero, whereas for (\cqlonebsT,\,\czbsT) the predicted $A_7$ is compatible with zero.
For the 1D scenarios \cqlonebsT{} and \cqlthreebsT, the central values of $A_7$ and $A_8$ remain at the per-mille level, but their $1\sigma$ ranges extend up to $\pm10\%$ and $\pm4.5\%$, respectively.
For $A_8$, the scenario (\cqlonebsT,\,\czbsT) gives $|A_8|\simeq(1.6$--$5.1)\%$, largest in the $[1.1,2.5]\,\mathrm{GeV}^2$ bin. In contrast, \czbsT{} yields only $|A_8|\lesssim\mathcal{O}(10^{-3})$, since this scenario modifies predominantly $C_{10}$, to which $A_8$ is insensitive at leading order.
Finally, $A_9$ remains below $10^{-2}$ in all scenarios with left-handed quark currents. The scenarios \cdlbsT{} and \czpbsT, which induce chirality-flipped $b\to s$ currents, allow $|A_8|$ and $|A_9|$ up to $(2.5$--$5)\%$ at $1\sigma$, with the largest effects in the $[6,8]\,\mathrm{GeV}^2$ bin.

The pattern of these results can be understood from the phase structure of the observables. Both $\mathcal{A}_{\mathcal{CP}}$ and $S_{7,8,9}$ require nonvanishing strong phases and therefore remain small as long as only perturbative strong phases are present, irrespective of the size of the NP weak phases. In contrast, the T-odd $\mathcal{CP}$ asymmetries $A_{7,8,9}$ are directly sensitive to the imaginary parts of the Wilson coefficients. Moreover, their different chirality structure, with $A_7$ probing ${\rm Im}\,C_{10}$, $A_8$ probing ${\rm Im}\,C_9$, and $A_9$ requiring right-handed currents, allows them to discriminate between the scenarios considered here.

Although the predicted direct $\mathcal{CP}$ asymmetries remain at the per-mille level, reaching at most $\simeq1\%$ in the preferred NP scenarios, they can still provide useful information on possible new weak phases in rare semileptonic decays. The angular $\mathcal{CP}$ asymmetries $A_7$ and $A_8$ exhibit much larger NP-induced effects, up to $\mathcal{O}(10\%)$ in the low-$q^2$ region, and therefore constitute particularly sensitive probes of complex NP couplings as in this region, strong phases arising from nonfactorizable hadronic effects are expected to be small.
These nonfactorizable effects instead generate the strong phases entering $\mathcal{A}_{\mathcal{CP}}$ and $S_{7,8,9}$ near charmonium resonance regions, and a more quantitative interpretation of the latter requires improved theoretical control over them, as they may enhance these observables in specific $q^2$ regions. On the experimental side, the current measurements remain statistically limited, but future high-statistics data from LHCb and Belle~II will significantly enhance the sensitivity to these observables and provide valuable tests of the complex flavor structure of NP. Recently, LHCb performed the first direct determination of the phases of $C_9$ and $C_{10}$ from an fit to the angular distribution in the four-dimensional space in the signal region of $B^0\to K^{*0}\mu^+\mu^-$, where interference with the charmonium resonances provides large strong phases~\cite{LHCb:2026ayl}. The extracted values show no significant CP violation, with ${\rm Im}(C_9)=-0.23\pm0.12$ and ${\rm Im}(C_{10})=-0.17\pm0.15$, about an order of magnitude more precise ($\sigma_{\rm Im}\simeq0.5$--$1$ for the 1D scenarios) than the imaginary parts obtained in our fits to CP-averaged angular observables. Interestingly, despite the significant differences between the two fits in their methodologies and in the number and types of observables used to constrain the Wilson coefficients, the LHCb extraction, \({\rm Re}(C_9)=3.50\pm0.34\), is consistent with our \cqlonebsT\ fit, which yields \({\rm Re}(C_9)={\rm Re}(C_9^{\rm SM}+\Delta C_9)\simeq3.30\). In contrast, LHCb obtains \({\rm Re}(C_{10})=-4.04\pm0.23\), which is close to the SM value but differs from our fit result, \({\rm Re}(C_{10})\simeq-3.55\). The limited availability of bin-integrated observables that are particularly sensitive to new-physics weak phases allows relatively large imaginary contributions to certain Wilson coefficients, such as \czbsT, in our scenarios. Consequently, these scenarios can predict sizeable values of \(|A_7|\).
Our predicted asymmetries, $|\mathcal{A}_{\mathcal{CP}}|\lesssim\mathcal{O}(10^{-2})$, which include only perturbative strong phases, are consistent with the $q^2$-dependent asymmetry reported in Ref.~\cite{LHCb:2026ayl}.

\section{Summary}
\label{sec:summary}
Motivated by the long-standing flavor anomalies, we have analyzed rare $B$- and $K$- meson decays involving $b \to s$ and $s \to d$ transitions in both the dilepton and dineutrino channels. The primary goal of this study is to investigate the implications of the latest $b/s \to s/d\,(\mu^+ \mu^-, \nu \bar{\nu})$ measurements within a model-independent framework. To this end, we employ the SMEFT formalism, which relates the dilepton and dineutrino modes through $SU(2)_L$ symmetry, incorporating all relevant dimension-six operators with general Wilson coefficients. We systematically study a wide range of observables associated with these decay modes, both within the SM and in the presence of SMEFT contributions, considering several one and two-dimensional benchmark scenarios.

We perform a \(\chi^2\) fit to a selected set of \(b\to s\) observables exhibiting tensions with the SM predictions in the \(B\to K^{(*)}\mu^+\mu^-\) and \(B\to K^{(*)}\nu\bar{\nu}\) channels. The fit determines the preferred values of the relevant SMEFT Wilson coefficients in one- and two-dimensional NP scenarios, allowing for complex couplings. Among the dimension-six SMEFT operators, the Wilson coefficients \cqlonebsT and \cqlthreebsT, corresponding to four-fermion operators, provide the preferred description of the current $b \to s$ data, including differential branching fractions, angular observables, and helicity fractions. In particular, the tensions in the differential distributions of $B \to K \mu^+ \mu^-$ are significantly reduced, while the discrepancy in the angular observable $P_5^\prime$ in $B \to K^* \mu^+ \mu^-$ is alleviated. These operators also help to reduce the tension in the branching fraction of $B \to K \nu \bar\nu$. Interestingly, the electroweak operator \czbsT is found to play an important role, especially when combined with either \cqlonebsT or \cqlthreebsT in the 2D fit scenarios.

We further show that assuming flavor-universal SMEFT couplings in the $b \to s$ and $s \to d$ sectors leads to excessively large predictions for the branching ratios of $K_{(L)} \to \pi^{(0)} \mu^+\mu^-$ and $K_{(L)} \to \pi^{(0)} \nu \bar \nu$, thereby violating the current experimental bounds. This observation strongly motivates the implementation of MFV in order to preserve consistency with existing flavor constraints. We therefore extend the analysis within the $U(3)^5$ and $U(2)^5$ flavor-symmetric frameworks, where the required CKM hierarchies between $b \to s$ and $s \to d$ transitions are naturally restored. After implementing the corresponding CKM-aligned scaling relations for the fitted SMEFT coefficients \cqloneT, \cqlthreeT, and \czT, the predicted branching ratios for rare kaon decays become fully consistent with the current experimental limits.

In addition, we present the differential distributions with respect to both $q^{2}$ and the reconstructed variable $q^{2}_{\mathrm{rec}}$, highlighting the effects of different NP Wilson coefficients in $B \to K^{(*)}\nu\bar{\nu}$ decays. These observables are particularly relevant for Belle~II, LHCb, and future flavor facilities, which will be able to probe the $b \to s\nu\bar{\nu}$ transitions with increasing precision and thereby provide stringent tests of the SMEFT framework.

Finally, allowing for complex Wilson coefficients in the fit leads to sizable imaginary components in several preferred NP scenarios. Motivated by this, we systematically investigate the resulting direct $\mathcal{CP}$ asymmetries in the $B \to K \mu^+\mu^-$ and $B \to K^* \mu^+\mu^-$ channels, as well as the $\mathcal{CP}$-sensitive angular observables $S_{7,\,8,\,9}$ and $A_{7,\,8,\,9}$ in $B \to K^* \mu^+\mu^-$ across different $q^2$ bins. We find that direct \(\mathcal{CP}\) asymmetries remain below the percent level, whereas \(A_7\) can reach \(1\%-10\%\) in certain \(q^2\) regions. These observables therefore provide sensitive probes of new weak phases and can be tested in future high-precision measurements at LHCb and Belle~II.

Future measurements of both $b \to s$ and $s \to d$ transitions in the $\mu^+\mu^-$ and $\nu\bar{\nu}$ modes will play a crucial role in uncovering the Lorentz structure of possible NP interactions underlying the observed flavor anomalies. Precise measurements of branching ratios, angular observables, and dineutrino distributions in both $B$ and $K$ decays at Belle~II, LHCb, and NA62 will provide stringent constraints and serve as powerful tests of the SMEFT framework employed in this work. Furthermore, the preferred operators identified in this analysis may provide important guidance toward constructing viable theories beyond the SM.

\ack{We thank Gino Isidori for useful comments on the Minimal Flavor Violation framework. N.D. would like to thank M. S. A. Alam Khan for discussions during the initial days of the project. P.P.\ acknowledges the use of Claude Opus~5.5 (Anthropic) as an assistant in setting up and running the \texttt{flavio}-based numerical evaluations and in cross-checking the resulting predictions. All results obtained in this way were independently verified by the authors, who take full responsibility for the content of this work.}

 \funding{N.D. and R.M. acknowledge support from the
SERB grant SPG/2022/001238. R.M.\ further acknowledges support from the DAE-BRNS YSRP grant No.~57/20/02/2024.}




\appendix

\section{Expressions for Wilson coefficients}
\label{app:SM_WCs}
The SM contribution to the effective Wilson coefficients entering transversity amplitudes Eq.~\eqref{eq:transversity_amplitudes} is obtained from the Wilson coefficients of the $\Delta B=1$ effective Hamiltonian evaluated at $\mu_b=5~\mathrm{GeV}$, including NNLO QCD and NLO QED corrections~\cite{Mahmoudi:2024zna}, combined with the matrix elements of the four-quark operators $\mathcal{O}_{1-6}$ and of the chromomagnetic operator $\mathcal{O}_8$ at leading order~\cite{Beneke:2001at} and NLO in $\alpha_s$~\cite{Asatrian:2019kbk}. We write
\begin{align}
C^{32}_{9,\rm SM}(q^2) &= C_9 + Y_t(q^2) + \frac{\lambda_u^{32}}{\lambda_t^{32}}\,Y_u(q^2)
+ \Delta C_9^{t}(q^2) + \frac{\lambda_u^{32}}{\lambda_t^{32}}\,\Delta C_9^{u}(q^2)\,,
\label{eq:C9SM}\\[4pt]
C^{32}_{7,\rm SM}(q^2) &= C_7^{\rm eff} + \Delta C_7^{t}(q^2) + \frac{\lambda_u^{32}}{\lambda_t^{32}}\,\Delta C_7^{u}(q^2)\,,
\label{eq:C7SM}
\end{align}
where the leading-order quark-loop contributions read~\cite{Altmannshofer:2008dz}
\begin{align}
Y_t(q^2) &= h(q^2, m_c)\left(\frac{4}{3} C_1 + C_2 + 6C_3 + 60C_5 \right)
- \frac{1}{2} h(q^2, m_b)\left(7C_3 + \frac{4}{3} C_4 + 76C_5 + \frac{64}{3} C_6 \right) \nonumber \\
&\quad - \frac{1}{2} h(q^2, 0)\left(C_3 + \frac{4}{3} C_4 + 16C_5 + \frac{64}{3} C_6 \right)
+ \frac{4}{3} C_3 + \frac{64}{9} C_5 + \frac{64}{27} C_6 \,, \\[4pt]
Y_u(q^2) &= \left(\frac{4}{3}C_1+C_2\right)\Big(h(q^2, m_c)-h(q^2, 0)\Big)\,,
\end{align}
with the quark-loop function ($x\equiv 4 m_q^2/q^2$)
\begin{align}
h(q^2,m_q) &= -\frac{4}{9}\left( \ln\frac{m_q^2}{\mu^2} - \frac{2}{3} - x \right)
- \frac{4}{9}(2+x)\sqrt{|x-1|}\,
\begin{cases}
\arctan\!\left(\dfrac{1}{\sqrt{x-1}}\right), & x > 1, \\[10pt]
\ln\!\left(\dfrac{1+\sqrt{1-x}}{\sqrt{x}}\right) - \dfrac{i\pi}{2}, & x \leq 1,
\end{cases}
\\[4pt]
h(q^2,0) &= \frac{8}{27} - \frac{4}{9}\ln\frac{q^2}{\mu^2} + \frac{4}{9}\,i\pi \,.
\end{align}
The $\mathcal{O}({\alpha_s})$ corrections are
\begin{align}
\Delta C_{7,9}^{t}(q^2) &= -\frac{\alpha_s(\mu_b)}{4\pi}\Big[C_1 F_1^{(7,9)}(q^2) + C_2 F_2^{(7,9)}(q^2) + C_8^{\rm eff} F_8^{(7,9)}(q^2)\Big]\,,\\[4pt]
\Delta C_{7,9}^{u}(q^2) &= -\frac{\alpha_s(\mu_b)}{4\pi}\Big[C_1 \big(F_1^{(7,9)}(q^2)-F_{1,u}^{(7,9)}(q^2)\big) + C_2 \big(F_2^{(7,9)}(q^2)-F_{2,u}^{(7,9)}(q^2)\big)\Big]\,,
\end{align}
where the two-loop functions $F_{1,2}^{(7,9)}$ of the charm-quark loops are taken from the analytic results of Ref.~\cite{Asatrian:2019kbk}, $F_8^{(7,9)}$ from Ref.~\cite{Beneke:2001at}, and the massless-quark functions $F_{1,2,u}^{(7,9)}$ from Ref.~\cite{Seidel:2004jh}.
The effective dipole coefficients are defined as
\begin{align}
C_7^{\rm eff} = C_7 - \frac{1}{3}C_3 - \frac{4}{9}C_4 - \frac{20}{3}C_5 - \frac{80}{9}C_6\,,\qquad
C_8^{\rm eff} = C_8 + C_3 - \frac{1}{6}C_4 + 20C_5 - \frac{10}{3}C_6\,.
\end{align}
In the loop functions we use the pole masses $m_c=1.681~\mathrm{GeV}$ and $m_b=4.780~\mathrm{GeV}$. The Wilson coefficient at $\mu_b$ are listed in Table~\ref{tab:new_SM_WCs}.
\begin{table}
\caption{Wilson coefficients at the scale $\mu_b=5\,\mathrm{GeV}$, including NNLO QCD and NLO QED corrections~\cite{Mahmoudi:2024zna}, and the effective dipole coefficients defined in the text.}
\centering
\begin{tabular}{c c c c c}
\hline
$C_1(\mu_b)$ & $C_2(\mu_b)$ & $C_3(\mu_b)$ & $C_4(\mu_b)$ & $C_5(\mu_b)$ \\
\hline
$-0.2507$ & $1.0136$ & $-0.0049$ & $-0.0766$ & $0.0003$ \\
\hline
$C_6(\mu_b)$ & $C_7(\mu_b)$ & $C_8(\mu_b)$ & $C_9(\mu_b)$ & $C_{10}(\mu_b)$ \\
\hline
$0.0009$ & $-0.3143$ & $-0.1710$ & $4.0459$ & $-4.2939$ \\
\hline
\end{tabular}
\label{tab:new_SM_WCs}
\end{table}

\section{Theoretical expressions for the observables}
\label{app:obs_exp}
In this appendix, we collect the theoretical expressions for the observables entering the fit, as well as those used for the predictions presented in Sec.~\ref{sec:results}. For completeness and to ensure the consistency of the numerical analysis, we summarize the relevant decay distributions, angular observables, and branching-ratio formulas for rare semileptonic and dineutrino $B$ and $K$ decays.

\subsection{$B \to K \mu^+\mu^-$}
\label{app:obs_B2Kmumu}
The differential distribution in dilepton invariant mass squared $q^2$ for the $B\to K \mu^+\mu^-$ decay can be written as~\cite{Bobeth:2007dw}

\begin{equation}
\frac{d\Gamma}{dq^2}(B^-\to K^- \mu^+\mu^-) = 2a_\mu + \frac{2}{3}c_\mu\,,
\label{eq:dGK}
\end{equation}
where 
\begin{eqnarray}
a_\mu &=& \frac{G_F^2 \alpha_{\rm em}^2 |V_{tb} V_{ts}^*|^2}{2^9\pi^5 m_{B}^3}\beta_\mu\, \lambda^{1/2}(m_B^2,m_K^2,q^2)\Big[ q^2 |F_P|^2 + \frac{\lambda(m_B^2,m_K^2,q^2)}{4}(|F_A|^2 + |F_V|^2) + 4m_\mu^2 m_{B}^2 |F_A|^2 \nonumber \\
&& \hspace{6.4cm} + 2m_\mu (m_{B}^2 - m_{K}^2 + q^2){\rm Re}(F_PF_A^*) \Big], \\
c_\mu &=& -\frac{G_F^2 \alpha_{\rm em}^2 |V_{tb} V_{ts}^*|^2}{2^{11}\pi^5 m_{B}^3}\beta_\mu^3 \,\lambda^{3/2}(m_B^2,m_K^2,q^2)(|F_A|^2 + |F_V|^2).
\end{eqnarray}
Here the kinematical factor $\lambda(x,y,z)=x^2 + y^2+z^2 - 2(xy+yz+zx)$ is the Kall\'en function and the muon mass dependent factor $\beta_\mu=\sqrt{1-4m_\mu^2/q^2}$.

The explicit expressions of the $q^2$ dependent functions $F_P$, $F_V$ and $F_A$ are given as follows:
\begin{eqnarray}
F_P &=& -m_\mu \big(C_{10}^{32,22}+C_{10}^{\prime \, 32,22}\big)\Big[ f_+(q^2) - \frac{m_{B}^2-m_{K}^2}{q^2}(f_0(q^2)-f_+(q^2)) \Big], \\
F_V &=& \big(C^{32,22}_9(q^2)+C^{\prime \, 32,22}_9(q^2)\big)\, f_+(q^2) + \frac{2\,\overline{m}_b(\mu_b)}{m_B + m_K}\, \big(C^{32}_{7,\rm SM}(q^2)+C^{\prime \, 32}_{7}(q^2)\big)\, f_T(q^2)\,, \\
F_A &=& \big(C_{10}^{32,22}+C_{10}^{\prime \, 32,22}\big)\,f_+(q^2).
\end{eqnarray}
The inputs for the form factors $f_i(q^2)$ are taken from lattice QCD computations~\cite{Parrott:2022rgu}.
The rate $d\bar\Gamma/dq^2$ of the $\mathcal{CP}$-conjugate decay $B^+\to K^+\mu^+\mu^-$ follows from Eq.~\eqref{eq:dGK} by complex conjugating all weak phases, i.e.\ $\lambda_u^{32}/\lambda_t^{32}$ and the NP Wilson coefficients, while leaving the strong phases (the absorptive parts of the quark-loop functions) unchanged. The $\mathcal{CP}$-averaged binned branching fraction compared with data is
\begin{equation}
\left\langle\frac{d\mathcal{B}}{dq^2}\right\rangle=\frac{\tau_{B^+}}{2\,(q^2_{\rm max}-q^2_{\rm min})}\int_{q^2_{\rm min}}^{q^2_{\rm max}}dq^2\left(\frac{d\Gamma}{dq^2}+\frac{d\bar\Gamma}{dq^2}\right).
\label{eq:BRK}
\end{equation}

\subsection{$B \to K^* \mu^+\mu^-$}
\label{app:obs_B2Kstmumu}

The differential decay distribution in $q^2$ and $\mathcal{CP}$ averaged angular observables for $B \to K^* \mu^+\mu^-$ are extracted from the angular coefficients of four-dimensional differential distributions expressed in terms of transversity amplitudes~\cite{Altmannshofer:2008dz}. From a total of eleven  angular coefficients, below we quote the ones that are relevant for our analysis.
\begin{align}
I_1^s &= 
\frac{(2+\beta_\mu^2)}{4} 
   \left(|A_\perp^L|^2 + |A_\parallel^L|^2 + (L \to R)\right)
 + \frac{4 m_\mu^2}{q^2} 
   \Re\!\left(A_\perp^{L} A_\perp^{R*} + A_\parallel^{L} A_\parallel^{R*}\right),
\\[0.4em]
I_1^c &= 
|A_0^L|^2 + |A_0^R|^2
 + \frac{4 m_\mu^2}{q^2} 
   \left(|A_t|^2 + 2\,\Re(A_0^{L} A_0^{R*}) \right),
\\[0.4em]
I_2^s &= 
\frac{\beta_\mu^2}{4}
\left(|A_\perp^L|^2 + |A_\parallel^L|^2 + (L \to R)\right),
\\[0.4em]
I_2^c &= 
-\,\beta_\mu^2 \left(|A_0^L|^2 + (L \to R)\right),
\\[0.4em]
I_5 &= 
\sqrt{2}\,\beta_\mu
\left[
\Re(A_0^{L} A_\perp^{L*}) - (L \to R)
\right],
\\[0.4em]
I_6^s &= 
2\beta_\mu
\left[
\Re(A_\parallel^{L} A_\perp^{L*}) - (L \to R)
\right]\,,
\\[0.4em]
I_7 &= \sqrt{2}\,\beta_\mu
\left[
\operatorname{Im}\!\left(A_0^L A_\parallel^{L*}\right)
-(L\to R)
\right],
\\[0.4em]
I_8 &= \frac{1}{\sqrt{2}}\,\beta_\mu^2
\left[
\operatorname{Im}\!\left(A_0^L A_\perp^{L*}\right)
+(L\to R)
\right],
\\[0.4em]
I_9 &= \beta_\mu^2
\left[
\operatorname{Im}\!\left(A_\parallel^{L*} A_\perp^L\right)
+(L\to R)
\right].
\end{align}
where the transversity amplitudes are given by
\begin{align}
A_\perp^{L,R}  &=  N \sqrt{2}\, \tilde{\lambda}^{1/2} \bigg[
\Big( \big(C_9^{32,22}+C_9^{\prime\,32,22}\big)  \mp \big(C_{10}^{32,22}+C_{10}^{\prime\,32,22}\big)  \Big) \frac{ V(q^2) }{ m_B  + m_{K^*}}
\nonumber\\
&\hspace{3cm}+ \frac{2m_b}{q^2} \big(C_7^\text{eff.}+C_7^{\prime\,\text{eff.}}\big)  T_1(q^2)
\bigg],\\[2ex]
A_\parallel^{ L,R}  & = - N \sqrt{2}\,(m_B ^2 - m_{K^*}^2) \bigg[ \Big( \big(C_9^{32,22}-C_9^{\prime\,32,22}\big) \mp \big(C_{10}^{32,22}-C_{10}^{\prime\,32,22}\big)  \Big)
\frac{A_1(q^2)}{m_B -m_{K^*}}
\nonumber\\
&\hspace{3cm}+\frac{2 m_b}{q^2} \big(C_7^\text{eff.}-C_7^{\prime\,\text{eff.}}\big)  T_2(q^2)
\bigg],\\[2ex]
A_0^{L,R}  &=  - \frac{N}{2 m_{K^*} \sqrt{q^2}}  \bigg\{
 \Big( \big(C_9^{32,22}-C_9^{\prime\,32,22}\big) \mp \big(C_{10}^{32,22}-C_{10}^{\prime\,32,22}\big)  \Big)
\nonumber\\
&\qquad\times\bigg[ (m_B ^2 - m_{K^*}^2 - q^2) ( m_B  + m_{K^*}) A_1(q^2)
 - \frac{\tilde{\lambda} A_2(q^2)}{m_B  + m_{K^*}}
\bigg]
\nonumber\\
& \qquad + {2 m_b} \big(C_7^\text{eff.}-C_7^{\prime\,\text{eff.}}\big)  \bigg[
 (m_B ^2 + 3 m_{K^*}^2 - q^2) T_2(q^2)
-\frac{\tilde{\lambda}}{m_B ^2 - m_{K^*}^2} T_3(q^2) \bigg]
\bigg\}, \\[2ex]
 A_t  &= \frac{2 N}{\sqrt{q^2}}\,\tilde{\lambda}^{1/2} \big(C_{10}^{32,22}-C_{10}^{\prime\,32,22}\big)  A_0(q^2) \,.
 \label{eq:transversity_amplitudes}
\end{align}
with  
\begin{equation}
N \equiv \vert V_{tb}V_{ts}^*\vert \left[\frac{G_F^2 \alpha_{\rm em}^2}{3\cdot 2^{10}\pi^5 m_B ^3}
 q^2 \tilde{\lambda}^{1/2}
\beta_\mu \right]^{1/2}~~{\rm and}~~ \tilde{\lambda}\equiv \lambda(m_B^2,m_{K^*}^2,q^2).
\end{equation}
The inputs for the form factors $V(q^2),\,A_i(q^2),\,T_i(q^2)$ are taken from light-cone sum rule computations~\cite{Bharucha:2015bzk}.

The differential decay rate for $\bar B^0\to\bar K^{*0}\mu^+\mu^-$ follows from the angular coefficients as
\begin{equation}
\frac{d\Gamma}{dq^2}=\frac{3}{4}\left(2I_1^s+I_1^c\right)-\frac{1}{4}\left(2I_2^s+I_2^c\right),
\label{eq:dGKst}
\end{equation}
and the corresponding rate $d\bar\Gamma/dq^2$ of the $\mathcal{CP}$-conjugate decay $B^0\to K^{*0}\mu^+\mu^-$ is obtained by replacing $I_j \to \Bar{I}_j$, where $\Bar{I}_j$ follows from by complex conjugating all weak phases, i.e.\ $\lambda_u^{32}/\lambda_t^{32}$ and the NP Wilson coefficients, while the strong phases are left unchanged~\cite{Kruger:1999xa}. All observables are $\mathcal{CP}$ averaged and integrated over a $q^2$ bin $[q^2_{\rm min},q^2_{\rm max}]$, with $\int^\prime\equiv\int_{q^2_{\rm min}}^{q^2_{\rm max}}dq^2$. The branching fraction is
\begin{equation}
\left\langle\frac{d\mathcal{B}}{dq^2}\right\rangle=\frac{\tau_{B^0}}{2\,(q^2_{\rm max}-q^2_{\rm min})}\int^\prime \left(\frac{d\Gamma}{dq^2}+\frac{d\bar\Gamma}{dq^2}\right),
\label{eq:BRKst}
\end{equation}
and the longitudinal polarization fraction, the forward-backward asymmetry, the optimized observable $P_5^{\prime}$ and the $\mathcal{CP}-$averaged T-odd observables $S_j(j=7,8,9)$ are defined as\cite{Altmannshofer:2008dz}

\begin{align}
\langle F_L\rangle &= -\frac{\int^\prime\left(I_2^c+\bar I_2^c\right)}{\int^\prime\left(d\Gamma/dq^2+d\bar\Gamma/dq^2\right)}\,, \qquad
\langle A_{FB}\rangle = -\frac{\int^\prime\left[\frac{3}{4}\left(I_6^s+\bar I_6^s\right)+\frac{3}{8}\left(I_6^c+\bar I_6^c\right)\right]}{\int^\prime\left(d\Gamma/dq^2+d\bar\Gamma/dq^2\right)}\,, \nonumber\\[4pt]
\langle P_5^{\prime}\rangle &= \frac{\int^\prime\left(I_5+\bar I_5\right)}{2\sqrt{-\int^\prime\left(I_2^s+\bar I_2^s\right)\,\int^\prime\left(I_2^c+\bar I_2^c\right)}}\,, \qquad
\langle S_j\rangle = \eta_j\,\frac{\int^\prime\left(I_j+\bar I_j\right)}{\int^\prime\left(d\Gamma/dq^2+d\bar\Gamma/dq^2\right)}\,,
\label{eq:obsB2Kst}
\end{align}
together with the corresponding CP asymmetries
\begin{equation}
\langle A_j\rangle = \eta_j\,\frac{\int^\prime\left(I_j-\bar I_j\right)}{\int^\prime\left(d\Gamma/dq^2+d\bar\Gamma/dq^2\right)}\,,\qquad j=7,8,9\,,
\label{eq:AjKst}
\end{equation}
where $\eta_{7}=\eta_{9}=-1$ and $\eta_8=+1$. The signs in $\langle A_{FB}\rangle$ and $\eta_j$ correspond to the convention used by LHCb~\cite{LHCb:2025mqb}.\footnote{Relative to the convention of Ref.~\cite{Altmannshofer:2008dz} for the angular coefficients, the LHCb convention corresponds to $S_6^{s,c}\to-S_6^{s,c}$ and $S_{7,9}\to-S_{7,9}$ (and likewise $A_{7,9}\to-A_{7,9}$), while $S_{5}$ and $S_8$ ($A_8$) are unchanged.}

\subsection{$B_s \to  \mu^+\mu^-$}
\label{app:obs_Bs2mumu}
In the absence of scalar and pseudoscalar operators, the branching ratio is determined by the axial-vector quark current, i.e. by $C_{10}- C^{\prime}_{10}$:
\begin{equation}
\mathcal{B}(B_s\to\mu^+\mu^-) = \tau_{B_s}\,\frac{G_F^2\,\alpha_{\rm em}^2}{16\pi^3}\,\big|\lambda_t^{32}\big|^2 f_{B_s}^2\, m_{B_s}\, m_\mu^2
\sqrt{1-\frac{4m_\mu^2}{m_{B_s}^2}}\;\Big|C_{10}^{32,22}-C_{10}^{\prime\,32,22}\Big|^2 .
\label{eq:Bsmumu}
\end{equation}

Owing to the sizeable width difference of the $B_s$ system, the measured time-integrated branching ratio is related to Eq.~\eqref{eq:Bsmumu} by~\cite{DeBruyn:2012wk}
\begin{equation}
\overline{\mathcal{B}}(B_s\to\mu^+\mu^-) = \frac{1+\mathcal{A}_{\Delta\Gamma}\,y_s}{1-y_s^2}\;\mathcal{B}(B_s\to\mu^+\mu^-) ,
\qquad
y_s\equiv\frac{\tau_{B_s}\Delta\Gamma_s}{2} ,
\label{eq:Bsmumu-bar}
\end{equation}
with 
\begin{equation}
\mathcal{A}_{\Delta\Gamma}=\cos 2\varphi_P ,
\qquad
\varphi_P\equiv\arg\!\left(\frac{C_{10}^{32,22}-C_{10}^{\prime\,32,22}}{C_{10,\rm SM}^{32}}\right) ,
\label{eq:ADG}
\end{equation}
so that $\mathcal{A}_{\Delta\Gamma}=1$ in the SM.

\subsection{$K \to \pi \mu^+\mu^-$}
For the flavor changing neutral current rare decay in kaon sector $K^+ \to \pi^+ \mu^+\mu^-$,
the differential decay distribution in rescaled dilepton invariant mass squared $(z\equiv q^2/m_K^2)$ is given by
\begin{align}
\label{eq:Kpill}
\frac{d\Gamma}{dz}\,
&=\; \frac{G_F^2\alpha_{\rm em}^2m_K^5}{12\pi(4\pi)^4}\,
\sqrt{\bar{\lambda}}\;
\sqrt{1-4\frac{r_\mu^2}{z}}
\nn \\ 
&\times\; \Bigg\{
\bar{\lambda}\,\bigg(1+2\frac{r_\mu^2}{z}\bigg)
\bigg|V_+(z)+\sqrt{2}\,\lambda_t\,
f_+^{K\pi}(z)\,(C_{9,\,{\rm NP}}^{21,22}+C_{9,\,{\rm NP}}^{\prime\,21,22})\bigg|^2
\nn \\
&+2\,|\lambda_t|^2\,{|C_{10,\,{\rm NP}}^{21,22}+C_{10,\,{\rm NP}}^{\prime\,21,22}|}^2
\bigg[
\bar{\lambda}\,\bigg(1-4\frac{r_\mu^2}{z}\bigg)
\,|f_+^{K\pi}(z)|^2
+\frac{6r_\mu^2}{z}\,(1-r_\pi^2)^2
\,|f_0^{K\pi}(z)|^2
\bigg]
\Bigg\}\,.
\end{align}
where 
$\bar{\lambda} \equiv \lambda(1, z, r_\pi^2)$, with $r_i = m_i / m_K$. The Wilson coefficients can be defined as
\begin{equation}
    \begin{aligned}
        C_{9,\,{\rm NP}}^{\,21,22}
        &= [\widetilde{c}_{qe}]^{21,22}
        + [\widetilde{c}_{ql}^{(1)}]^{21,22}
        + [\widetilde{c}_{ql}^{(3)}]^{21,22}
        - \zeta\, [\widetilde{c}_{Z}]^{21} \,,
        \\
     C_{9,\,{\rm NP}}^{\prime\,21,22}
        &= [\widetilde{c}_{de}]^{21,22}
        + [\widetilde{c}_{dl}]^{21,22}
        - \zeta\, [\widetilde{c}_{Z}^{\prime}]^{21} \,, \\
        C_{10,\,{\rm NP}}^{21,22}
        &= [\widetilde{c}_{qe}]^{21,22}
        - [\widetilde{c}_{ql}^{(1)}]^{21,22}
        - [\widetilde{c}_{ql}^{(3)}]^{21,22}
        + [\widetilde{c}_{Z}]^{21} \,,
        \\
        C_{10,\,{\rm NP}}^{\prime\,21,22}
        &= [\widetilde{c}_{de}]^{21,22}
        - [\widetilde{c}_{dl}]^{21,22}
        + [\widetilde{c}_{Z}^{\prime}]^{21} \,.
    \end{aligned}
\end{equation}
In the limit where only one NP operator is present at a time, the expression for \(d\Gamma/dz\) in Eq.~\eqref{eq:Kpill} reduces to that obtained in Ref.~\cite{Mandal:2019gff}. The parametrization of the hadronic matrix elements and the kinematic conventions adopted here follow Ref.~\cite{Mandal:2019gff}.

The SM contribution to the decay $K^+ \to \pi^+ \mu^+ \mu^-$ is dominated by the
$\mathcal{CP}$-conserving one-photon exchange $K^+ \to \pi^+ \gamma^*$ and is described by a
vector form factor $V_+(z)$,
\begin{equation}
V_+(z) = a_+ + b_+\, z + V_{\pi\pi}(z)\,,
\end{equation}
where $V_{\pi\pi}(z)$ encodes the pion-loop contribution evaluated at $\mathcal{O}(p^6)$ in
$\chi$PT~\cite{Ecker:1987hd}. In our analysis we follow Ref.~\cite{DAmbrosio:1998gur} for the
parameterization of $V_{\pi\pi}(z)$. The parameters $a_+$ and $b_+$ can, in principle, be estimated within $\chi$PT in a model-dependent manner. In practice, however, they are typically extracted directly from experimental data. Accordingly, we adopt the values determined from the NA62 measurements of the branching fraction in the muon channel for $a_+$ and $b_+$~\cite{NA62:2022qes},
\begin{align}
a_+ = -0.575 \pm 0.012\,, \qquad
b_+ = -0.722 \pm 0.040\,.
\end{align}
The estimates for the $K_{3\ell}$ form factors $f^{K\pi}_{+,0}(z)$ are taken from Ref.~\cite{FlaviaNetWorkingGrouponKaonDecays:2008hpm}.
The branching fraction is then obtained by integrating Eq.~\eqref{eq:Kpill} in the entire kinematical range for $z$ given by $4r_{\mu}^{2} \leq z \leq (1 - r_{\pi})^{2}$.

\subsection{$K_L \to \pi^0 \mu^+\mu^-$ }

The decay $K_L \to \pi^0 \mu^+ \mu^-$ receives contributions from both $\mathcal{CP}$-conserving and $\mathcal{CP}$-violating effects within the SM. The $\mathcal{CP}$-conserving contribution, arising from the long-distance process $K_L \to \pi^0 \gamma \gamma \to \pi^0 \mu^+ \mu^-$, is known to be subdominant compared to the $\mathcal{CP}$-violating contributions. The latter include indirect $\mathcal{CP}$-violation from $K^0$--$\bar{K}^0$ mixing, as well as direct $\mathcal{CP}$-violation within the SM and potential additional contributions from NP couplings. 
Given the relative suppression of the $\mathcal{CP}$-conserving component, we neglect it in our analysis and express the differential decay distribution for $K_L \to \pi^0 \mu^+ \mu^-$ as
\begin{align}
\label{eq:KLpill}
\frac{d\Gamma}{dz}\,
\big(K_L^0\to\pi^0\mu^+\mu^-\big)_{\rm CPV}
&=\; \frac{G_F^2\alpha_{\rm em}^2m_K^5}{12\pi(4\pi)^4}\,
\sqrt{\bar{\lambda}}\;
\sqrt{1-4\frac{r_\mu^2}{z}}
\nn \\
&\times\; \Bigg\{
\bar{\lambda}\,\bigg(1+2\frac{r_\mu^2}{z}\bigg)
\bigg|V_0(z)+i\sqrt{2}\,
f_+^{K\pi}(z)\,
\Im\!\bigg[
\lambda_t\,
\big(C_{9,\,{\rm SM}}^{\rm 21}+C_{9,\,{\rm NP}}^{21,22}
+C_{9,\,{\rm NP}}^{\prime\,21,22}\big)
\bigg]\bigg|^2
\nn \\
&+2\,\bigg\{
\Im\!\bigg[
\lambda_t\,
\big(C_{10,\,{\rm SM}}^{\rm 21}
+C_{10,\,{\rm NP}}^{21,22}
+C_{10,\,{\rm NP}}^{\prime\,21,22}\big)
\bigg]\bigg\}^{2}
\nn \\
&\times\; \bigg[
\bar{\lambda}\,\bigg(1-4\frac{r_\mu^2}{z}\bigg)
\,|f_+^{K\pi}(z)|^2
+\frac{6r_\mu^2}{z}\,(1-r_\pi^2)^2
\,|f_0^{K\pi}(z)|^2
\bigg]
\Bigg\}\,.
\end{align}

The $\mathcal{CP}$-violating contribution fully dominated by the vector component of $K_S^0 \to \pi^0 \mu^+ \mu^-$ can be parametrized as

\begin{align}
V_0(z) \approx |\epsilon_K| e^{i\pi/4} a_S\left(1 + \frac{z}{r_V^2} \right)\,,
\end{align}
with $|\epsilon_K| = (2.228 \pm 0.011) \times 10^{-3}$.
As mentioned earlier, the branching fraction can be obtained by integrating the following kinematical range $4r_{\mu}^{2} \leq z \leq (1 - r_{\pi})^{2}$.

\subsection{$B \to K^{(*)} \nu\bar{\nu} $ }
\label{app:BtoKnunuexp}
The set of observables in dineutrino modes is limited compared to the charged-lepton channels discussed in Appendices~\ref{app:obs_B2Kmumu} and \ref{app:obs_B2Kstmumu}. The differential decay distribution with respect to the dineutrino invariant mass squared ($q^2$) for $B \to K \nu \bar{\nu}$ is given by~\cite{Altmannshofer:2009ma, Buras:2014fpa}
\begin{equation}
\label{eq:diffdis_b2knunu}
    \frac{d\Gamma}{dq^2}(B \to K \nu \bar{\nu}) = \frac{G_{F}^{2}\alpha_{\rm em}^2}{256\times 3\pi^5 m_B^3} |V_{tb} V_{ts}^{*}|^{2} \lambda^{3/2}(m_B^2, m_K^2, q^2) [f_+ (q^2)]^2 \sum_{\alpha=1}^3 \left|C_L^{ {32,\alpha\alpha}}+C_R^{ {32,\alpha\alpha}}\right|^2 .
\end{equation}
Similarly, for $B \to K^*\, \nu\, \bar{\nu}$,
\begin{equation}
\label{eq:diffdis_b2kstnunu}
    \frac{d\Gamma}{dq^2}(B \to K^*\, \nu\, \bar{\nu}) = \sum_{\alpha=1}^3 \big(|A_{\perp}^{\alpha \alpha}|^2 + |A_{\parallel}^{\alpha \alpha}|^2 + |A_0^{\alpha \alpha}|^2\big),
\end{equation}
where $A_{\lambda}^{\alpha \alpha}$ denote the transversity amplitudes, which are expressed in terms of the $B \to K^*$ form factors introduced in Appendix~\ref{app:obs_B2Kstmumu}, together with the corresponding Wilson coefficients~\cite{Altmannshofer:2009ma,Buras:2014fpa}.
\begin{align}
A_\perp^{\alpha \alpha}  &=  2N \sqrt{2} \tilde{\lambda}^{1/2} \bigg[ 
\left( C_L^{32,\alpha\alpha}  + C_R^{32,\alpha \alpha}  \right) \frac{ V(q^2) }{ m_B  + m_{K^*}} 
\bigg],\\[2ex]
A_\parallel^{\alpha \alpha}  & = - 2N \sqrt{2}(m_B  + m_{K^*})  \left( C_L^{32,\alpha \alpha} - C_R^{32,\alpha \alpha}  \right) 
A_1(q^2) ,\\
A_0^{\alpha \alpha}  &=  - \frac{N}{m_{K^*} \sqrt{q^2}}  \bigg\{ 
 \left( C_L^{32,\alpha \alpha} - C_R^{32,\alpha \alpha}  \right)
\bigg[ (m_B ^2 - m_{K^*}^2 - q^2) ( m_B  + m_{K^*}) A_1(q^2) 
 - \frac{\tilde{\lambda} A_2(q^2)}{m_B  + m_{K^*}}
\bigg] 
\bigg\}. 
\end{align}

\subsection{$K_{(L)} \to \pi \nu\bar\nu$ }
\label{app:obs_Knunu}

The branching fractions can be written as
\begin{equation}
\begin{aligned}
    \mathcal{B}(K^+\to\pi^+\nu\bar\nu) &= \frac{1}{3}\mathcal{B}(K^+\to\pi^+\nu\bar\nu)\big|_{\rm SM}  \sum_{\alpha=1}^3 \left|1+ \frac{ C_{L,\,{\rm NP}}^{21,\alpha\alpha}
      + C_{R,\,{\rm NP}}^{21,\alpha\alpha}  }{C^{21}_{L,\,\rm SM}}\right|^2 , \\
    \mathcal{B}(K_L \to \pi^0 \nu \bar{\nu}) &= \frac{1}{3}\mathcal{B}(K_L \to \pi^0 \nu \bar{\nu})\big|_{\rm SM}  \sum_{\alpha=1}^3 \left[ 1 + \frac{ \mathrm{Im}\, [ \lambda_t \left( C_{L,\,{\rm NP}}^{21,\alpha\alpha}
      + C_{R,\,{\rm NP}}^{21,\alpha\alpha} \right)]  }{\mathrm{Im}\, [ \lambda_t C^{21}_{L,\,\rm SM}]} \right]^2 .
\end{aligned}
\end{equation}
where the SM Wilson coefficient is given in Eq.~\eqref{eq:CL21SM} and the NP Wilson coefficients can easily be read from Eqs.~\eqref{eq:CLb2snunu} and \eqref{eq:CRb2snunu} as
\begin{align}
C_{L,\,{\rm NP}}^{21,\alpha\alpha}&= [\widetilde{c}_{ql}^{(1)}]^{21,\alpha \alpha} 
   - [\widetilde{c}_{ql}^{(3)}]^{21,\alpha \alpha} 
   + [\widetilde{c}_{Z}]^{21}\,, \\
C_{R,\,{\rm NP}}^{21,\alpha\alpha}&  =  [\widetilde{c}_{dl}]^{21,\alpha \alpha} 
   + [\widetilde{c}^\prime_{Z}]^{21}\,.
\end{align}

\section{Numerical inputs}
\label{input-para}
In this section, we summarize the inputs utilized throughout our numerical analysis. Table~\ref{tab:WC_inputs} compiles the central values and the corresponding uncertainties for the relevant fundamental constants and input parameters. Unless otherwise specified, all values are adopted from the PDG~\cite{PDG2024}.

\begin{table}[ht!]
\caption{Input parameters used in the analysis. Unless stated otherwise, values are taken from PDG~\cite{ParticleDataGroup:2024cfk}.}
\centering
\renewcommand*{\arraystretch}{1.1}
\begin{tabular}{l c c}
\hline
Description & Parameter & Value \\
\hline
Bottom quark mass ($\overline{\text{MS}}$ Scheme) & $m_b(m_b)$ & $4.183 \pm 0.004\GeV$ \\
Bottom quark mass ($\overline{\text{MS}}$ Scheme) & $\overline{m}_b(\mu)$ & $4.052\GeV$ \\
charm quark mass & $m_c(m_c)$ & $1.2730 \pm 0.008\hs \GeV$ \\
Pole masses in quark-loop functions$^{a}$ & $m_c^{\rm pole},\ m_b^{\rm pole}$ & $1.681\hs\GeV,\ 4.780\hs\GeV$ \\
Fine-structure constant & $\alpha_{em}(M_Z)$ & $1/127.930$ \\
Fine-structure constant (normalisation of $O_{9,10}$, $O_{L,R}$) & $\alpha_{em}(\mu)$ & $1/133$ \\
Strong coupling constant & $\alpha_s(M_Z)$ & $0.1180 \pm 0.0009$ \\
Strong coupling constant & $\alpha_s(\mu)$ & $0.2131$ \\
Weak mixing angle ($\overline{\text{MS}}$) & $\sin^2\theta_W$ & $0.23116$ \\
Fermi Constant & $G_{F}$ & $1.1663788 \times 10^{-5}\,\mathrm{GeV}^{-2}$ \\
muon mass & $m_\mu$ & $0.1056584\hs \GeV$ \\
Renormalization scale & $\mu$ & $5\,\mathrm{GeV}$ \\
B-meson mass & $m_B$ & $5.27941\hs \GeV$ \\
$B_s$-meson mass & $m_{B_s}$ & $5.36693\hs \GeV$ \\
$B^+$-meson lifetime & $\tau_{B^+}$ & $1.637 \times 10^{-12}\,\mathrm{s}$ \\
$B^0$-meson lifetime & $\tau_{B^0}$ & $1.5113 \times 10^{-12}\,\mathrm{s}$ \\
$B_s$-meson lifetime & $\tau_{B_s}$ & $1.515 \times 10^{-12}\,\mathrm{s}$ \\
$B_s$ decay constant$^{b}$ & $f_{B_s}$ & $0.2303\,\mathrm{GeV}$ \\
$K$-meson lifetime & $\tau_K$ & $1.2380 \times 10^{-8}\,\mathrm{s}$ \\
$K$-meson mass & $m_K$ & $0.493677\,\mathrm{GeV}$ \\
$K^*$-meson mass & $m_{K^*}$ & $0.89555\,\mathrm{GeV}$ \\
Pion mass & $m_{\pi}$ & $0.13957\,\mathrm{GeV}$ \\
CKM element & $|V_{tb} V_{ts}^*|$ & $4.106\times 10^{-2}$ \\
CKM element & $|V_{td}|$ & $8.58\times 10^{-3}$ \\
CKM ratio & $V_{ub}V_{us}^*/(V_{tb}V_{ts}^*)$ & $-0.00875+0.01848\,i$ \\
\hline
\multicolumn{3}{l}{\footnotesize $^{a}$ Two-loop conversion from $m_c(m_c)$ and $m_b(m_b)$. \quad $^{b}$ FLAG. }
\end{tabular}
\label{tab:WC_inputs}
\end{table}

\section{Flavio results and Correlation matrices}
For comparison with our fit results, this appendix presents the results obtained using the \texttt{flavio} package~\cite{Straub:2018kue} for the same set of observables. The theoretical predictions and their covariance matrix are taken from \texttt{flavio}.
In addition to the perturbative contributions to the Wilson coefficients included in our analysis, \texttt{flavio} accounts for QCD-factorisation spectator-scattering contributions and parametrizes hadronic power corrections through nuisance parameters. The resulting larger theoretical uncertainties lead to a somewhat smaller tension with the SM and slightly reduced NP shifts, while the best-fit values remain consistent with ours within uncertainties, as shown in Table~\ref{tab_bestfit_flavio}. We note that the overall pattern of preferred scenarios remains unchanged.
\label{app:corr}
\begin{table}
\caption{Same fit as Table~\ref{tab_bestfit_extended}, with all predictions of observables and the theory covariance from \texttt{flavio}
(including QCDF spectator scattering and hadronic power correction contribution parametrization as nuisance parameters).}
\centering
\setlength{\tabcolsep}{8pt}
\renewcommand{\arraystretch}{1.8}
\resizebox{\textwidth}{!}{
\begin{tabular}{l c c c}
\hline
SMEFT couplings & Best fit $(x_i, y_i)$ & $\chi^2_{\rm min}/{\rm d.o.f}$ & Correlation \\
\hline
SM & $-$ & 2.401 & $-$ \\
\hline
\cqlonebsT   & $[-0.581\pm0.133,\ -0.022^{+0.816}_{-0.818}]$ & 1.768 & $0.043$ \\
\cqlthreebsT & $[-0.575\pm0.132,\ -0.020^{+0.793}_{-0.798}]$ & 1.783 & $0.037$ \\
\czbsT       & $[1.053\pm0.270,\ 1.982\pm0.388]$             & 2.032 & $0.675$ \\
\cdlbsT      & $[-0.321\pm0.091,\ -0.009^{+0.597}_{-0.584}]$ & 2.129 & $-0.011$ \\
\czpbsT      & $[0.380\pm0.129,\ 0.002\pm0.574]$             & 2.263 & $0.005$ \\
\cqebsT      & $[-0.131\pm0.208,\ 0.406\pm0.626]$            & 2.524 & $0.655$ \\
\cdebsT      & $[0.128\pm0.161,\ -0.004\pm0.289]$            & 2.550 & $0.004$ \\
\hline
(\cqlonebsT, \czbsT)   & $[(-1.048\pm0.314\,,~-0.518\pm0.718\,),~(-0.493\pm0.479\,,~-1.960\pm0.622\,)]$ & 1.393 & Eq.~\eqref{eq:rho_flavio_cql1_cz} \\
(\cqlthreebsT, \czbsT) & $[(-1.024\pm0.310\,,~-0.516\pm0.682\,),~(-0.455\pm0.477\,,~-1.978\pm0.608\,)]$ & 1.427 & Eq.~\eqref{eq:rho_flavio_cql3_cz} \\
\hline
\end{tabular}}
\label{tab_bestfit_flavio}
\end{table}

The correlation matrices obtained in our 2D-fit scenarios are given by 
\begin{align}
\rho_{\left(
[\widetilde{c}_{ql}^{(1)}]^{32,22},\,
[\widetilde{c}_{z}]^{32}
\right)}
&=
\begin{array}{c|cccc}
 & {\rm Re}\,[\widetilde{c}_{ql}^{(1)}]^{32,22}
 & {\rm Im}\,[\widetilde{c}_{ql}^{(1)}]^{32,22}
 & {\rm Re}\,[\widetilde{c}_{z}]^{32}
 & {\rm Im}\,[\widetilde{c}_{z}]^{32} \\ \hline
{\rm Re}\,[\widetilde{c}_{ql}^{(1)}]^{32,22}
& \phantom{-}1     & -0.030 & 0.936 & 0.749 \\
{\rm Im}\,[\widetilde{c}_{ql}^{(1)}]^{32,22}
& -0.030 & \phantom{-}1     & 0.153 & 0.379 \\
{\rm Re}\,[\widetilde{c}_{z}]^{32}
& \phantom{-}0.936 & \phantom{-}0.153 & 1     & 0.813 \\
{\rm Im}\,[\widetilde{c}_{z}]^{32}
& \phantom{-}0.749 & \phantom{-}0.379 & 0.813 & 1
\end{array}
\label{eq:corr_cql1_cz}
\\[12pt]
\rho_{\left(
[\widetilde{c}_{ql}^{(3)}]^{32,22},\,
[\widetilde{c}_{z}]^{32}
\right)}
&=
\begin{array}{c|cccc}
 & {\rm Re}\,[\widetilde{c}_{ql}^{(3)}]^{32,22}
 & {\rm Im}\,[\widetilde{c}_{ql}^{(3)}]^{32,22}
 & {\rm Re}\,[\widetilde{c}_{z}]^{32}
 & {\rm Im}\,[\widetilde{c}_{z}]^{32} \\ \hline
{\rm Re}\,[\widetilde{c}_{ql}^{(3)}]^{32,22}
& 1     & 0.796 & 0.799 & 0.674 \\
{\rm Im}\,[\widetilde{c}_{ql}^{(3)}]^{32,22}
& 0.796 & 1     & 0.468 & 0.597 \\
{\rm Re}\,[\widetilde{c}_{z}]^{32}
& 0.799 & 0.468 & 1     & 0.837 \\
{\rm Im}\,[\widetilde{c}_{z}]^{32}
& 0.674 & 0.597 & 0.837 & 1
\end{array}
\label{eq:corr_cql3_cz}
\end{align}

The correlation matrices for the same choice of 2D-fit scenarios obtained using \texttt{flavio} read 
\begin{align}
\rho^{\texttt{flavio}}_{\left(
[\widetilde{c}_{ql}^{(1)}]^{32,22},\,
[\widetilde{c}_{z}]^{32}
\right)}
&=
\begin{pmatrix}
\phantom{-}1 & -0.625 & \phantom{-}0.818 & -0.567 \\
-0.625 & \phantom{-}1 & -0.344 & \phantom{-}0.486 \\
\phantom{-}0.818 & -0.344 & \phantom{-}1 & -0.727 \\
-0.567 & \phantom{-}0.486 & -0.727 & \phantom{-}1
\end{pmatrix},
\label{eq:rho_flavio_cql1_cz}
\\[12pt]
\rho^{\texttt{flavio}}_{\left(
[\widetilde{c}_{ql}^{(3)}]^{32,22},\,
[\widetilde{c}_{z}]^{32}
\right)}
&=
\begin{pmatrix}
\phantom{-}1 & -0.610 & \phantom{-}0.819 & -0.556 \\
-0.610 & \phantom{-}1 & -0.338 & \phantom{-}0.482 \\
\phantom{-}0.819 & -0.338 & \phantom{-}1 & -0.720 \\
-0.556 & \phantom{-}0.482 & -0.720 & \phantom{-}1
\end{pmatrix},
\label{eq:rho_flavio_cql3_cz}
\end{align}

\bibliographystyle{iopart-num}
\bibliography{biblio}

\end{document}